\documentclass[useAMS,usenatbib,A4paper]{mn2e}

\usepackage{times}
\usepackage{natbib}
\usepackage{lscape}
\usepackage[usenames]{color}
\usepackage{graphicx}
\usepackage{amssymb}
\usepackage{wasysym}
\usepackage{pifont}
\usepackage{ulem}

\long\def\Ignore#1{\relax}

\newcommand{\degrees} {^\circ}
\newcommand{\kms}{\mbox{${\rm km\, s^{-1}}$}}
\newcommand{\Msun}{\mbox{$\rm M_{\odot}$}}

\newcommand{\atlastd}{ATLAS$^{3\rm{D}}$}

\newcommand{\ie}{{\it i.e.}}

\newcommand{\cut}[1]{}

\title[Formation of stellar nuclear discs]{The formation of stellar
  nuclear discs in bar-induced gas inflows} \author[D. R. Cole et
al.]{David R. Cole$^{1}$\thanks{E-mail: {\tt drdrcole@gmail.com}},
  Victor P. Debattista$^{1}$\thanks{E-mail: {\tt
      vpdebattista@gmail.com}}, Peter Erwin$^{2}$\thanks{E-mail: {\tt
      erwin@mpe.mpg.de}}, Samuel W. F. Earp$^{1}$\thanks{E-mail: {\tt
      swfearp@gmail.com}}, Rok Ro{\v s}kar$^{3}$\thanks{E-mail: {\tt
      roskar@physik.uzk.ch}} \\ $^{1}$Jeremiah Horrocks Institute,
  University of Central Lancashire, Preston, PR1 2HE, United Kingdom
  \\ $^{2}$Max-Planck-Insitut f{\"u}r extraterrestrische Physik,
  Giessenbachstrasse, D-85748 Garching, Germany;
  Universit{\"a}ts-Sternwarte M{\"u}nchen, \\ Scheinerstrasse 1,
  D-81679 M{\"u}nchen, Germany \\ $^{3}$Institute for Computational
  Science, University of Z{\"u}rich, Wintherthurerstrasse 190,
  Z{\"u}rich, CH-8057, Switzerland}
\begin{document}

\date{Accepted xxx Received xxx ; in original form \today}

\maketitle

\label{firstpage}


\begin{abstract}
  The role of gas in the mass assembly at the nuclei of galaxies is
  still subject to some uncertainty.  Stellar nuclear discs 
  bridge the gap between the large-scale galaxy and the central
  massive objects that reside there.  Using a high resolution
  simulation of a galaxy forming out of gas cooling and settling into
  a disc, we study the formation and properties of nuclear discs.
  Gas, driven to the centre by a bar, settles into a rotating
  star-forming nuclear disc (ND).  This ND is thinner, younger,
  kinematically cooler, and more metal-rich than the surrounding bar.
  The ND is elliptical and orthogonal to the bar.  The complex
  kinematics in the region of the ND are a result of the superposition
  of older stars streaming along the bar and younger stars circulating
  within the ND.  The signature of the ND is therefore subtle in the
  kinematics.  Instead the ND stands out clearly in metallicity and
  age maps.  We compare the model to the density and kinematics of
  real galaxies with NDs finding qualitative similarities.  Our
  results suggest that gas dissipation is very important for forming
  nuclear structures.
\end{abstract}

\begin{keywords}
  galaxies: bulges --- galaxies: evolution --- galaxies:kinematics and
  dynamics --- galaxies: nuclei ---galaxies: structure
\end{keywords}


\section{Introduction}  
\label{sec:intro}

The central regions of galaxies host a wide variety of structures.  On
the largest scales, pseudo-bulges are thought to form from the secular
evolution of discs \citep[see the review of][]{Kormendy2004}.  At the
opposite extreme are supermassive black holes and nuclear star
clusters.  The growth of these is still a matter of debate.  Nuclear
star clusters show multiple populations \citep{Schinnerer2003,
  Rossa2006, Walcher2006, Seth2010, Lyubenova2013} often with episodes
of star formation within the last 100 million years, suggesting that
gas is able to reach the centres of galaxies \citep{Seth2006,
  Seth2008, Hartmann2011, DeLorenzi2013}.

Disc-like structures (by their discy isophotes) are
  observed in the centres of galaxies at a wide range of scales. They
  are distinct in that they lie outside of the region where light from
  the main disc dominates. They are smaller than the bulge but often
  extending beyond the central nuclear star cluster
  \citep[e.g.][]{Balcells2007}. (See Section \ref{sec:compobs} for
  examples of nuclear discs (NDs) in real galaxies.) NDs are found in
spiral galaxies \citep{Zasov1999, Pizzella2002, Dumas2007,
  GarciaBurillo2012} and in 20 per cent of early-type galaxies
\citep{Scorza1998, Kormendy2001, deZeeuw2002, Emsellem2004,
  Trujillo2004, Krajnovic2008, Ledo2010}.  \citet{Krajnovic2008} found
NDs in early-type galaxies are associated with fast rotators.
Near-infrared {\it Hubble Space Telescope} ({\it HST}) imaging has
also provided indirect evidence for NDs as discy isophotes in the
nuclei of early-type galaxies \citep{Ravindranath2001} or
photometrically distinct exponential components in bulges
\citep{Balcells2003}. \citet{Erwin2014b} also found evidence for NDs
which they call ``discy pseudobulges''.

Understanding the formation of NDs is very important for a clearer
picture of the assembly of mass at the centres of galaxies.  The
formation of NDs is generally thought to require in-situ star
formation. A number of ideas about how gas can be funnelled to the
central regions of galaxies have been proposed.  Observations of
luminous and ultra-luminous infrared galaxies have found NDs in their
nuclei.  These have masses in the range $10^8$ to $10^{10}\Msun$,
effective radii of a few hundred parsecs and $v/\sigma = 1-5$
\citep{Medling2014}.  \citet{Medling2014} conclude that mergers funnel
gas to the centre of the galaxy prior to star formation.  Likewise,
simulations have predicted that NDs are able to form in gas-rich
galaxy mergers \citep{Mayer2008, Mayer2010, Chapon2013}.
\citet{McDermid2006} found NDs and counter-rotating cores in the
central kpc of early-type galaxies and evidence for recent
circum-nuclear star formation.  The presence of dynamically decoupled
features such as a counter-rotating disc in NGC 4458
\citep{Morelli2004, Morelli2010} and a disc rotating perpendicular to
the main galactic disc \citep{Bertola1999b, Pizzella2002, Corsini1999,
  Corsini2012} can be explained by the capture of external gas.
However interactions may not be the main mechanisms driving gas to
nuclei in more isolated galaxies, where observations show that there
is continuing star formation in some NDs e.g. in NGC 5845
\citep{Kormendy1994} and NGC 4486A \citep{Kormendy2005} and their
stellar populations have both young and old components
\citep{vandenBosch1998, Krajnovic2004, Morelli2004, Corsini2012}.
Mechanisms that have been proposed for driving gas to small radii in
such systems include nested bars \citep{Shlosman1989}, the
magnetorotational instability \citep{Milosavljevic2004}, and
cloud-cloud mergers \citep{Bekki2007}.  \citet{Agarwal2011} proposed
that NDs can form out of the debris of infalling star clusters and
\citet{Portaluri2013} showed that the available photometric and
kinematic data are still consistent with this idea.  However, detailed
modelling \citep{DeLorenzi2013} and comparison to simulations
\citep{Hartmann2011} of the kinematics of the nuclear star cluster in
NGC~4244 \citep{Seth2008} reveal that gas dissipation had to have
played a major role in the formation of its nuclear cluster,
indicating that some gas must be able to reach the inner $\sim 10$
parsecs.

Possibly associated with NDs are $\sigma$-drops which are galaxies
having a significant drop in velocity dispersion at the nucleus
\citep[see for example][]{Emsellem2001}. One explanation for these
$\sigma$-drops is that there is an inflow of gas to the nuclear region
which creates a dynamically cool disc where stars form, reducing the
central velocity dispersion \citep{Wozniak2003,Comeron2008}.  In
face-on galaxies, \citet{MendezAbreu2014} imaged small NDs co-spatial
with the region of the $\sigma$-drops.

Bars are thought to be a mechanism for driving gas to small radii.
\citet{FalconBarroso2004} found that this was a natural explanation
for the presence of a ND in the edge-on S0 galaxy NGC 7332,
  where the presence of a bar is inferred from a boxy/peanut shape
\citep{Seifert1996}.  Barred spiral galaxies have more molecular gas
in their central kiloparsec than unbarred galaxies
\citep{Sakamoto1999,Sheth2005}.  \citet{Wang2012} examined a sample of
over 3700 face-on disc galaxies and found that there is a correlation
between the presence of strong bars and centrally enhanced star
formation. They also found that the increase in star formation depends
primarily on the ellipticity of the bar and not on the size of the bar
or on the mass and structure of the host galaxy.
\citet{Dumas2007}, studying active and non-active
galaxies, found that NDs are associated with enhanced AGN
activity.  \citet{Hao2009} found that the bar fraction was higher in
galaxies hosting an AGN and in star forming galaxies than in inactive
galaxies.  \citet{Hicks2013} showed that Seyfert galaxies have a more
centrally concentrated nuclear stellar and molecular gas surface
brightness profile probably due to a ND-like structure composed of gas
and stars in a region out to 250 pc.

Here we study the morphology and evolution of a simulation of a disc
galaxy using smooth particle hydrodynamics (SPH) at high resolution.
The model of this galaxy developed in isolation and so the resulting
features are purely due to internal origin. We concentrate on the
central region of the model.  Section \ref{sec:simulation} describes
the simulation methods, Section \ref{sec:description} presents the
global morphology and star formation history of the model, Section
\ref{sec:cdisc} then focusses on the properties of the nuclear stellar
disc, while Section \ref{sec:gasdist} discusses the properties of the
gas in the nuclear region. Section \ref{sec:compobs} compares our
model to several early-type galaxies.  Finally Section
\ref{sec:discuss} presents our conclusions.


\section{The simulation}
\label{sec:simulation}

The simulation we consider here is the same as model HG1 of
\citet{Gardner2014} and the model studied by \citet{Ness2014}.  The
model has a disc galaxy forming inside a corona of pressure-supported
gas embedded in a dark matter halo, a technique we have used
extensively for studying disc galaxy evolution \citep{Roskar2008,
  Roskar2012}.  The main computational difference here with respect to
that work is that we have employed higher mass resolution, which
allows a higher star formation threshold and an increased supernova
feedback coupling.  The dark matter halo consists of $5 \times 10^6$
dark matter particles in two mass species: $4.5\times 10^6$ particles
of mass $m_p = 8.5 \times 10^4 \Msun$ while the remainder have a mass
of $1.7 \times 10^6 \Msun$.  The dark matter particles are arranged in
two mass shells, with particles initially inside 56 kpc having the low
mass and those outside having the larger mass. This arrangement allows
us to increase resolution inside 56 kpc. These particles all have a
force softening of $\epsilon = 103$ pc.  The halo has virial radius
$r_{200} = 198$ kpc, concentration $c = 19$ and and virial mass
$M_{200} = 9.0 \times 10^{11} \Msun$.  Mixed in with the dark matter
is a hot gas corona, consisting of $5 \times 10^6$ particles,
initially in pressure equilibrium.  Gas particles have $\epsilon = 50$
pc and, initially, all have equal mass $2.7 \times 10^4 \Msun$.  The
corona has the same density profile normalised to a total mass 11 per
cent that of the dark halo.  We give the gas angular momentum with
$L_z \propto R$ such that $\lambda\approx0.041$.  No stellar
particles are present at the start of the simulation since all stars form
out of gas that cools and reaches a density high enough to trigger
star formation.

The simulation was evolved with the $N$-body$+$smooth particle
hydrodynamics (SPH) code {\sc gasoline} \citep{Wadsley2004}.  We use a
base timestep of 10 Myr with a refinement parameter $\eta = 0.175$,
and an opening angle of $\theta = 0.7$.  The timestep of gas particles
also satisfies the condition $\delta t_{gas} = \eta_{courant}
h/[(1+\alpha)c + \beta\mu_{max}]$, where $\eta_{courant} = 0.4$, $h$
is the SPH smoothing length, $\alpha$ is the shear coefficient, which
is set to 1, $\beta=2$ is the viscosity coefficient and $\mu_{max}$ is
described in \citet{Wadsley2004}. $\eta_{courant}$ is the refinement
parameter for SPH particles and controls their timestep size. The SPH
kernel is defined using the 32 nearest neighbours.  Gas cooling is
calculated without taking into account the gas metallicity.  

We use the gas cooling, star formation and stellar feedback
  prescriptions of \citet{Stinson2006}. A gas particle undergoes star
formation if it has number density $n > 100$ $\mathrm{cm}^{-3}$,
temperature $T < 15,000$ K and is part of a converging flow;
efficiency of star formation is 0.1, \ie\ 10 per cent of gas eligible
to form stars spawn stars per dynamical time.  Star particles form
with an initial mass of 35 per cent that of the gas particle, which at
our resolution corresponds to $9.4 \times 10^3 \Msun$.  Gas particles
can spawn multiple star particles but once they drop below 21 per cent
of their initial mass the remaining mass is distributed amongst the
nearest neighbors, leading to a decreasing number of gas particles.
Each star particle represents an entire stellar population with a
Miller-Scalo \citep{Miller1979} initial mass function. The evolution
of star particles includes feedback from type II and type Ia
supernovae, with their energy injected into the interstellar medium
(ISM). The effect of the supernovae explosions is modeled at the
sub-grid level as a blastwave propagating through the ISM
\citep{Stinson2006}. As in \citet{Governato2010}, we assume that $0.4
\times 10^{51}$ ergs of energy per supernova couple to the ISM.  We
also include feedback from AGB stellar winds.  The gas corona has zero
metallicity to start with; we track the production of iron and oxygen
in the simulation using the yields of \citet{Woosley1995}.  Diffusion
of metals \citep[e.g.][]{Loebman2011} between gas particles was not
used in this simulation.  We also do not include any feedback from an
AGN since the model does not contain a supermassive black hole.


\section{Global properties of the model}
\label{sec:description}

By the end of the simulation, at $\sim 10$ Gyr, the stellar disc
consists of $\sim 1.1 \times 10^7$ particles.  The total stellar mass
of the galaxy is $\sim 6.5\,\times\,10^{10} \Msun$, typical of an
$L_*$ galaxy such as the Milky Way. It is not, however, an exact
analogue of the Milky Way as the scale-length of the disc is smaller,
at $R_{\rm d} \simeq 1.7$ kpc, whereas the Milky Way has $1.8 \leq
R_{\rm d} \leq 4.0$ kpc \citep[e.g.][]{Ojha2001, Chang2011,
  McMillan2011}.  As shown in \citet{Gardner2014}, the bar forms a
clear box/peanut shape as seen from the side, and an X-shape along the
line-of-sight when observed like the Milky Way.  \citet{Ness2014}
showed further that the bulge stars have a range of ages, as observed
in the Milky Way \citep{Bensby2011, Bensby2013, Valenti2013}.

\begin{figure}
\centering
\begin{tabular}{c}
\includegraphics[width=0.7\hsize,angle=-90]{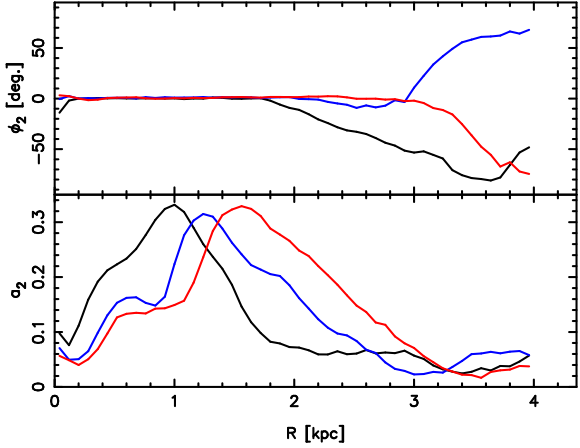} 
\end{tabular}
\caption{The $m=2$ Fourier amplitude, $a_2$, (bottom) and relative
  phase, $\phi_2$ (top) of the stellar density distribution at 6
  (black lines), 8 (blue lines) and 10 Gyr (red lines).  }
\label{fig:barm2}
\end{figure}

\begin{figure}
\centering
\begin{tabular}{c}
\includegraphics[width=0.95\hsize,angle=0]{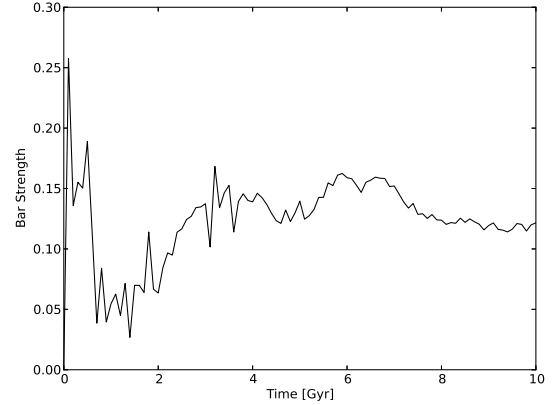} 
\end{tabular}
\caption{Evolution of the bar strength, defined as the amplitude of
  the $m=2$ Fourier moment. }
\label{fig:barstrength}
\end{figure}

The amplitude $a_2$ and relative phase $\phi_2$ of the $m=2$
Fourier component are shown in Fig.  \ref{fig:barm2}. The peak of
$a_2$, and the location where $\phi_2$ deviates from a constant,
increase with time, indicating that the bar is growing longer.  By 10
Gyr the bar is $\sim 3$ kpc long, increasing from $\sim 2$ kpc at 6
Gyr.  The evolution of the bar strength, defined as the amplitude of
the global $m=2$ Fourier moment \citep[e.g.][]{Debattista2000}, is
shown in Fig.  \ref{fig:barstrength}. The bar forms at $t \sim 3.2$
Gyr; its strength starts to grow at 5.2 Gyr and peaks at 5.8 Gyr and
again at 6.5 Gyr.

\begin{figure}
\centering
\begin{tabular}{c}
\includegraphics[width=0.95\hsize,angle=0]{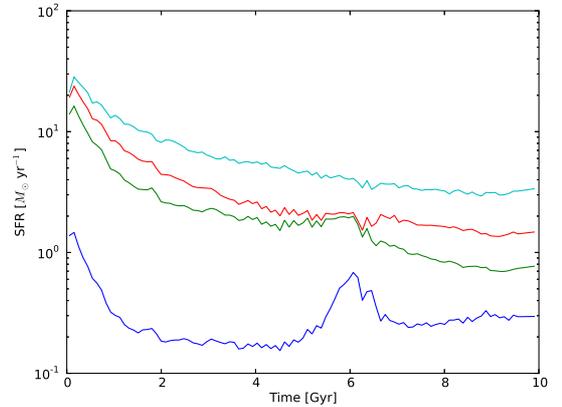} 
\end{tabular}
\caption{ Star formation history of the model. The cyan line is the
  global history, the red, green and blue lines are the star formation
  histories inside 1 kpc, 500 pc and 100 pc, respectively. Note the
  peaks at $\sim 6$ Gyr for the inner two regions.}
\label{fig:sfh}
\end{figure}

Fig. \ref{fig:sfh} shows the global star formation history as well as
that inside a number of radii.  The global star formation rate is
initially high, dropping rapidly to $\sim 3\Msun$ per year.  It does
not show any maxima, in contrast to the region inside 500 pc, which
has an episode of enhanced star formation at $\sim 6$ Gyr.

\section{Nuclear stellar disc}
\label{sec:cdisc}

\begin{figure*}
\centering
\begin{tabular}{c}
\includegraphics[width=0.33\hsize,angle=0]{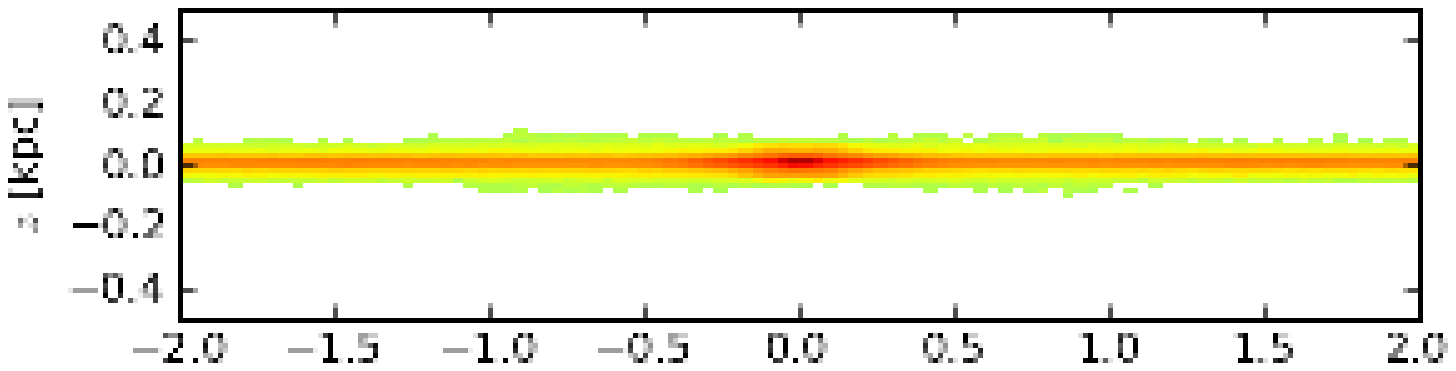} 
\includegraphics[width=0.33\hsize,angle=0]{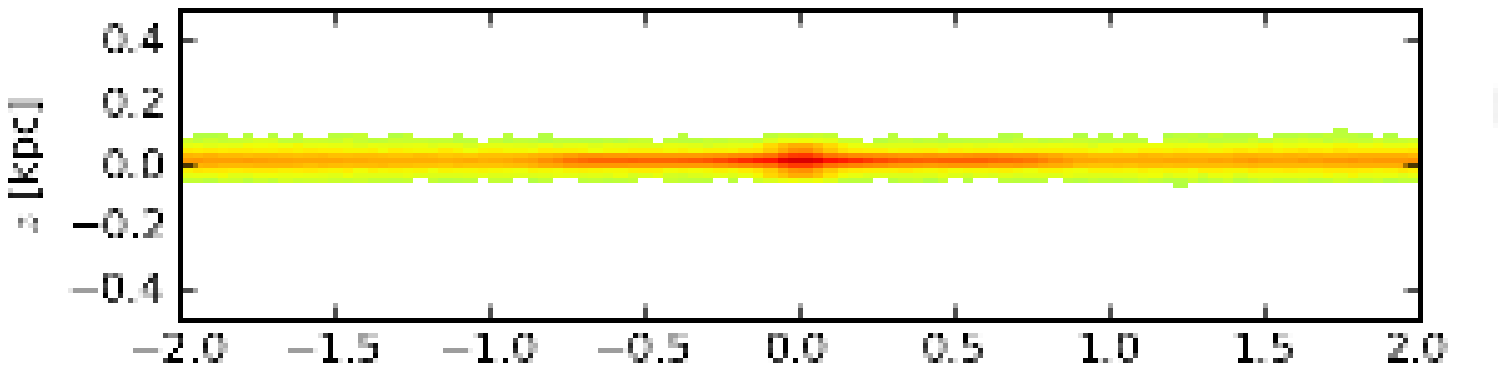} 
\includegraphics[width=0.33\hsize,angle=0]{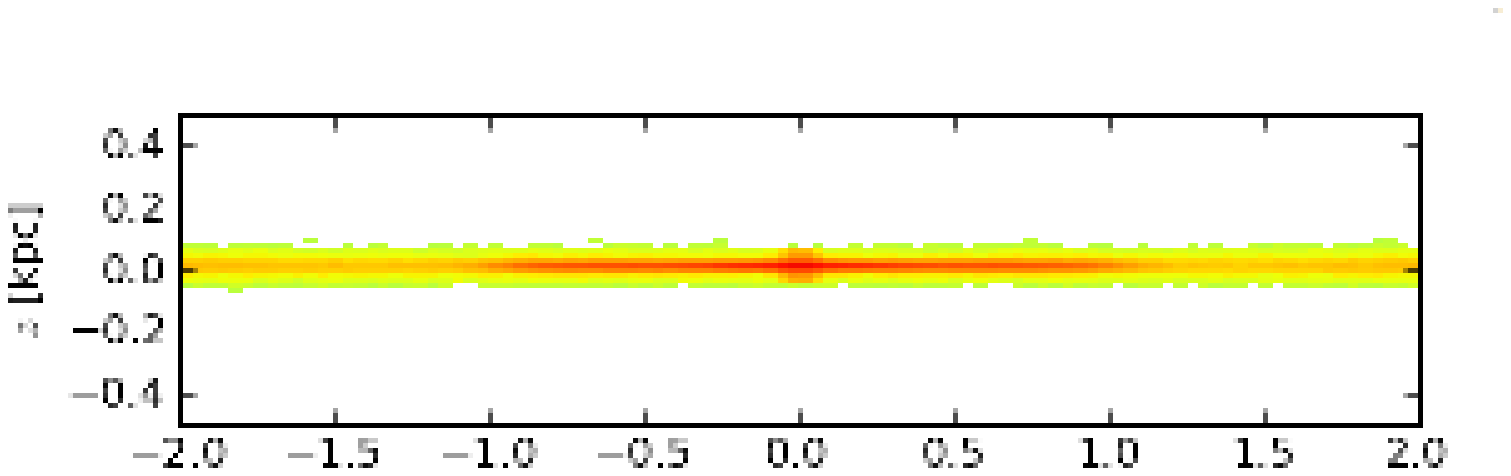} \\
\includegraphics[width=0.33\hsize,angle=0]{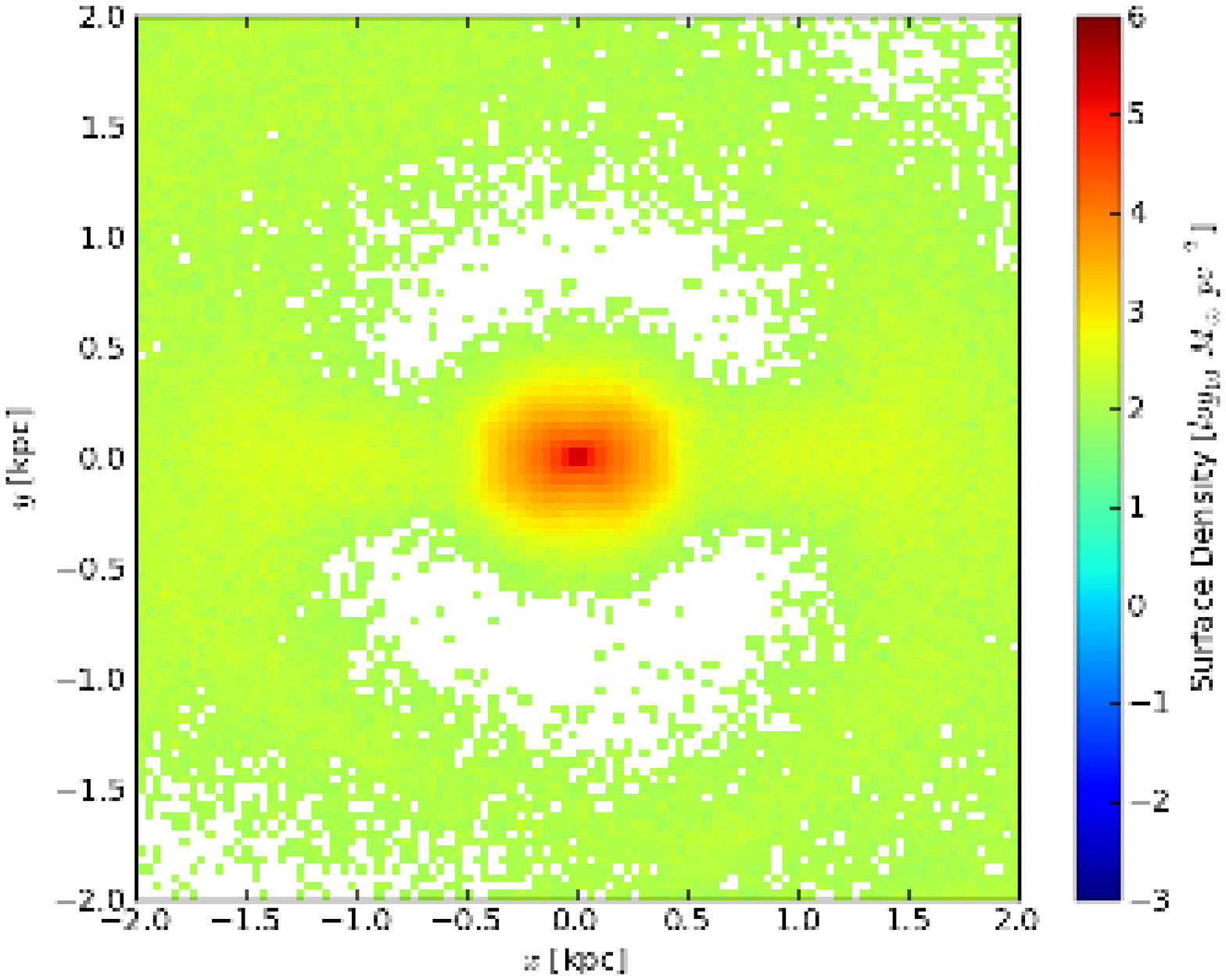} 
\includegraphics[width=0.33\hsize,angle=0]{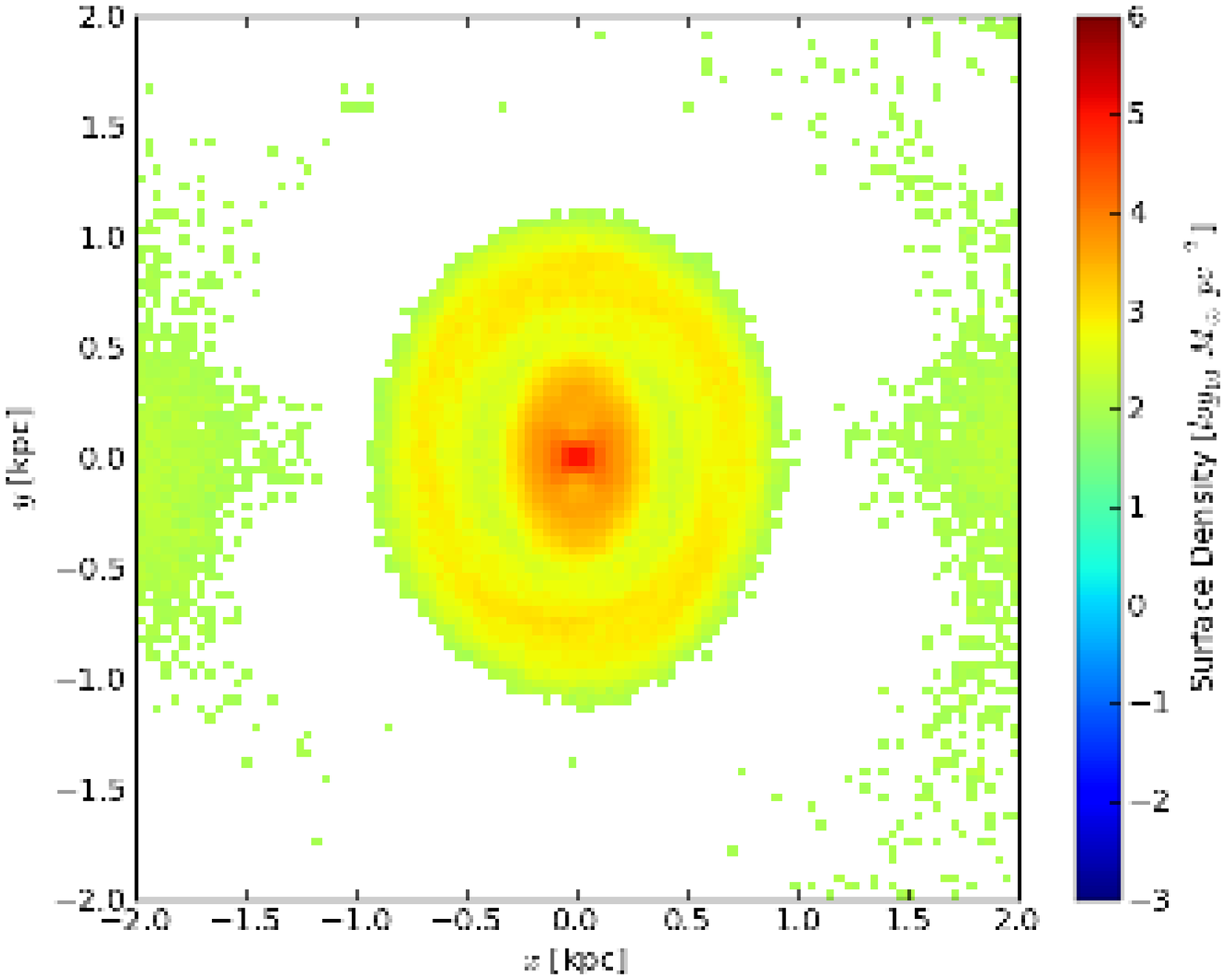} 
\includegraphics[width=0.33\hsize,angle=0]{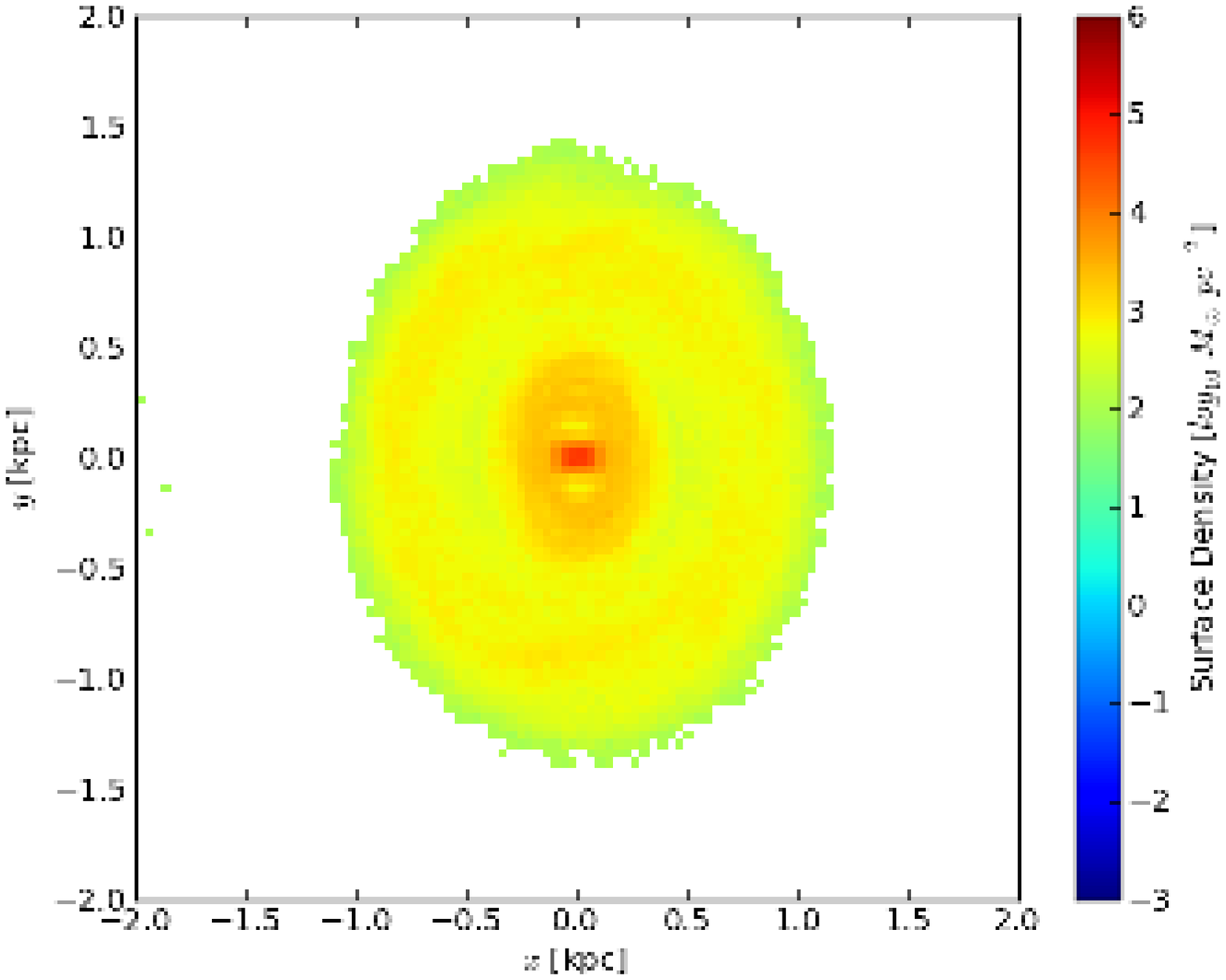} 
\end{tabular}
\caption{ Face-on (bottom) and edge-on (top) surface density for the
  stars younger than 2 Gyr at 6 Gyr (left), 8 Gyr (centre) and 10 Gyr
  (right).  The bar is along the $x$-axis in all panels.}
\label{fig:ageden}
\end{figure*}

Fig. \ref{fig:sfh} showed that the central 500 pc experiences
increased star formation activity around $\sim 6$ Gyr, compared with
the rest of the simulation.  We now turn our attention to the young
stars within this central region. Fig.  \ref{fig:ageden} shows the
surface density of stars younger than 2 Gyr at 6, 8 and 10 Gyr.  At 10
Gyr an elliptical, thin disc of stars is present inside $\sim 1.5$
kpc. The ND is perpendicular to the main bar, which is
horizontal\footnote{In this paper we always show the system with the
  disc in the $(x,y)$ plane and the bar rotated into the $x$-axis,
  except where noted.} in Fig.  \ref{fig:ageden} and is therefore most
likely supported by the x2 family of bar orbits, which is elongated in
this way \citep{Skokos2002}.  (The x3 orbits, which are also elongated
perpendicular to the bar, are generally unstable
\citep[e.g.][]{Sellwood1993} and very unlikely to be populated.)  The
ND can also be seen at 8 Gyr, but not at 6 Gyr. {At 10 Gyr the
  ND is bounded by a nuclear ring which is part of the ND. There is a
  second nuclear ring at semi-major axis of $\sim 300$ parsec which is
  not very well resolved on its minor axis. These nuclear rings can be
  seen in the figures comparing our simulation to real galaxies in
  Section \ref{sec:compobs}.

A lower limit on the mass of the ND at 10 Gyr is derived by
considering the mass of stars younger than 3 Gyr enclosed within 1.5
kpc.  This mass is $5.5 \times 10^9 \Msun$.  The bottom panel of Fig.
\ref{fig:profiles} plots the azimuthally-averaged surface-density
profile of stars.  The profile is well-approximated by an exponential
with scale-length 220 parsecs at 8 Gyr, increasing to 250 parsecs at
10 Gyr.  The corresponding ND mass is $2.5 \times 10^{10} \Msun$ at 8
Gyr and $3.1 \times 10^{10} \Msun$ at 10 Gyr.  For later comparison,
we have also measured the density profiles for slits along the
line-of-nodes for the system orientated the same way as NGC~3945
(see Section \ref{sec:n3945} comparing our model to NGC 3945); these
profiles are shown in the top panel of Fig.  \ref{fig:profiles}.
Doing this we obtain a mass of $1.6 \times 10^{10} \Msun$ at 8 Gyr and
$1.9 \times 10^{10} \Msun$ at 10 Gyr, the latter number corresponding
to 29 per cent of the stellar mass of the model.

\begin{figure}
\centering
\begin{tabular}{c}
\includegraphics[width=\hsize,angle=0]{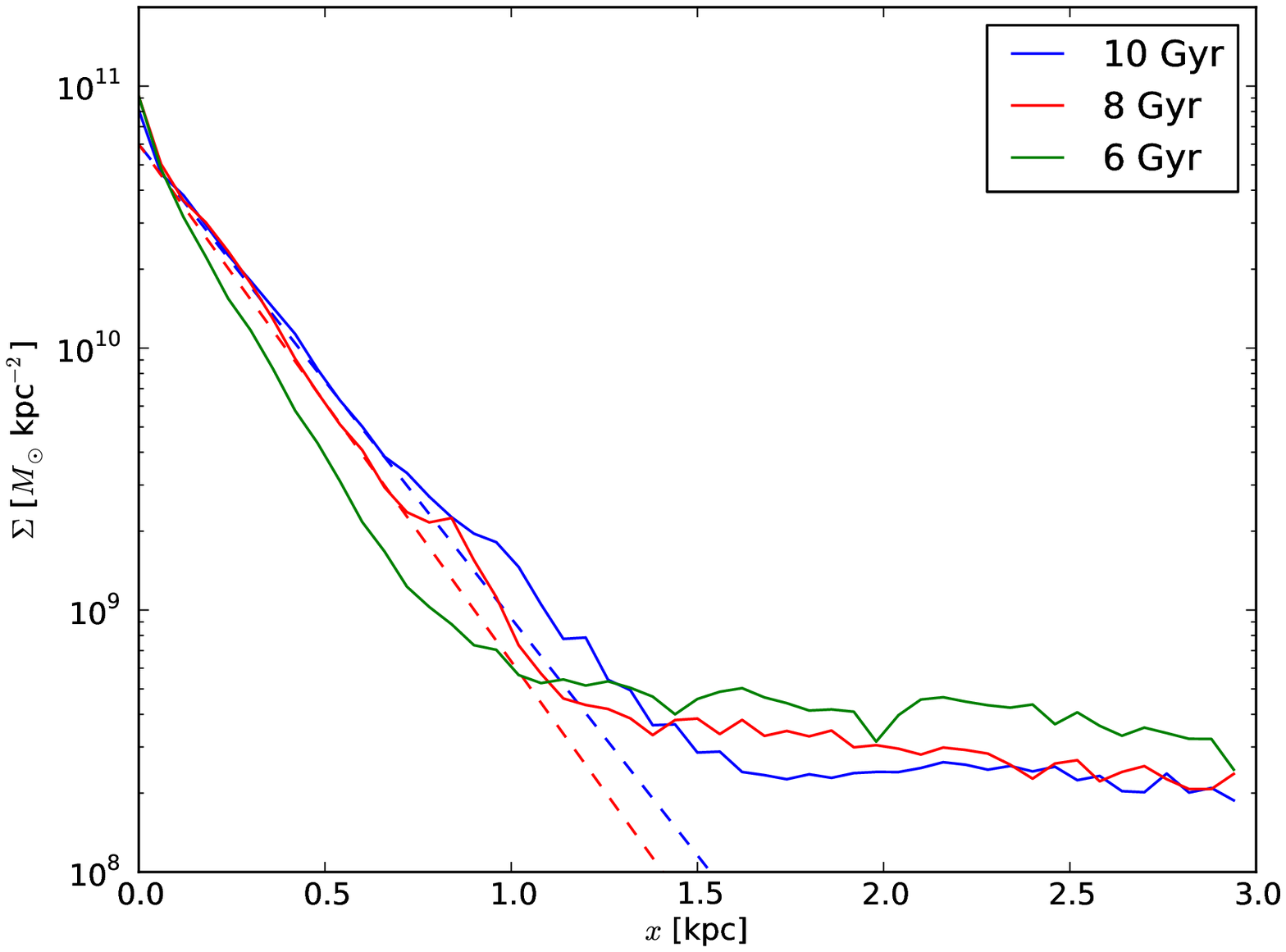}  \\
\includegraphics[width=\hsize,angle=0]{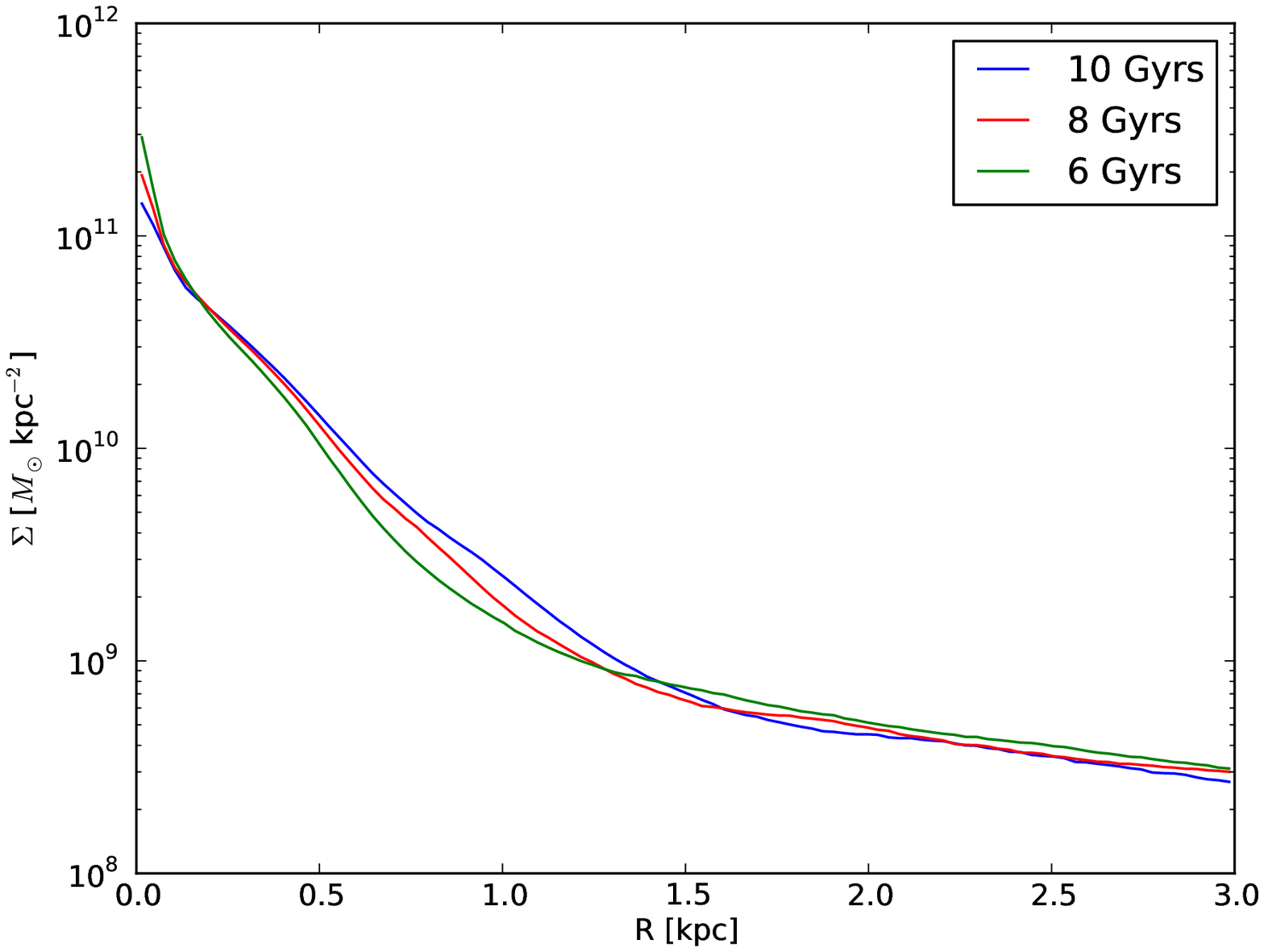} \\
\end{tabular}
\caption{Stellar surface density profiles of the central regions at 6,
  8 and 10 Gyr as indicated.  Top: Surface densities along the
  major-axis of the ND with our model inclined at 55 degrees the same
  as NGC 3945.  The dashed lines indicate exponential profiles.
  Bottom: Azimuthally averaged surface density profiles for the
  face-on orientation. }
\label{fig:profiles}
\end{figure}

\subsection{Kinematics}
\label{ssec:kine}

\begin{figure*}
\centering
\begin{tabular}{c}
\includegraphics[width=0.5\hsize,angle=0]{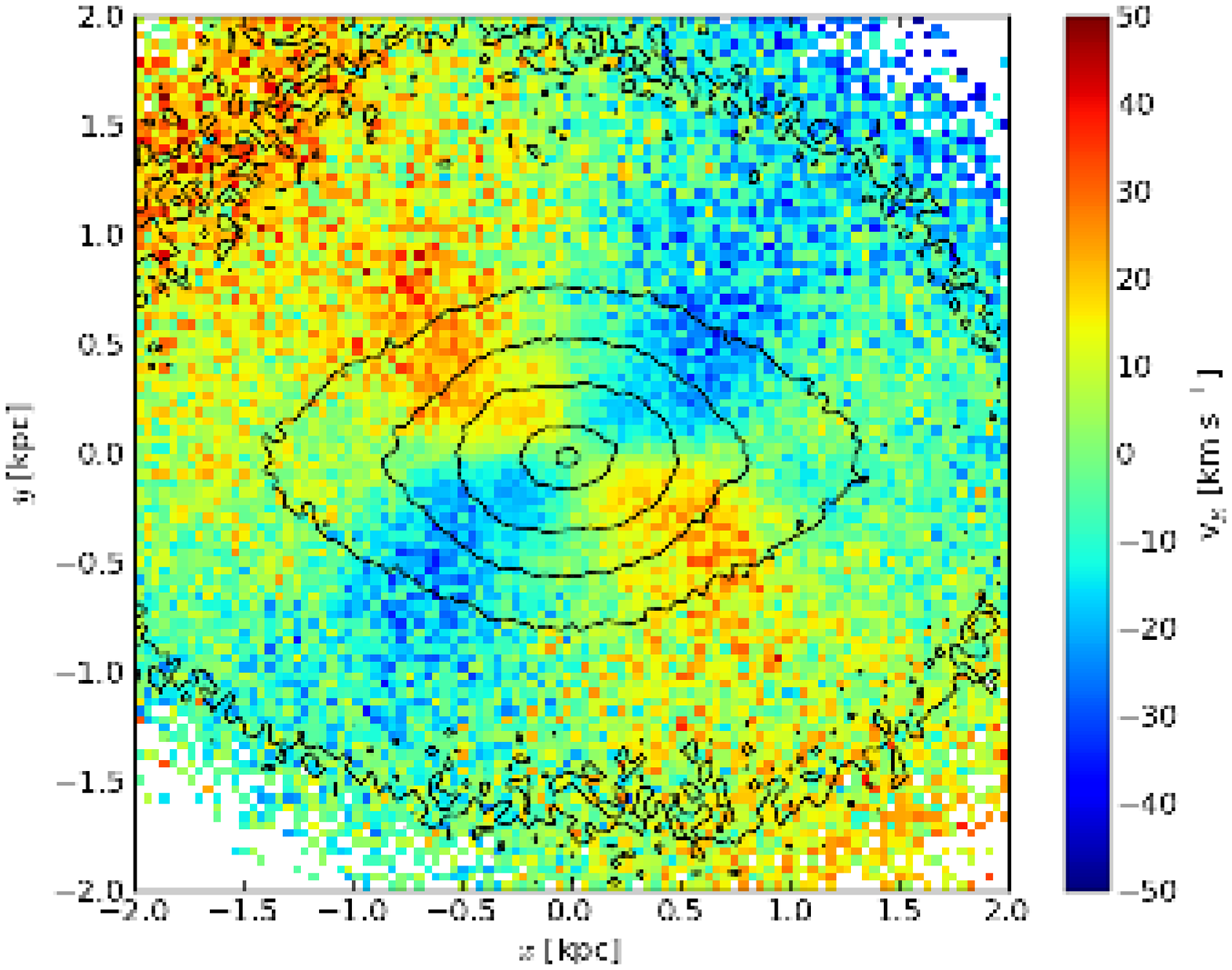} 
\includegraphics[width=0.5\hsize,angle=0]{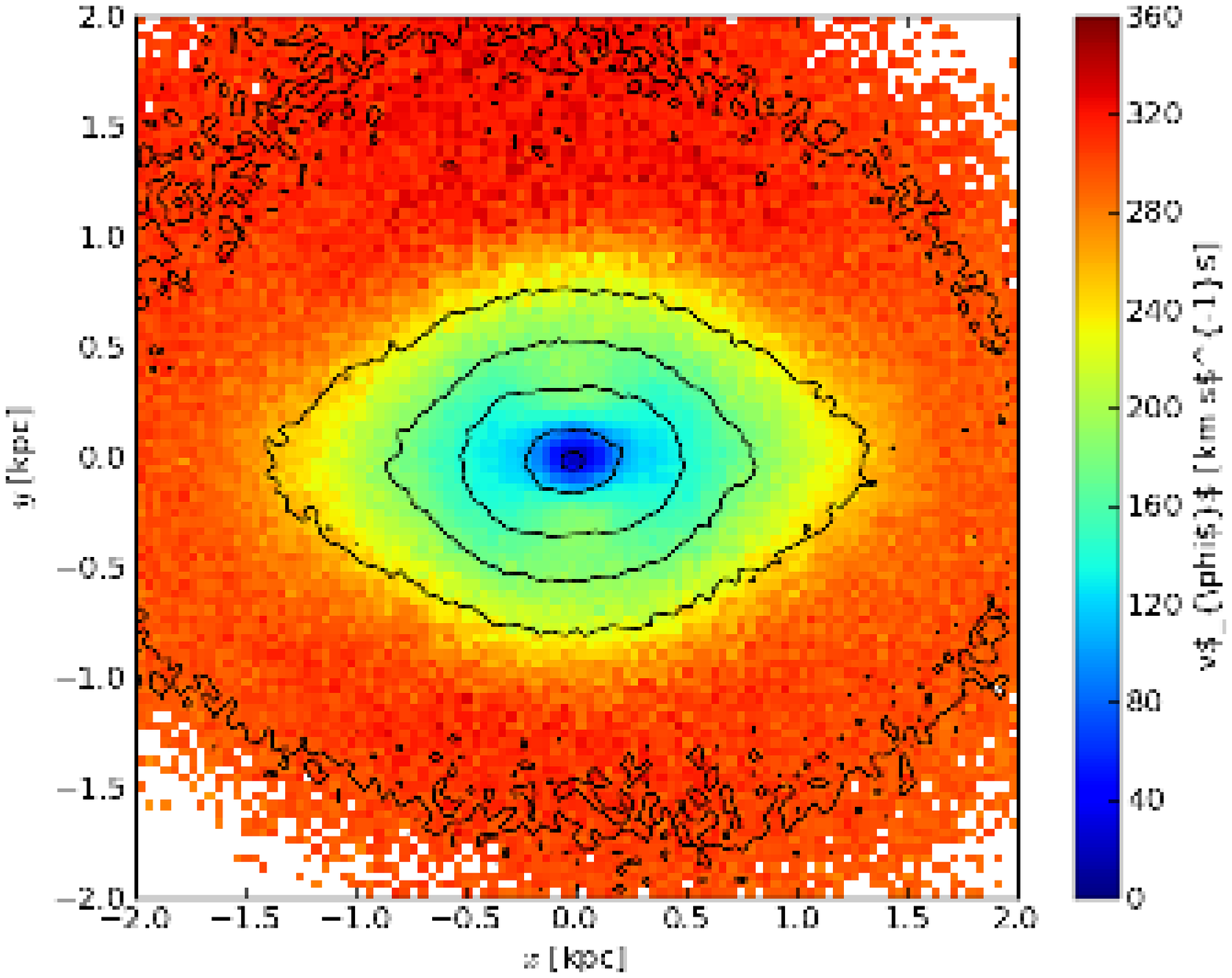} \\
\includegraphics[width=0.5\hsize,angle=0]{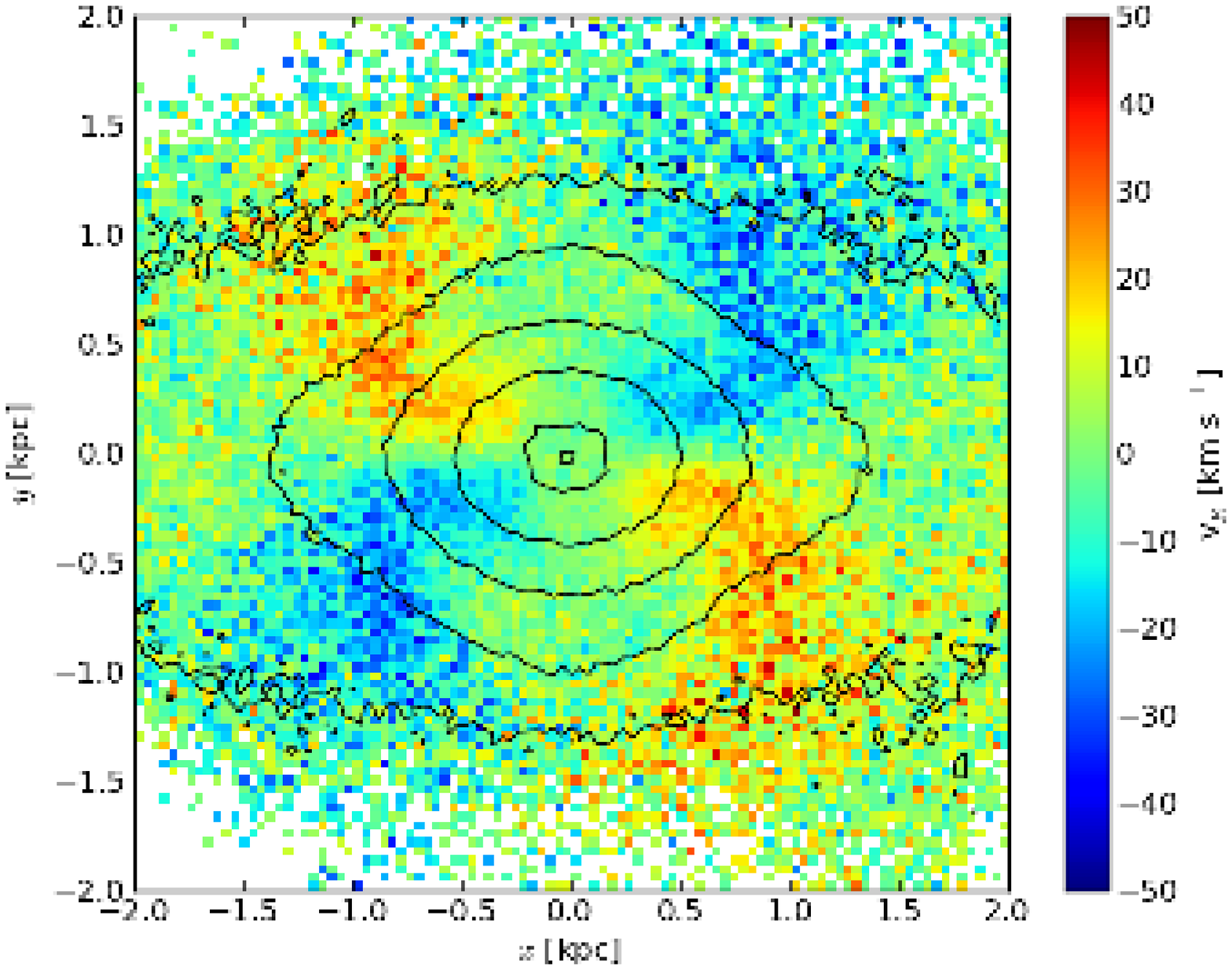} 
\includegraphics[width=0.5\hsize,angle=0]{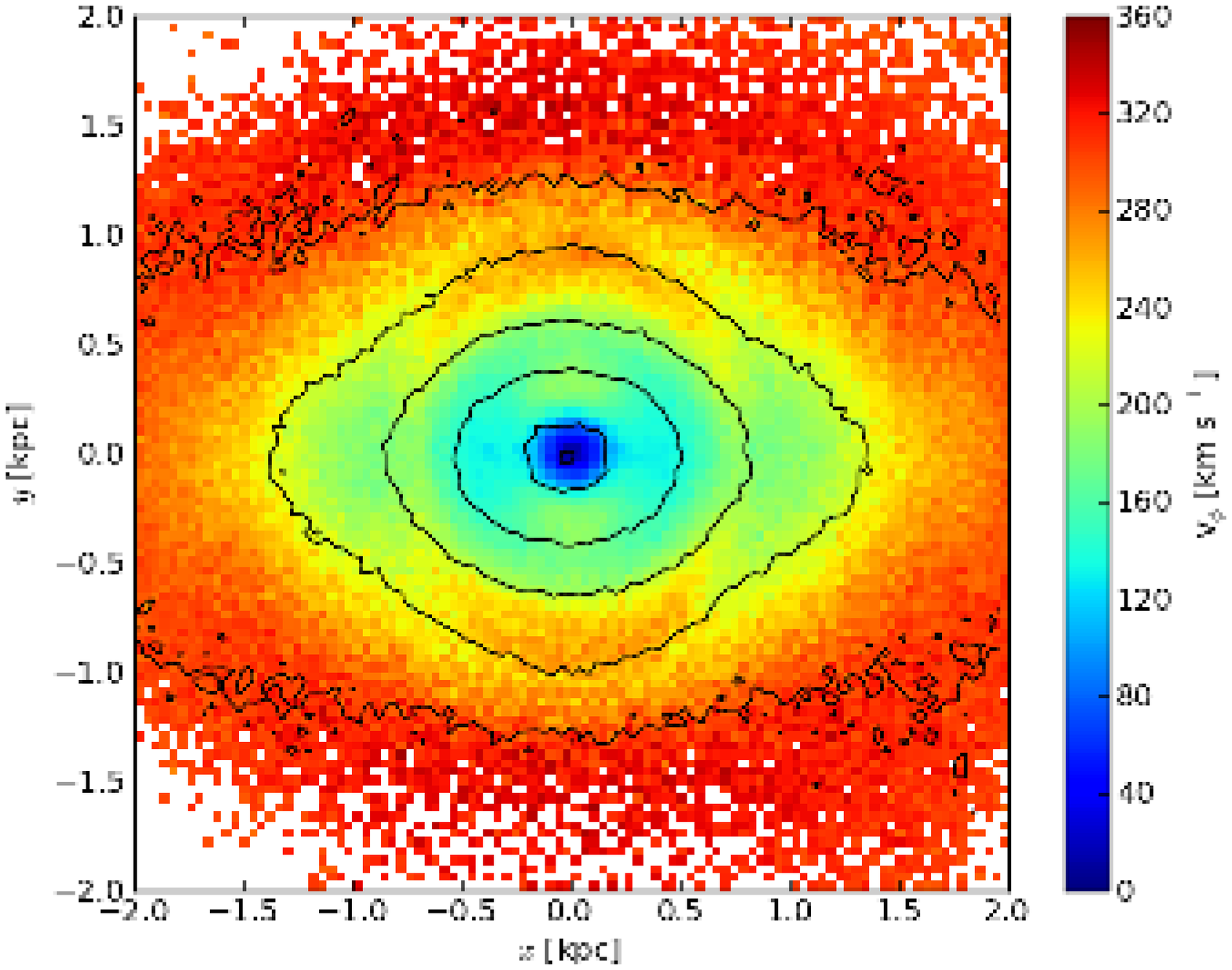} \\
\includegraphics[width=0.5\hsize,angle=0]{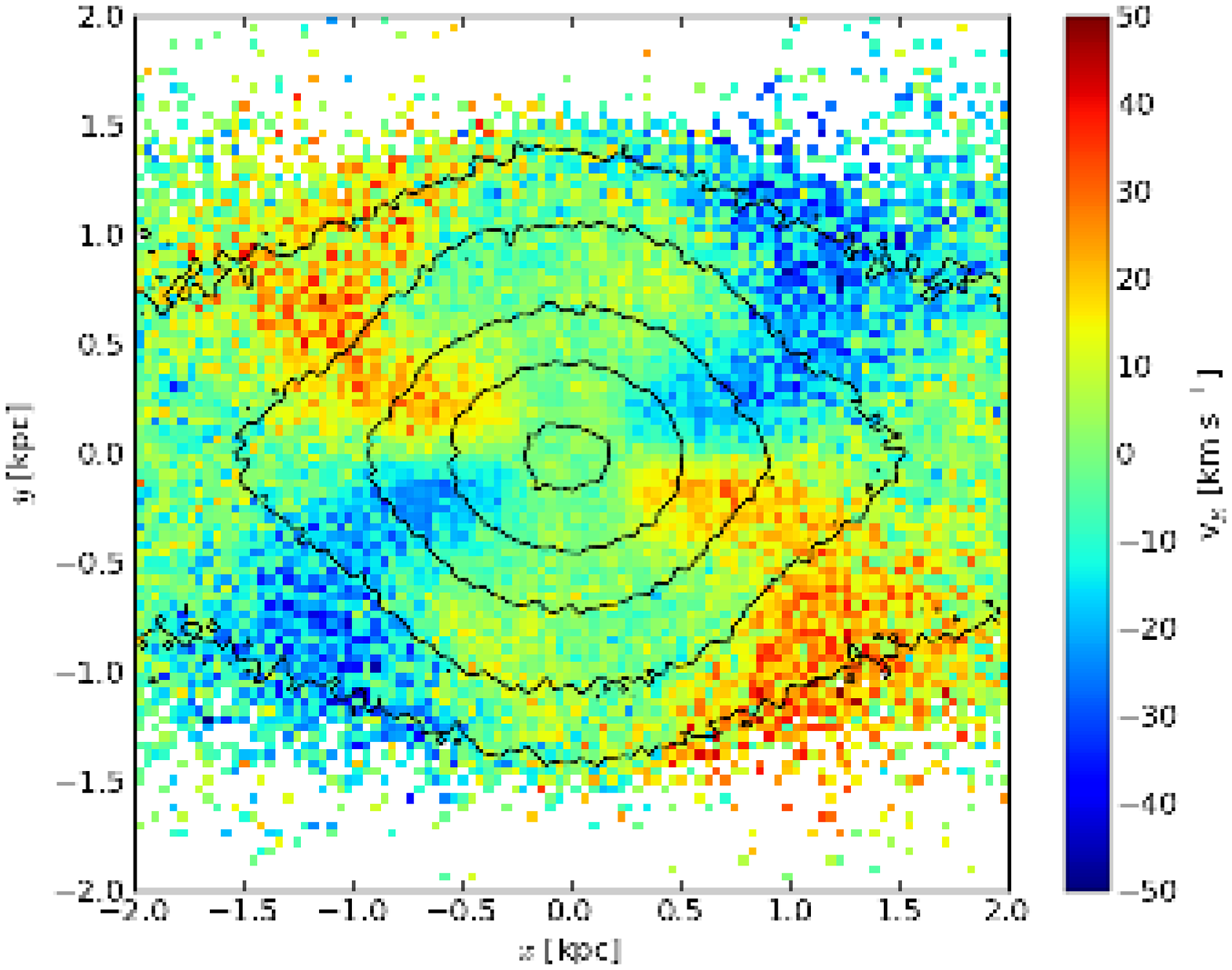}
\includegraphics[width=0.5\hsize,angle=0]{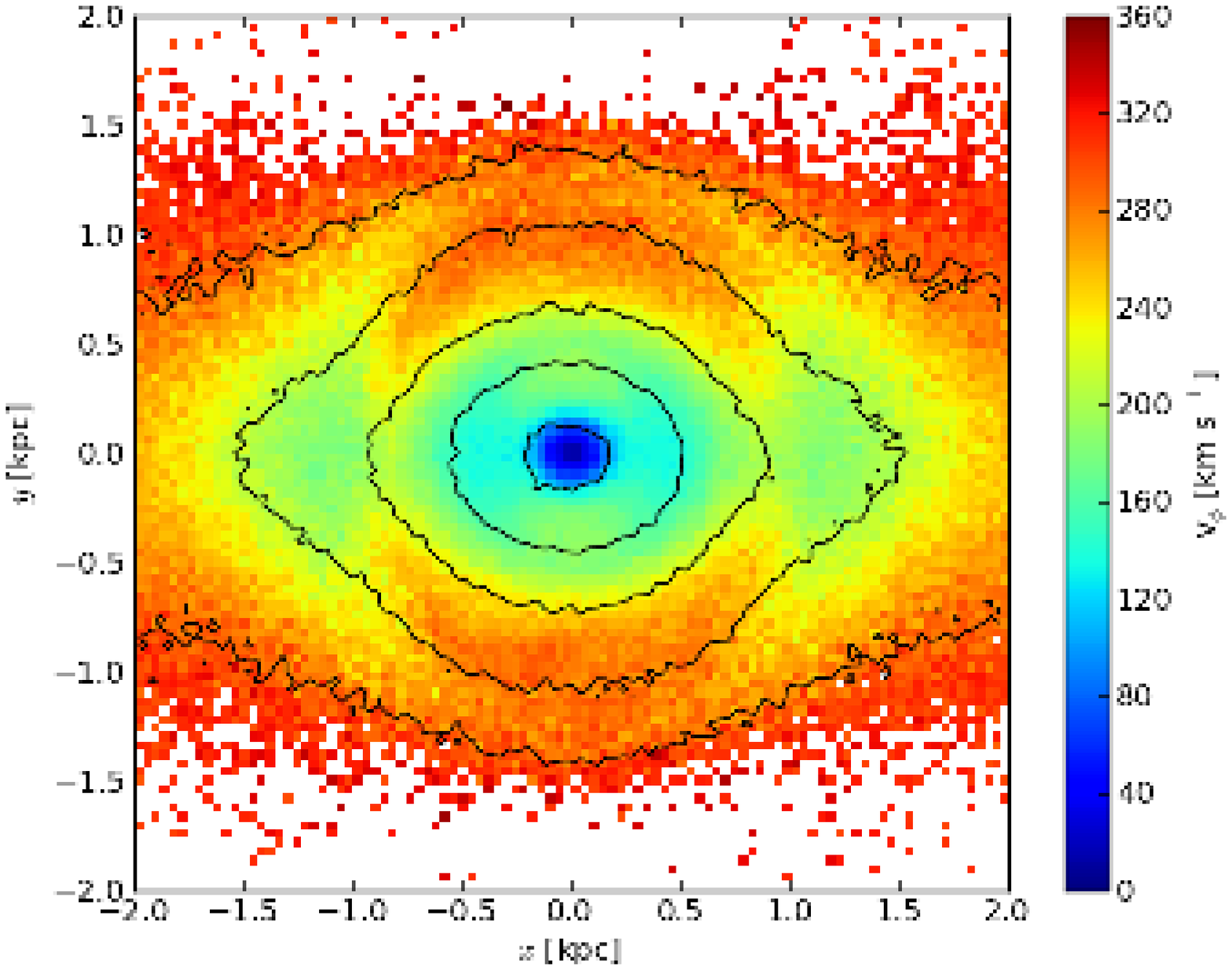} \\
\end{tabular}
\caption{Mean stellar velocities at 6 Gyr (top row), 8 Gyr (middle
  row) and 10 Gyr (bottom row). $v_R$ is shown in the left column,
  while $v_\phi$ is in the right column.  Contours indicate the
  stellar surface density.}
\label{fig:starvels}
\end{figure*}

We adopt cylindrical coordinates $(R,\phi,z)$ to study the stellar
kinematics.  The left column of Fig. \ref{fig:starvels} shows the
mass-weighted average radial velocity, $v_R$, at 6, 8 and 10 Gyr.  At
all three times there are two diametrically opposed quadrants where
$v_R$ is negative (inward motion) and two where $v_R$ is positive
(outward motion). These radial motions of $\sim 50$ \kms\ are
indicative of non-circular motion in this inner region.  While
qualitatively similar, these maps exhibit a twisting of the ridge of
peak $|v_R|$ towards the major axis of the bar with time.  For
instance, the peak $|v_R|$ in the first quadrant is at $\sim
45\degrees$ to the bar at 6 Gyr slowly decreasing to $\sim 25\degrees$
by 10 Gyr.  At larger radii, beyond the ND ($\gtrsim$ 1 kpc), the
ridge of peak $|v_R|$ twists away from the bar's major axis.  This
evolution can be understood by considering separately the motion of
the old and young stars, as we do in Fig.  \ref{fig:starvels10Gyr}.
The ridge of peak $|v_R|$ for the old ($> 6$ Gyr) stars is quite
similar to that at 6 Gyr.  For the young stars ($< 3$ Gyr), however,
the peak $v_R$ is both larger and rotated by $\sim 90\degrees$
relative to the old stars.  The radial motions at larger radii are
caused by stars moving along the length of the bar, such that the
negative $v_R$ is on the leading side of the bar and the positive
$v_R$ is on the trailing side of the bar.  The combination of these
two independent motions leads to the twist of the ridge of peak
$|v_R|$ in the total population.  The bottom-left panel of Fig.
\ref{fig:starvels10Gyr} also shows clearly the difference between the
motions of stars in the ND and those in the bar, with $v_R$ changing
direction abruptly just outside the ND.

The right column of Fig.  \ref{fig:starvels} shows the mass-weighted
average tangential velocity, $v_\phi$.  At 6 Gyr the contours of
$v_\phi$ are parallel to the bar.  After the ND forms, strong $v_\phi$
peaks are located on the minor axis of the bar, i.e.  along the major
axis of the ND.  Between 8 Gyr and 10 Gyr the peak $v_\phi$ on the
major-axis of the ND also increases.  The motions of the young stars,
shown in Fig.  \ref{fig:starvels10Gyr}, show the largest $v_\phi$ is
on the minor-axis of the ND.  The combination of the velocities from
the main bar and the ND, however, lead to the strongest $v_\phi$ on
the ND's major's axis while masking the ND kinematics on the bar's
major-axis.

\begin{figure*}
\centering
\begin{tabular}{c}
\includegraphics[width=0.5\hsize,angle=0]{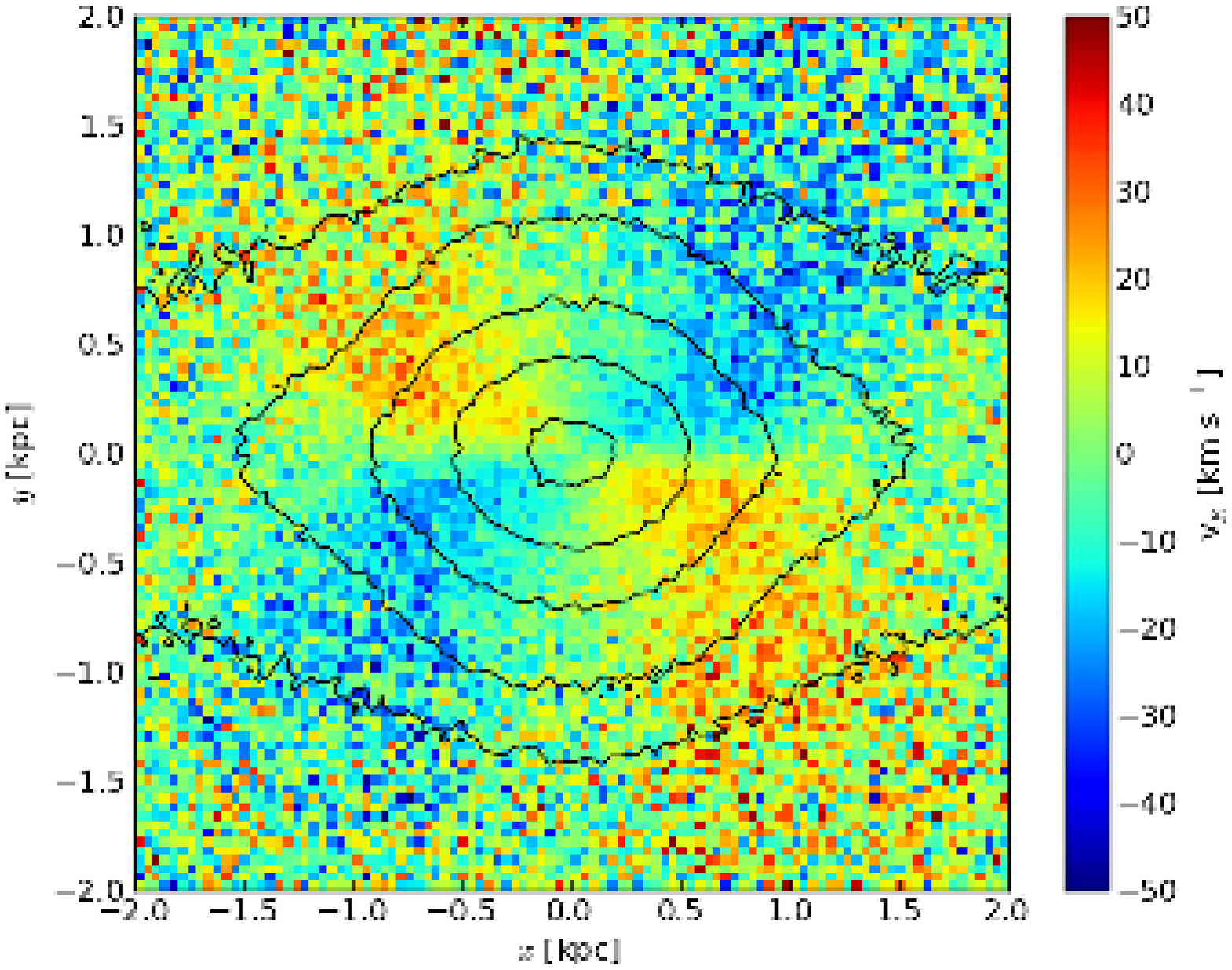} 
\includegraphics[width=0.5\hsize,angle=0]{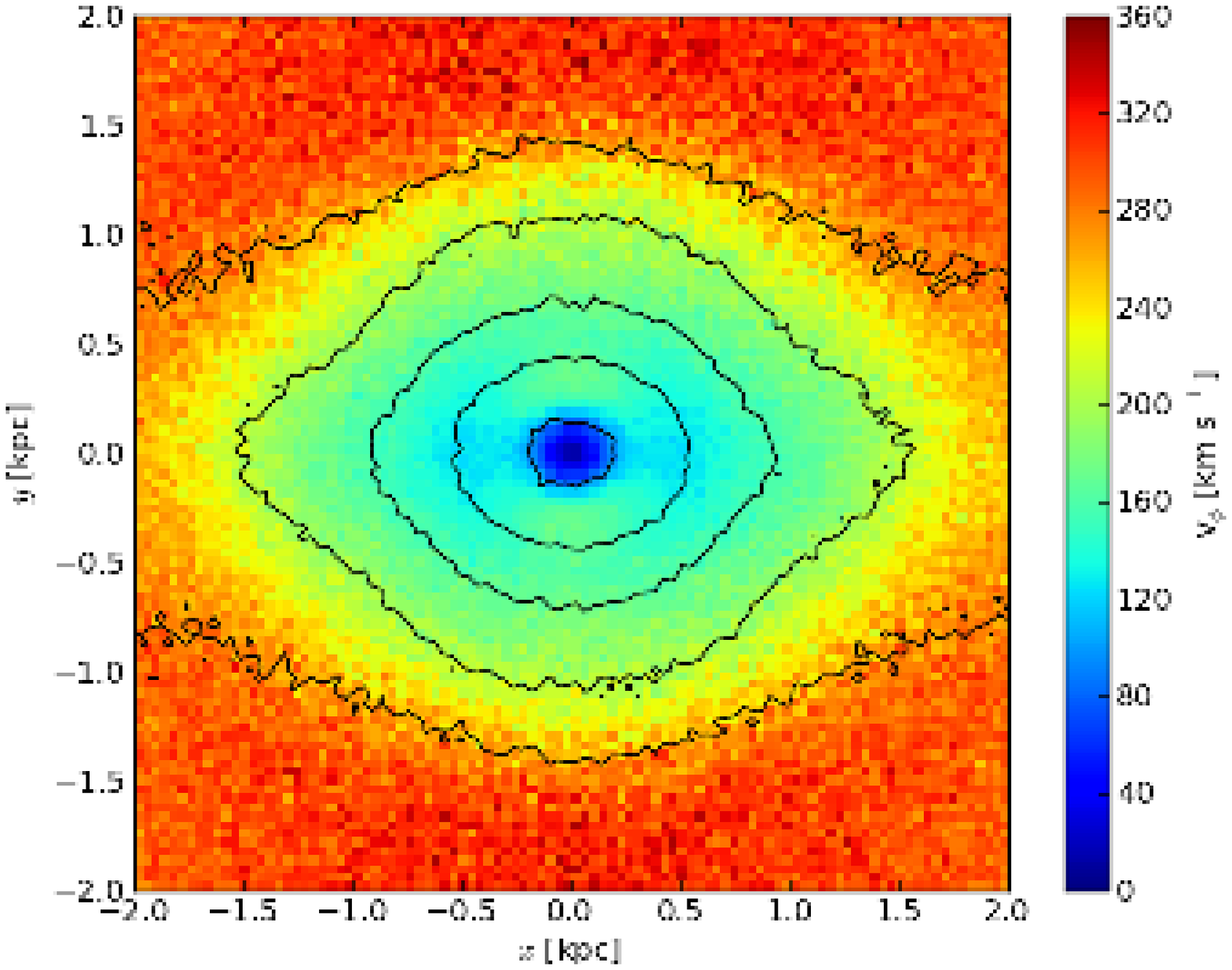} \\
\includegraphics[width=0.5\hsize,angle=0]{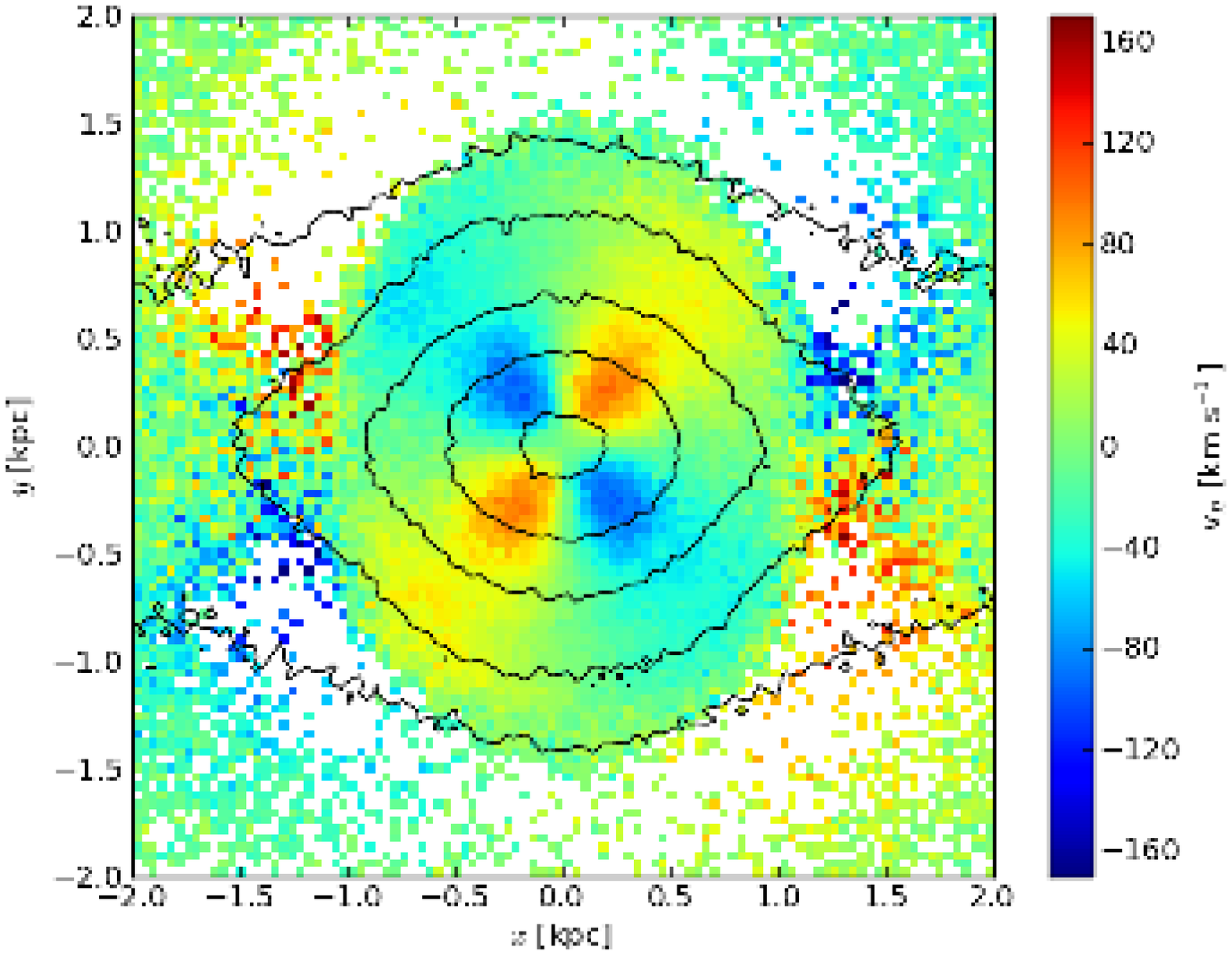} 
\includegraphics[width=0.5\hsize,angle=0]{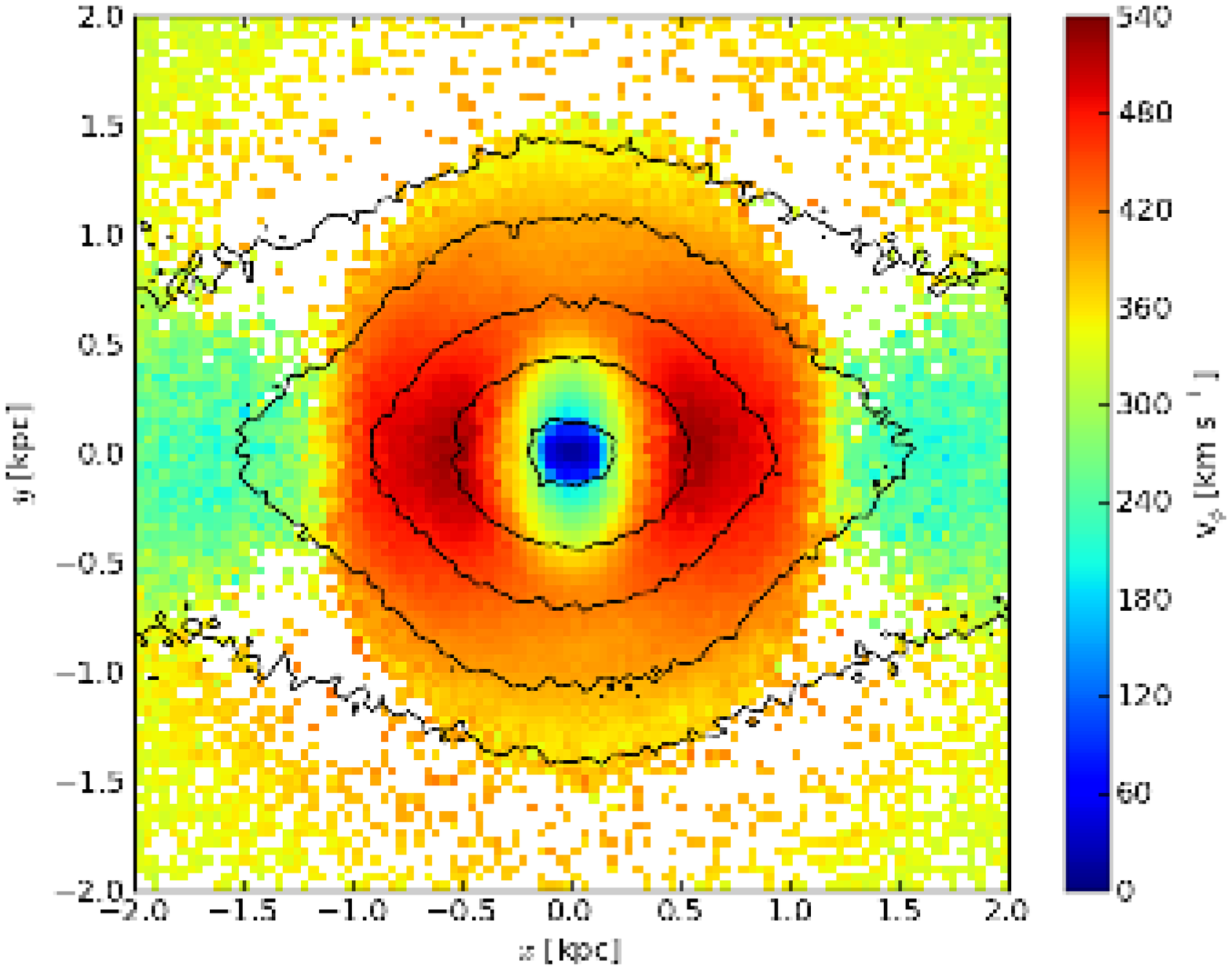} \\
\end{tabular}
\caption{Mass-weighted mean stellar velocities at 10 Gyr broken down
  by age. $v_R$ is shown in the left column, while $v_\phi$ is in the
  right column.  Stars older than 6 Gyr are shown in the top row while
  stars younger than 3 Gyr are shown in the bottom row.  Note the
  different scales between the top and bottom rows.  Contours indicate
  the total stellar surface density.}
\label{fig:starvels10Gyr}
\end{figure*}

\begin{figure*}
\centering
\begin{tabular}{c}
\includegraphics[width=0.5\hsize,angle=0]{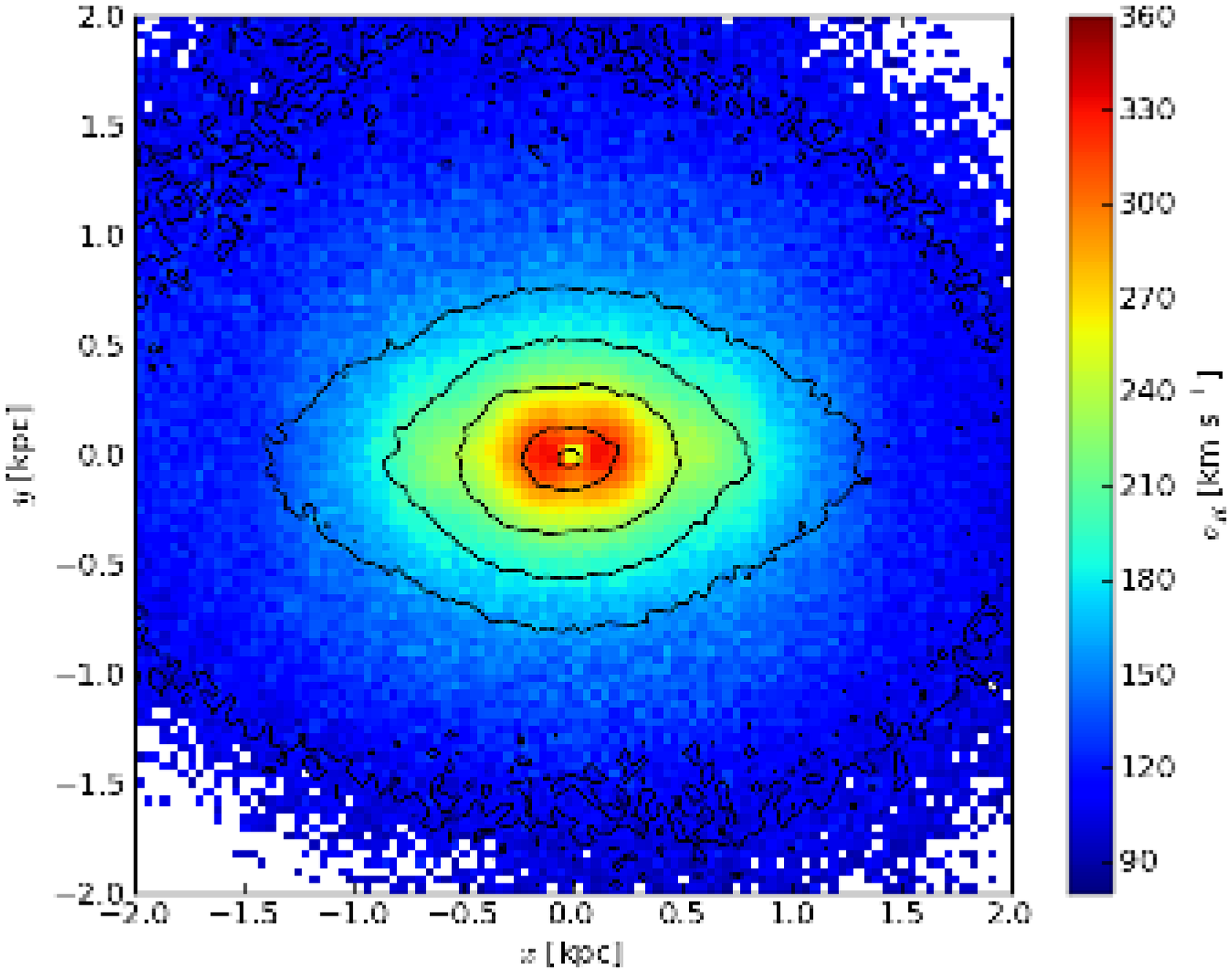} 
\includegraphics[width=0.5\hsize,angle=0]{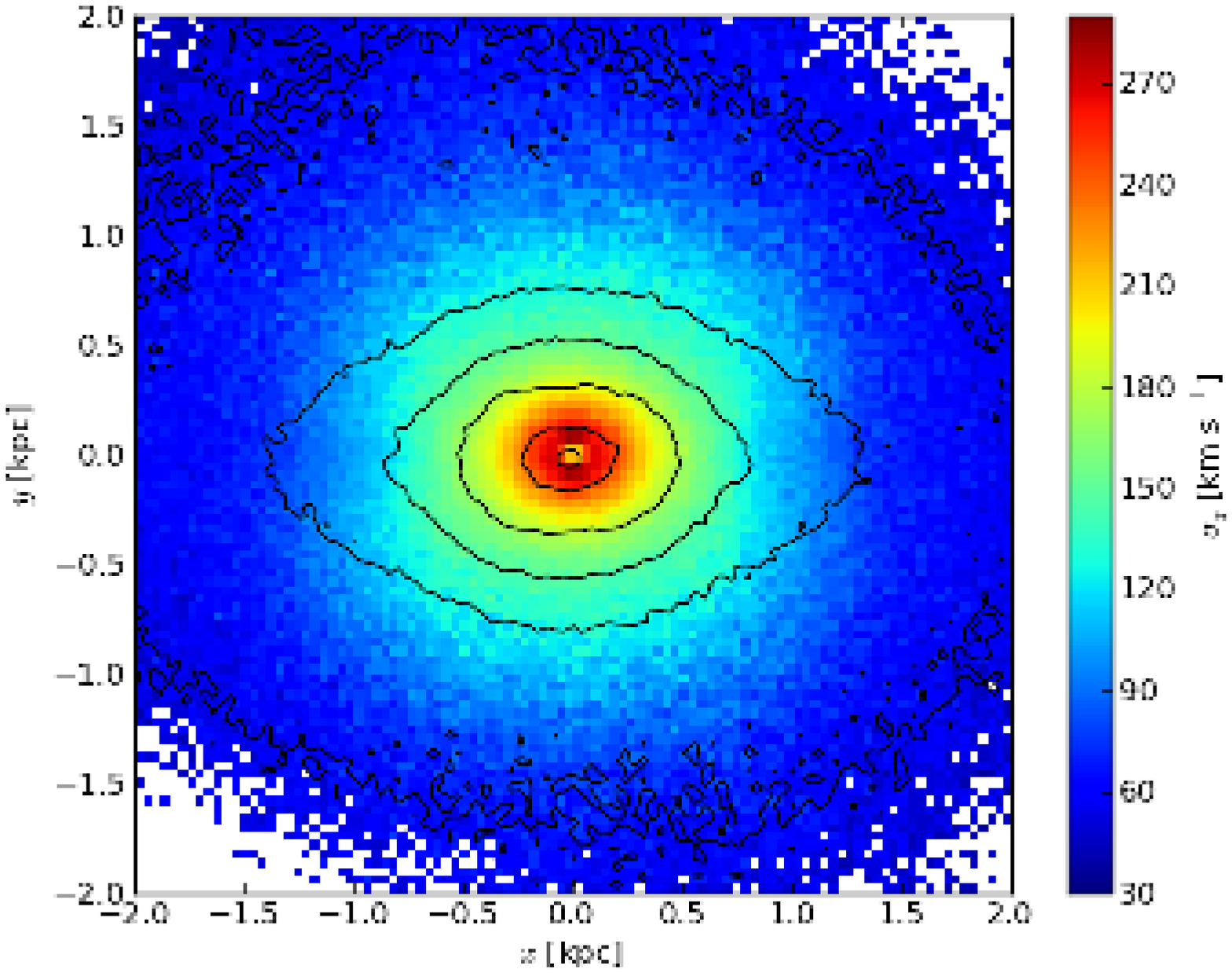} \\
\includegraphics[width=0.5\hsize,angle=0]{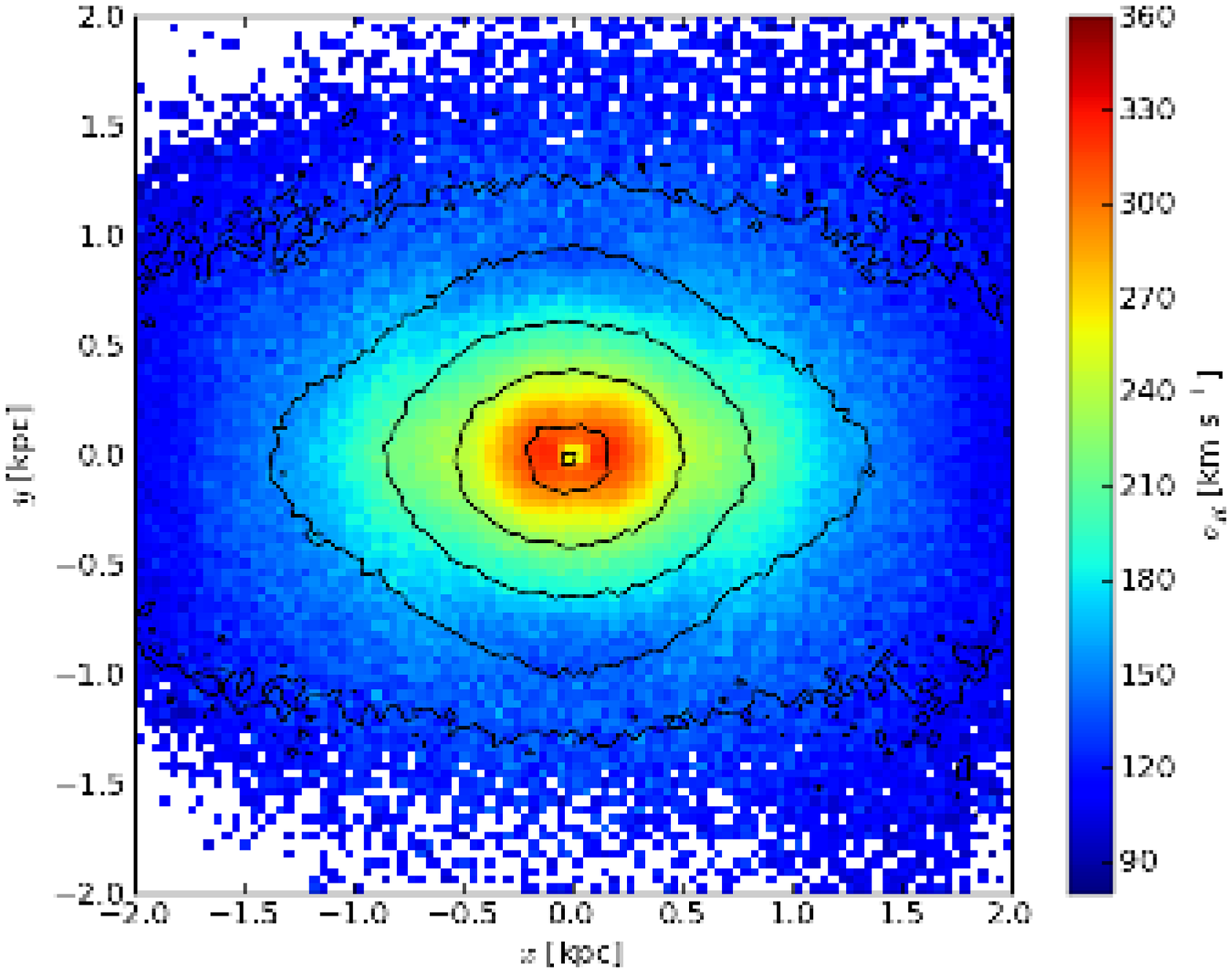} 
\includegraphics[width=0.5\hsize,angle=0]{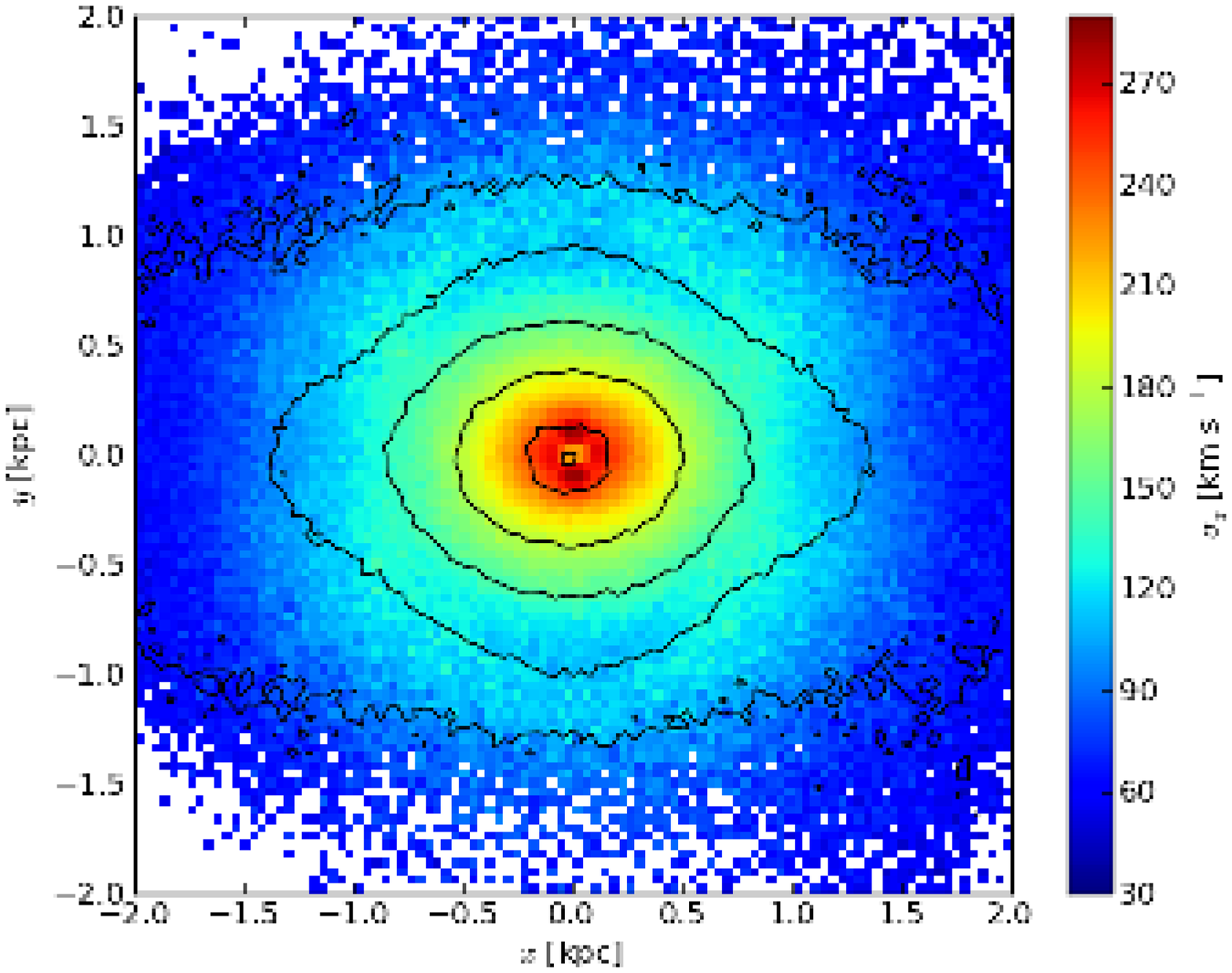} \\
\includegraphics[width=0.5\hsize,angle=0]{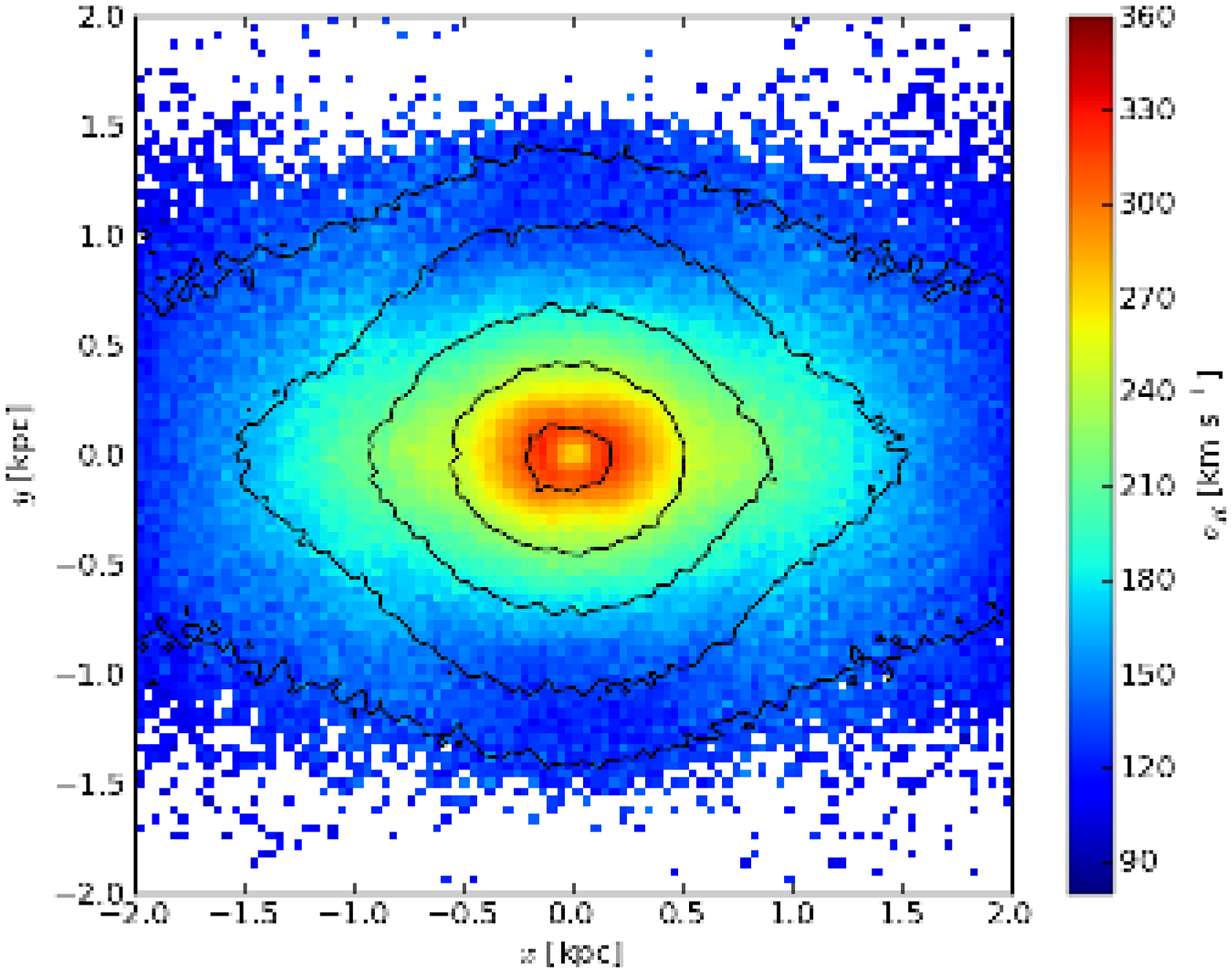}
\includegraphics[width=0.5\hsize,angle=0]{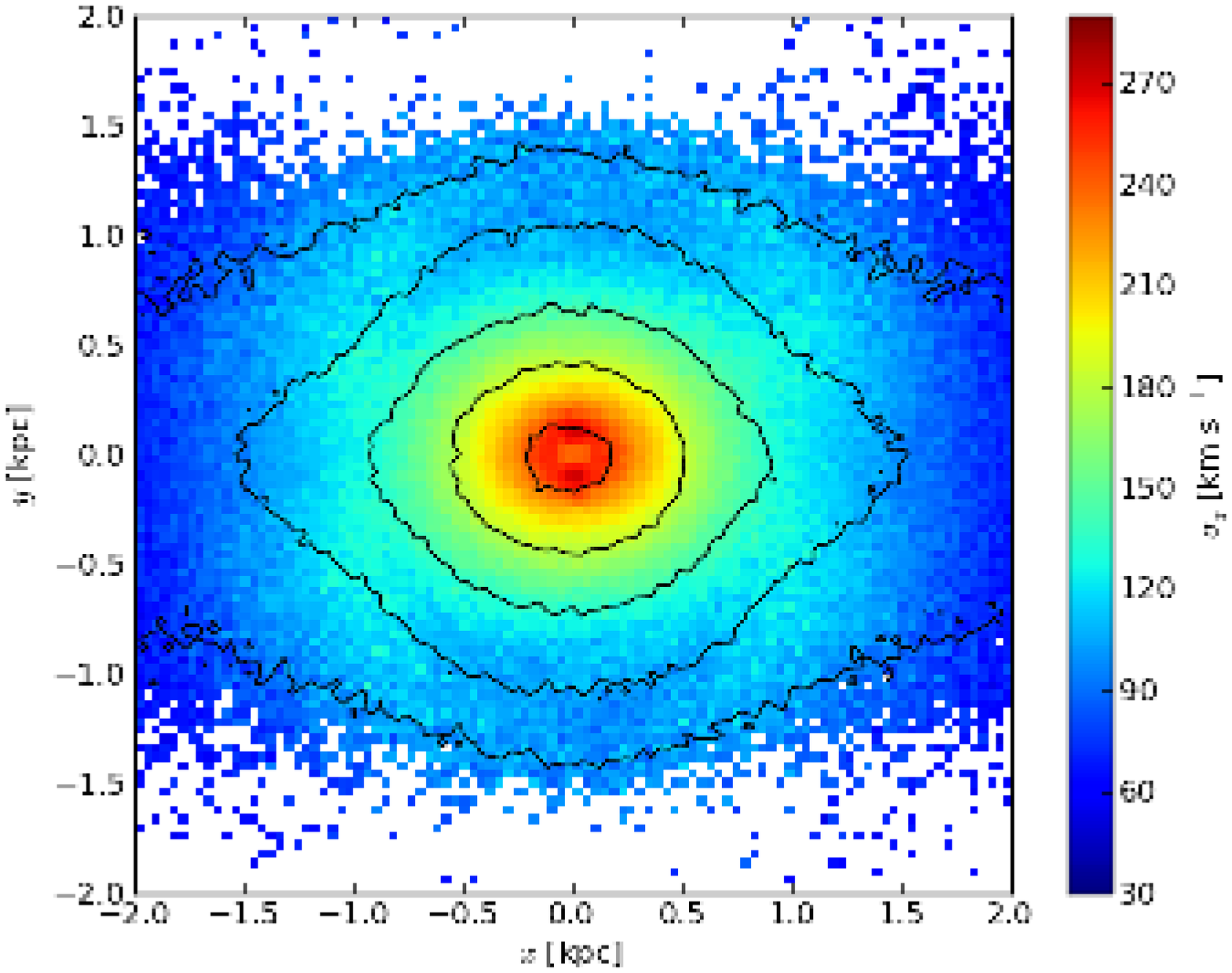}\\
\end{tabular}
\caption{Mass-weighted stellar velocity dispersions at 6 Gyr (top
  row), 8 Gyr (middle row) and 10 Gyr (bottom row). $\sigma_R$ is
  shown in the left column, while $\sigma_z$ is in the right column.
  Contours indicate the stellar surface density.}
\label{fig:stardisps}
\end{figure*}

The left column of Fig. \ref{fig:stardisps} shows the radial velocity
dispersion, $\sigma_R$, while the right column shows the vertical
velocity dispersion, $\sigma_z$.  After the ND forms both $\sigma_R$
and $\sigma_z$ along the major axis of the ND decline.  This cooling
occurs in an absolute sense, not just relative to the surrounding
hotter bar.  The decline in $\sigma_z$ reflects that the new stars are
forming in a thin population.

The left column of Fig. \ref{fig:h4ages} shows the face-on
mass-weighted Gauss-Hermite moment h4.  At 6 Gyr h4 has contours
elongated like the bar with a near zero value at the centre increasing
outwards.  With the formation of the ND h4 increases.  A strong peak
forms on the ND major-axis while minima are present on the bar's
major-axis.  By 10 Gyr the peak on the ND's major axis strengthens
further and the minima in the centre and along the bar's major axis
weaken.  The h4 minimum on the bar's major axis is a signature of the
growing peanut shape of the bulge \citep{Debattista2005,MendezAbreu2008}.

Together, $v_R$, $v_\phi$, $\sigma_R$, $\sigma_z$ and h4 all indicate
that a distinct, cool, thin ND, extending to $R \sim 1.5$ kpc has
formed by 10 Gyr.  The ND is elliptical and elongated perpendicular to
the main bar.

\begin{figure*}
\centering
\begin{tabular}{c}
\includegraphics[width=0.5\hsize,angle=0]{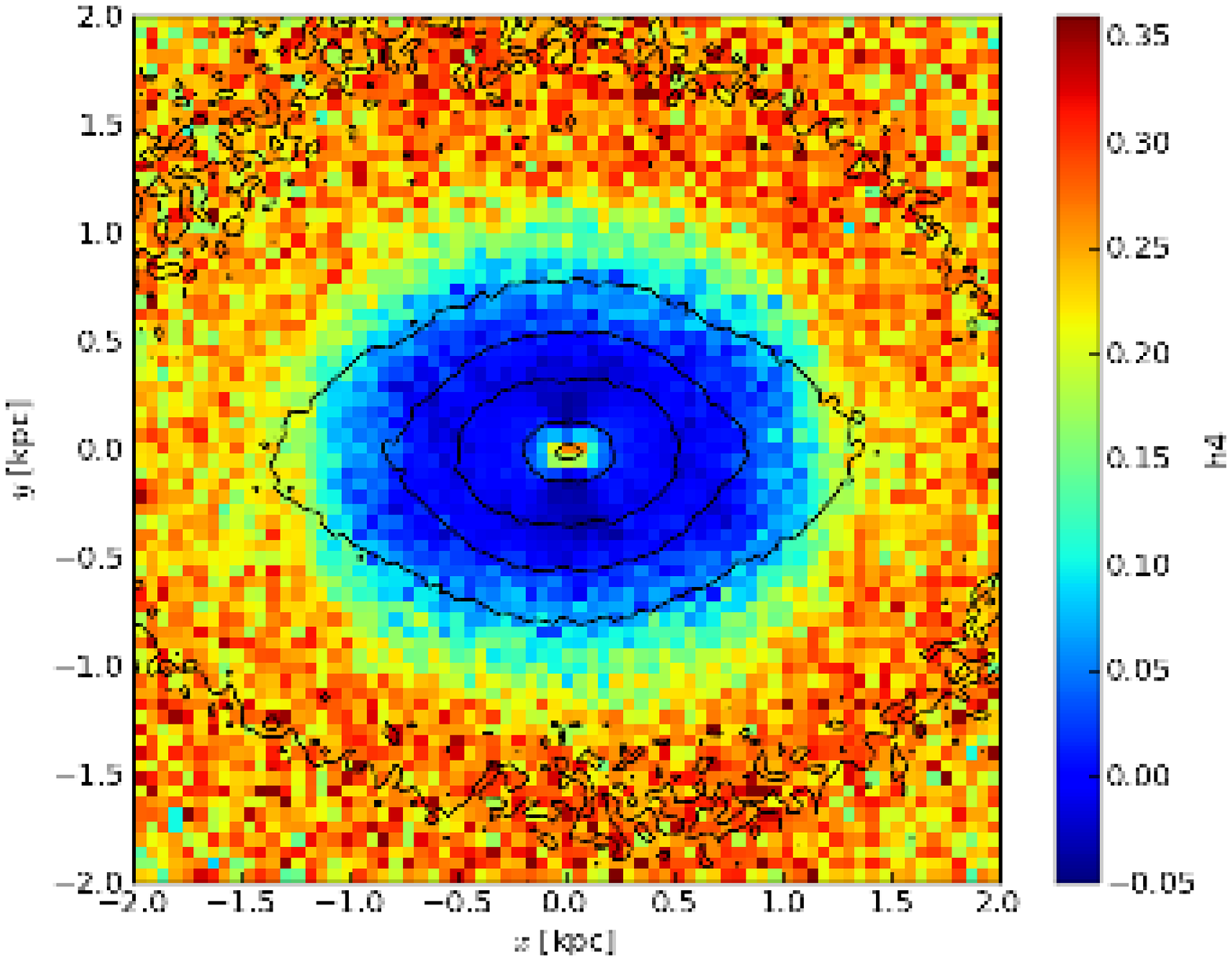}
\includegraphics[width=0.5\hsize,angle=0]{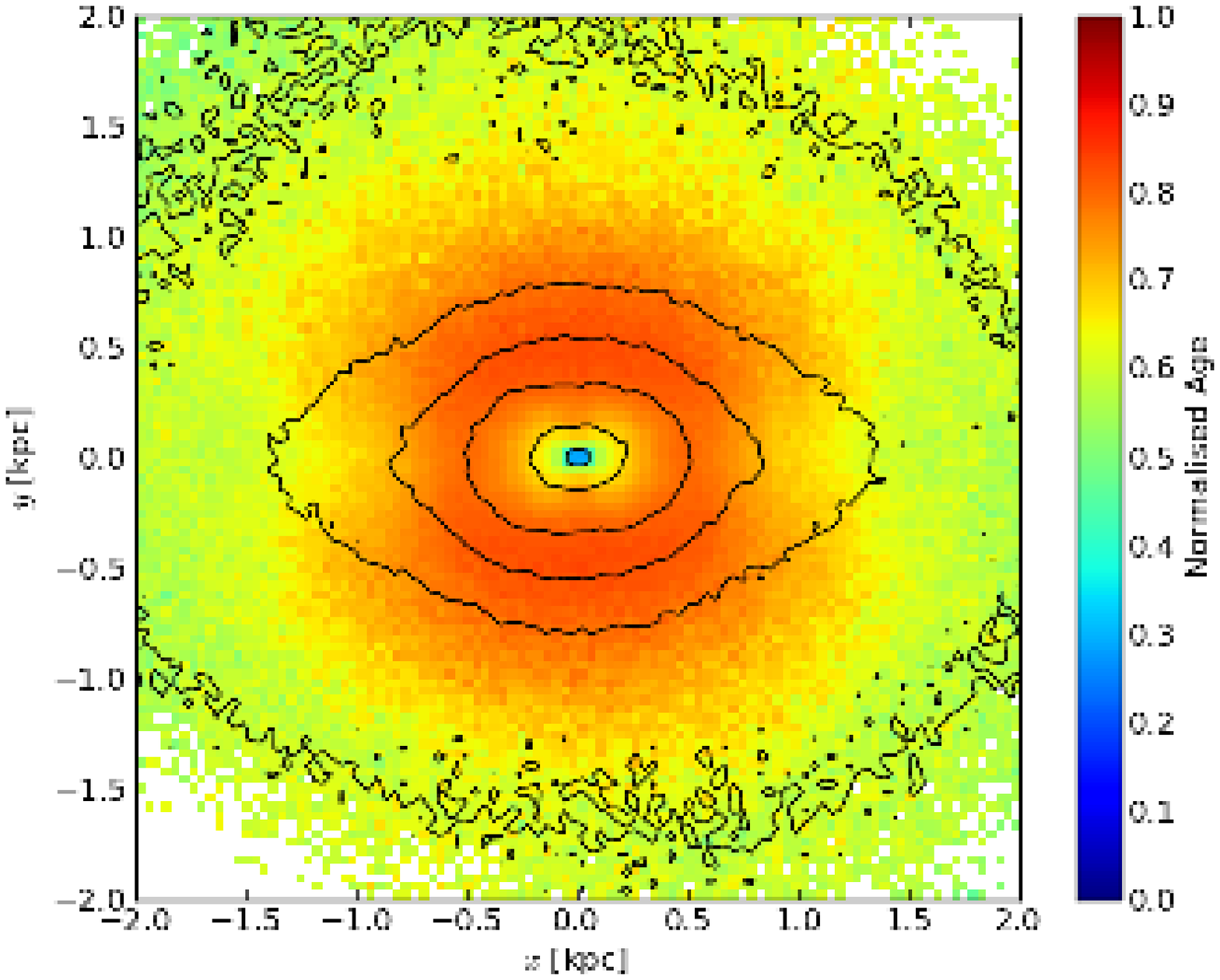} \\
\includegraphics[width=0.5\hsize,angle=0]{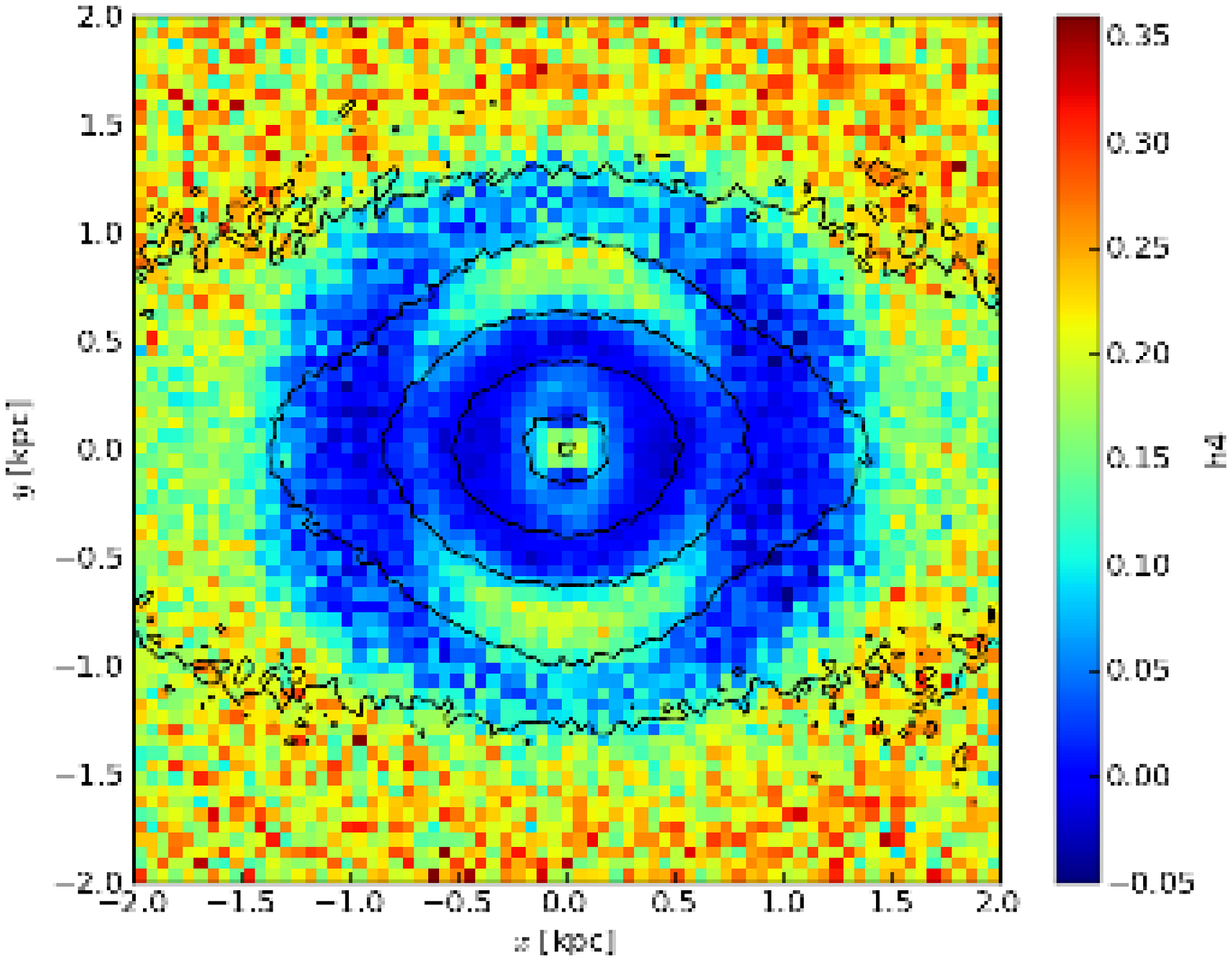}
\includegraphics[width=0.5\hsize,angle=0]{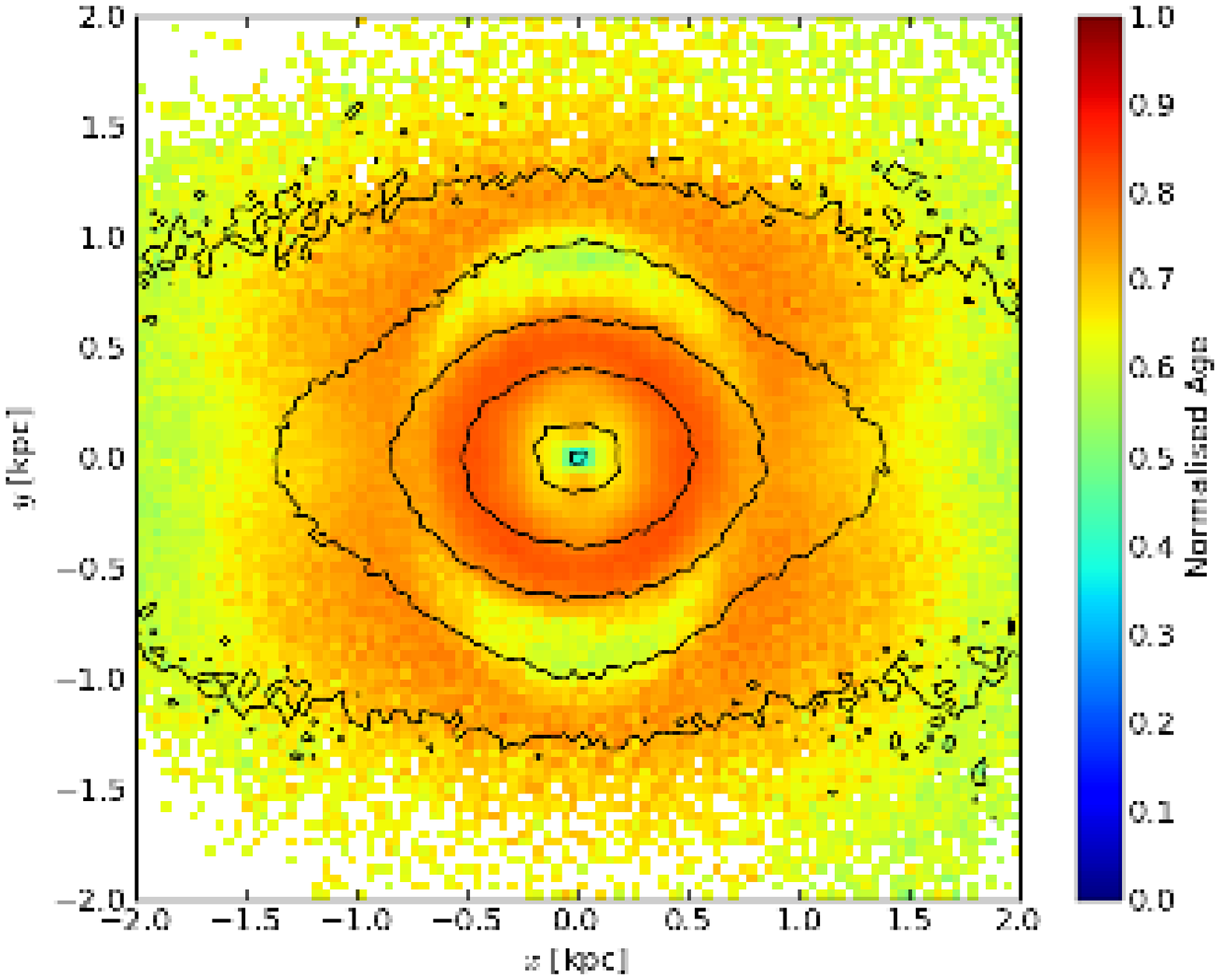} \\
\includegraphics[width=0.5\hsize,angle=0]{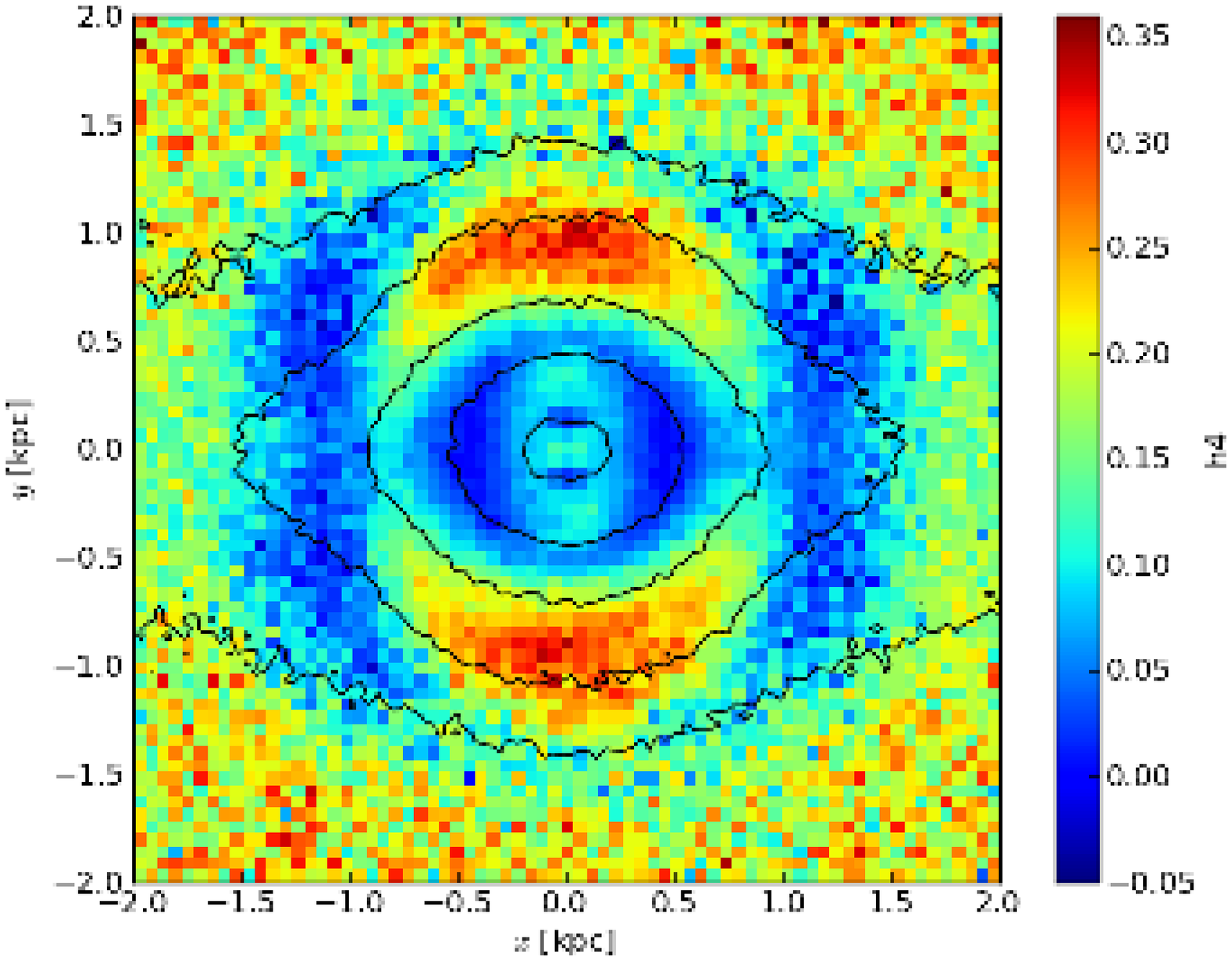} 
\includegraphics[width=0.5\hsize,angle=0]{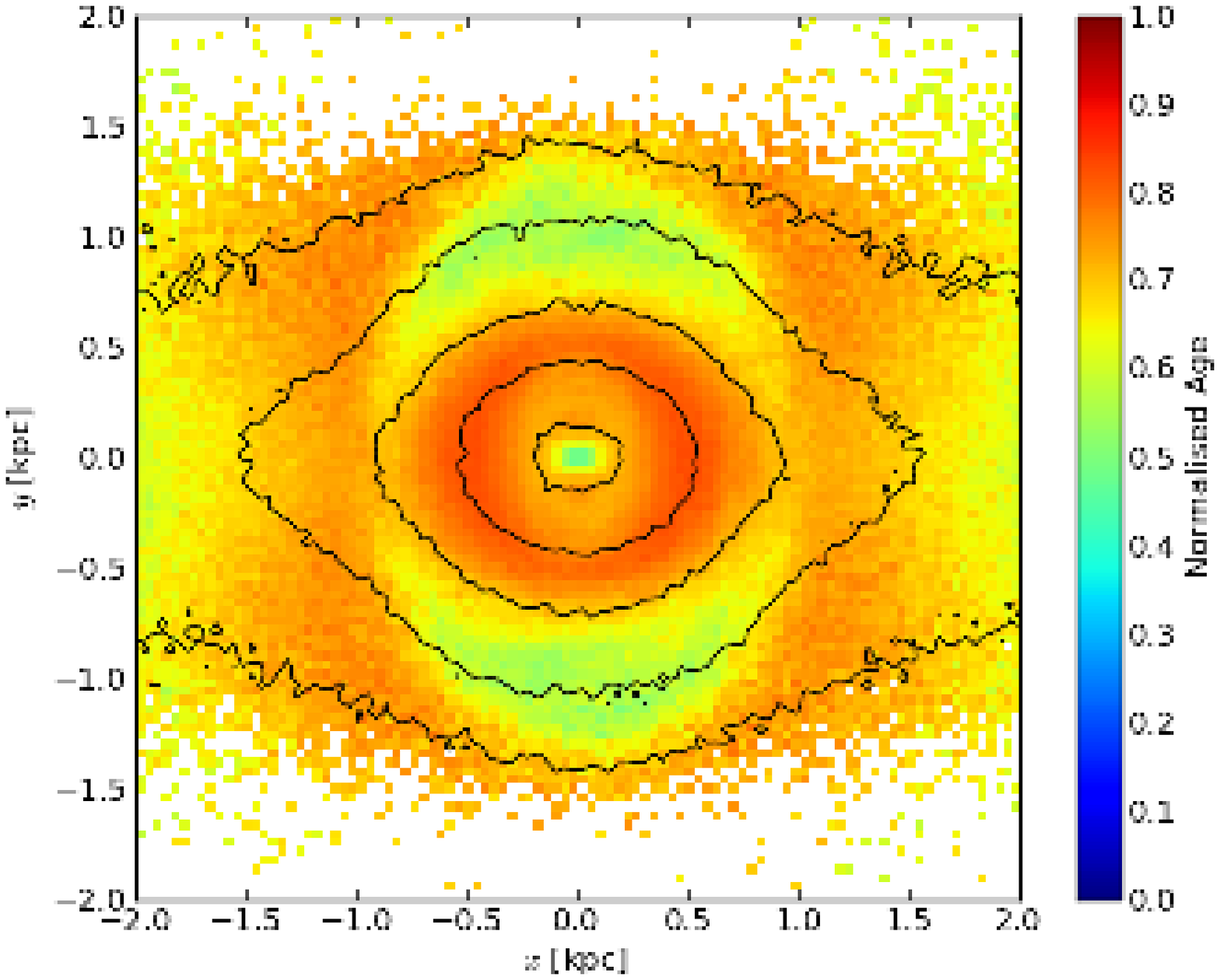} \\
\end{tabular}
\caption{Left: Face-on mass-weighted Gauss-Hermite vertical
  h4 kinematic moment.  Right: Mass-weighted mean stellar ages.  To
  facilitate comparison of the different panels, the ages have been
  normalized by the time of each snapshot i.e. the stellar ages are
  divided by the age of the simulation at the time of the snapshot: 6,
  8 or 10 Gyr. The stellar surface density is indicated by the
  contours.  The top, middle and bottom rows show 6, 8 and 10 Gyr.  }
\label{fig:h4ages}
\end{figure*}

\subsection{Ages}
\label{ssec:age}

The right column of Fig. \ref{fig:h4ages} shows maps of the
mass-weighted mean age of stars.  Old stars are present in the central
1 kpc, which is dominated by the bar; further out the stellar
populations are increasingly dominated by young stars formed in the
ND.  Perpendicular to the bar, a substantially younger population of
stars is present at $\sim 700$ parsec, concentrated at the sides of
the bar \citep[see also][]{Ness2014}.

The distribution of stellar ages in barred galaxy simulations was
studied by \citet{Wozniak2007} and \citet{Wozniak2008,Wozniak2009}.
They used models with a pre-existing stellar disc, adding a gas disc
with 10 per cent mass fraction, and evolved for 3 Gyr, at the end of which
46 per cent of the gas had been converted to stars.  \cite{Wozniak2007}
found concentrations of young stars at the ends of the bar in their
model.  Our simulation also has young stars at the end of the bar,
giving the mean age map a dimpled appearance along the bar's major
axis \citep[see also][]{Ness2014}.  

\citet{Wozniak2007} and \citet{Wozniak2008} did not report a ND in
their model (see Figs. 1 and 2 in \cite{Wozniak2007}), but did find a
young population in a thin disc corresponding to 17 per cent of the
stars in the inner 1 kpc. However \citet{Wozniak2009} found a ND
within a radius of 0.5 kpc. They conclude that their ND is
associated with a $\sigma$-drop \citep[see also][]{Wozniak2003}.

\subsection{Chemistry}
\label{ssec:chemistry}

\begin{figure*}
\centering
\begin{tabular}{c}
\includegraphics[width=0.5\hsize,angle=0]{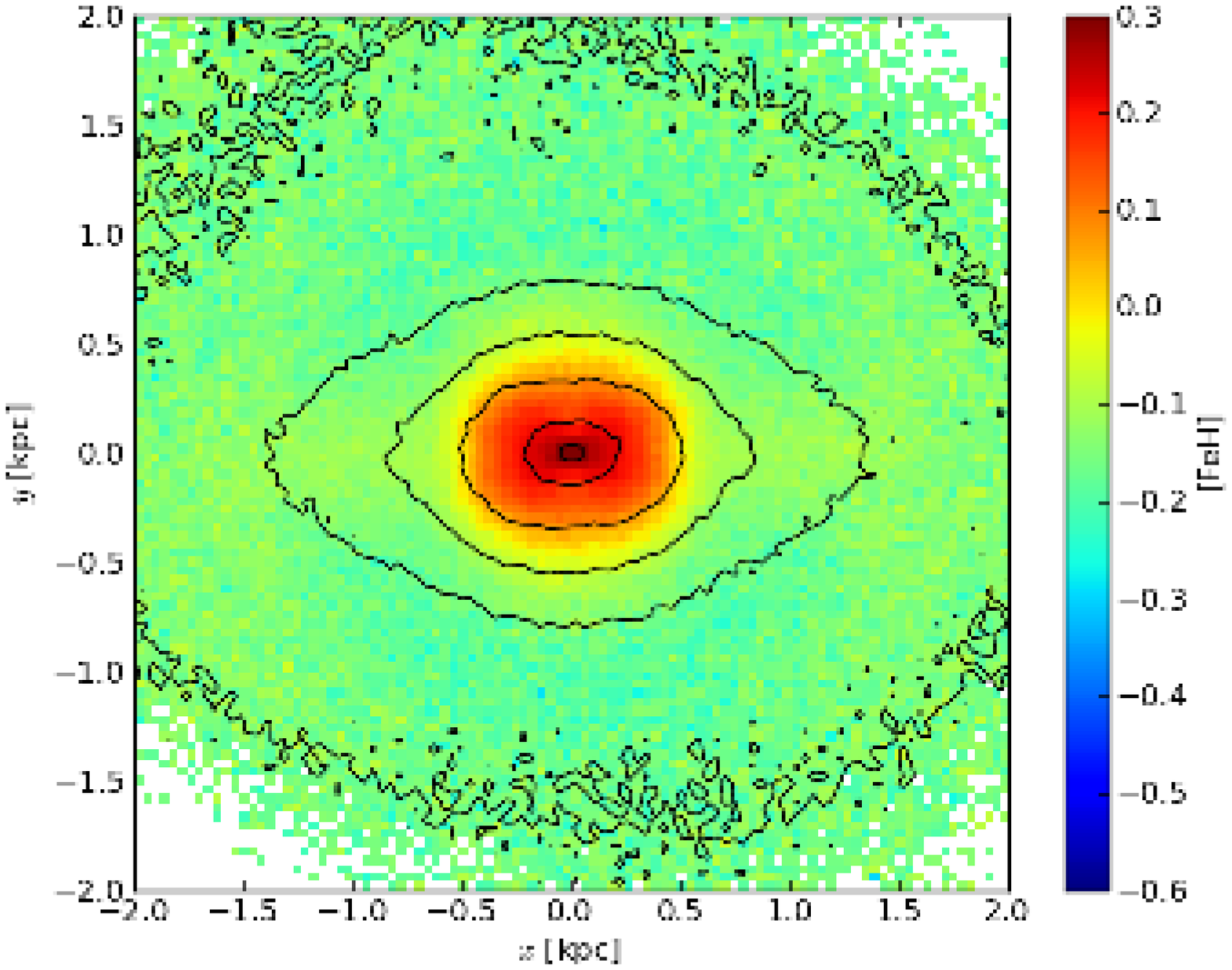} 
\includegraphics[width=0.5\hsize,angle=0]{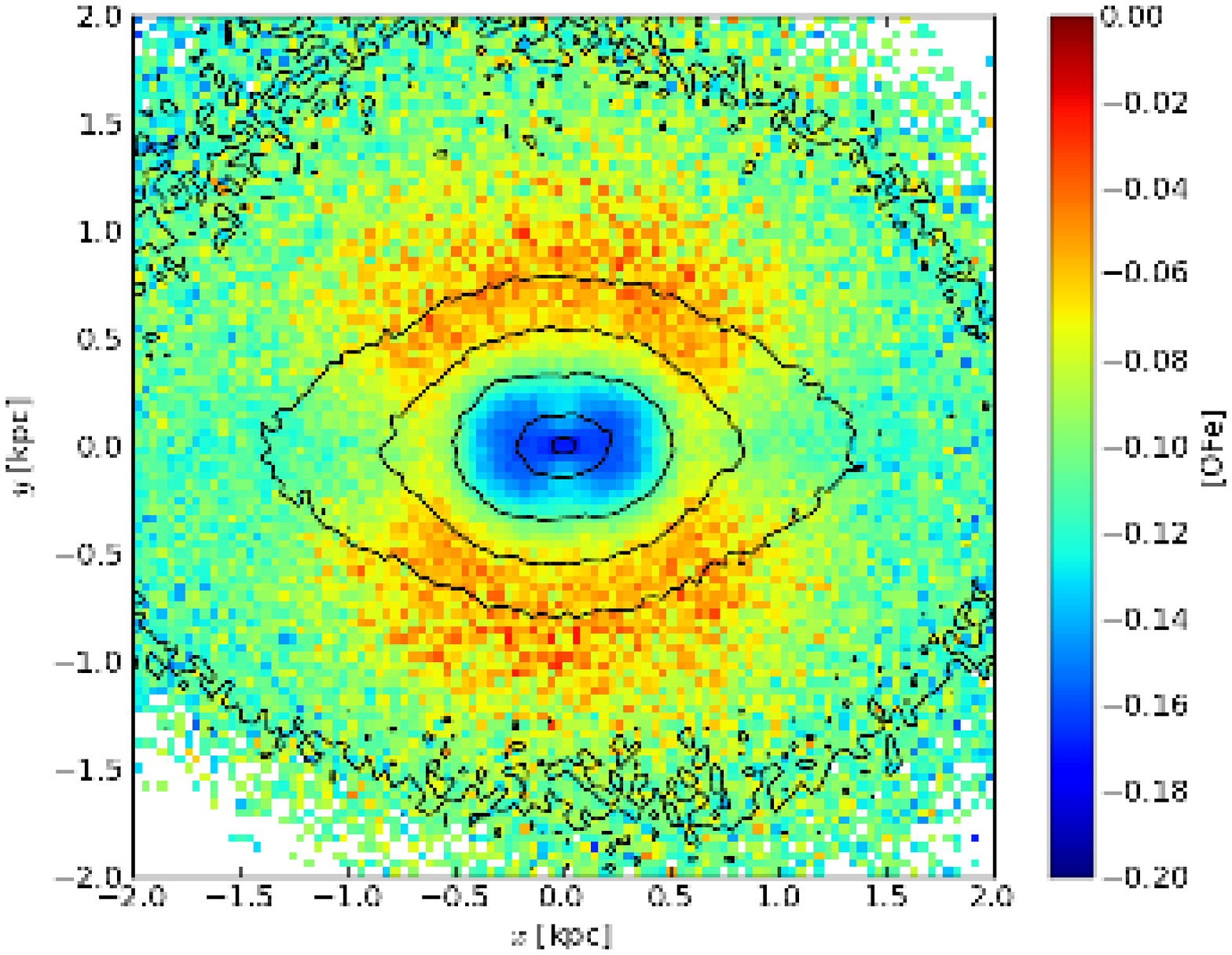} \\
\includegraphics[width=0.5\hsize,angle=0]{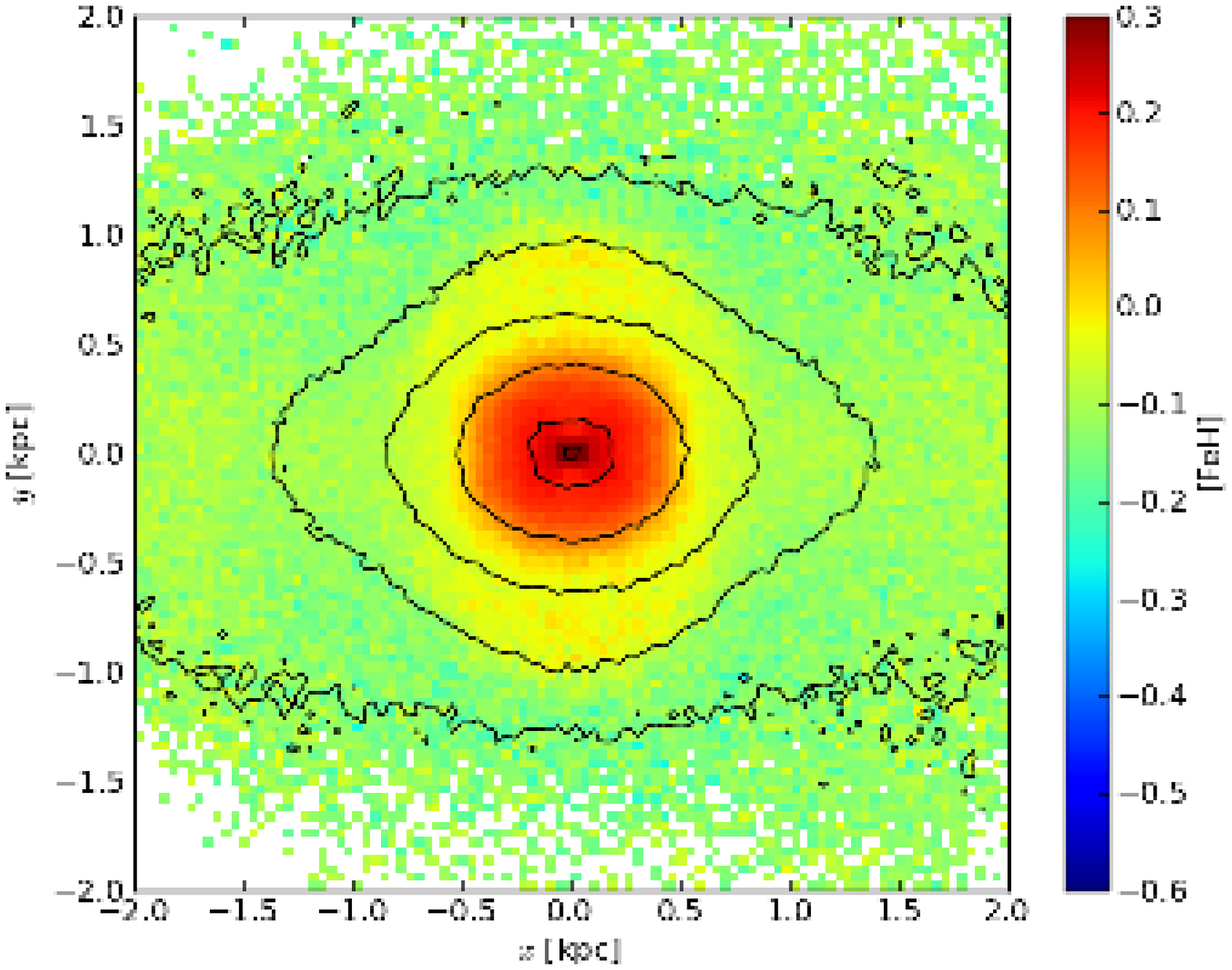} 
\includegraphics[width=0.5\hsize,angle=0]{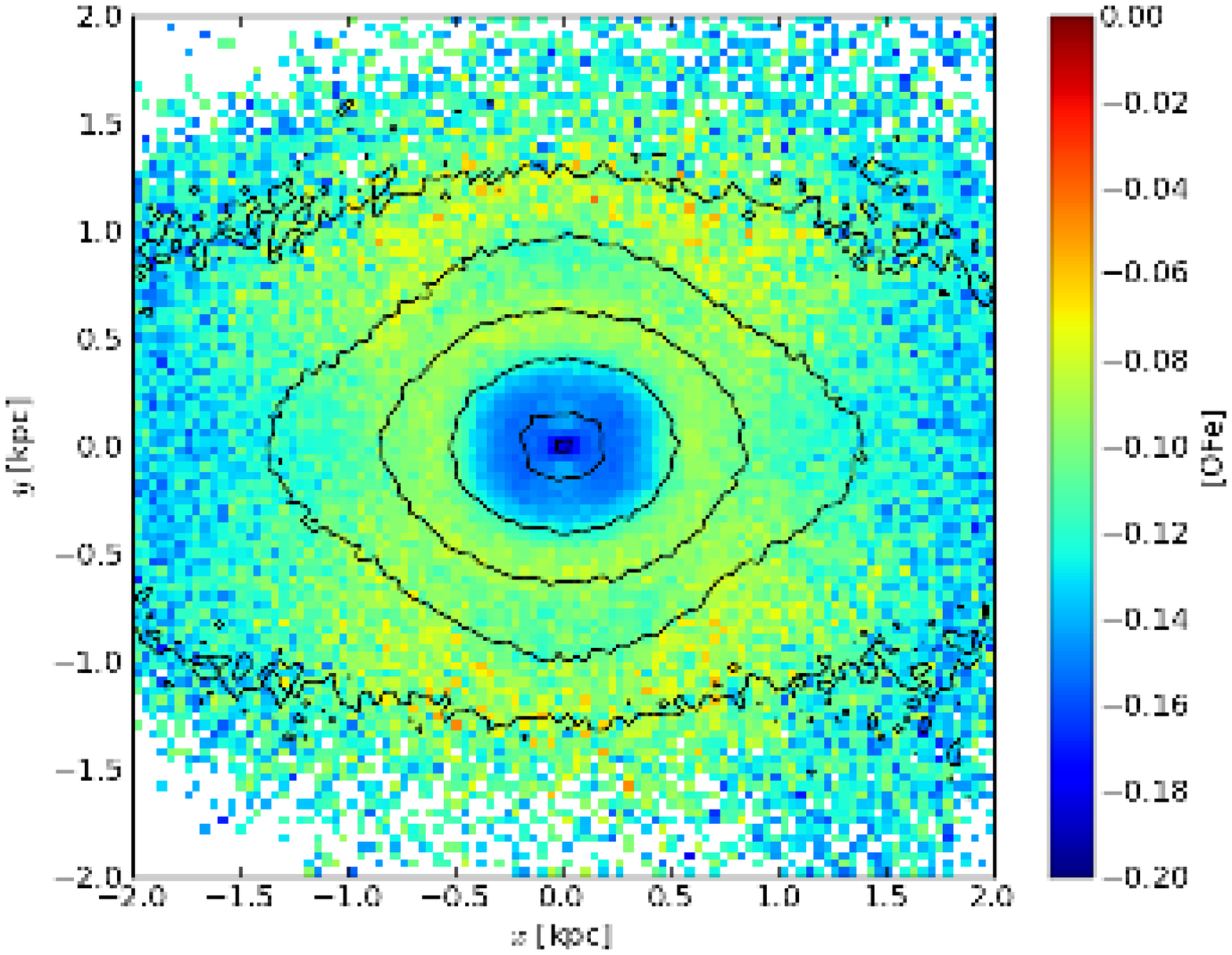} \\
\includegraphics[width=0.5\hsize,angle=0]{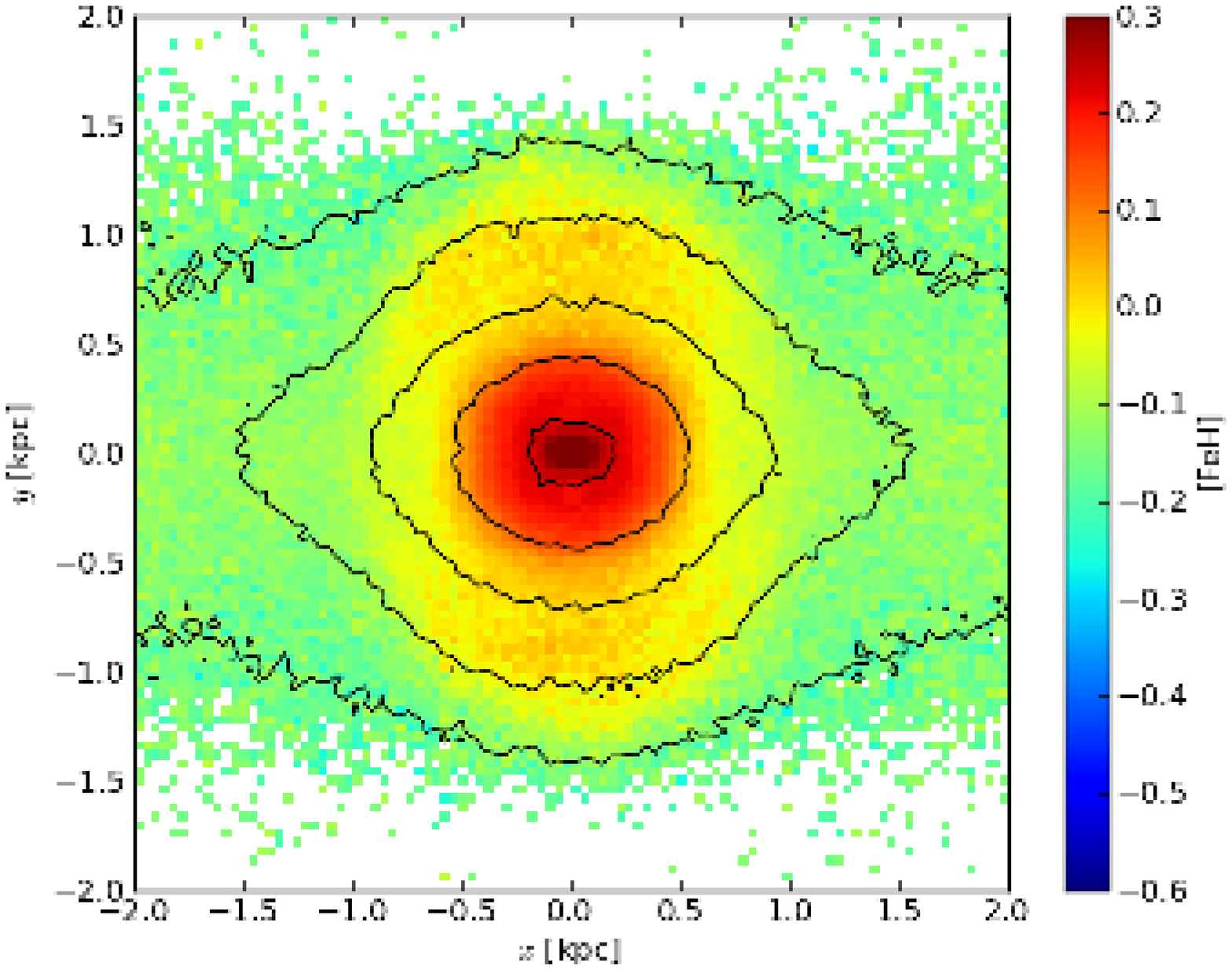} 
\includegraphics[width=0.5\hsize,angle=0]{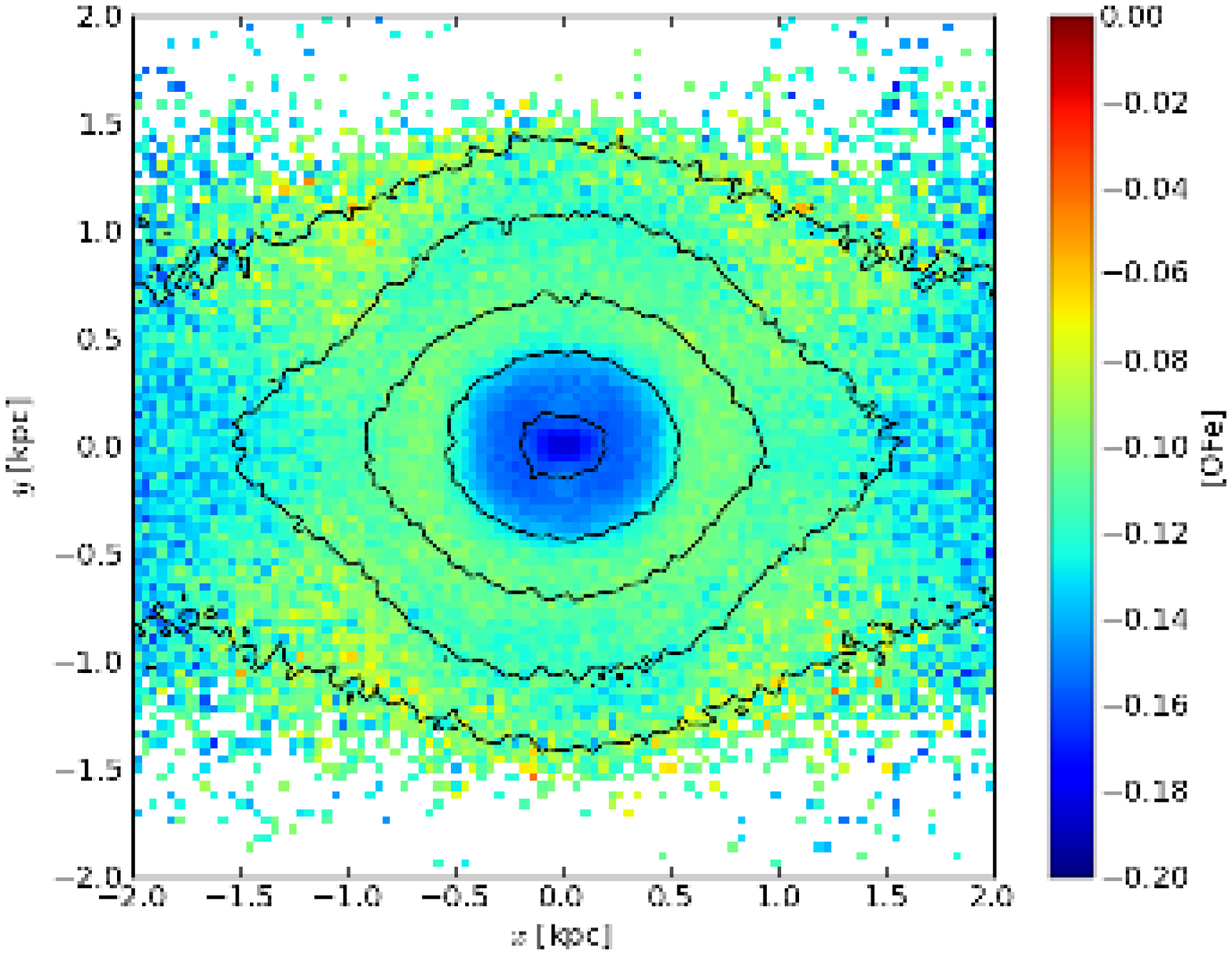} 
\end{tabular}
\caption{Mass-weighted mean stellar chemistry. The left column shows
  [Fe/H] while the right column shows [O/Fe]. The top row is at 6 Gyr,
  the middle is at 8 Gyr and the bottom is at 10 Gyr.  The stellar
  surface density is indicated by the contours.}
\label{fig:starchemistry}
\end{figure*}

Fig. \ref{fig:starchemistry} shows mass-weighted maps of the mean
stellar metallicity [Fe/H] and abundance [O/Fe].  The ND stands out
very clearly in the metallicity maps as a region that is more
metal-rich than the bar.  As we show below the star formation in this
region from metal-rich gas gives rise to this metallicity signature of
the ND, along the minor axis of the bar.  The [O/Fe] maps instead show
little direct evidence of the ND.  Before it forms, the inner 500 pc
region is $\alpha$-deficient but the surrounding region out to $\sim
1$ kpc is $\alpha$-enhanced along the bar minor axis.  Once the ND
forms, however, [O/Fe] drops by $\sim 0.05-0.1$ dex, and the ND [O/Fe]
blends in smoothly with that of the main bar.


\section{Cool Gas Properties}
\label{sec:gasdist}

\begin{figure}
\centering
\includegraphics[width=\hsize,angle=0]{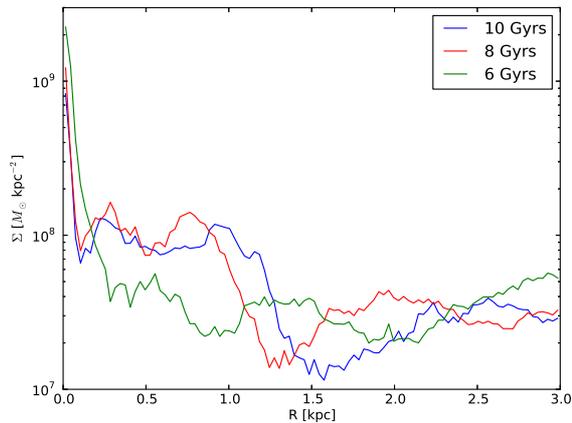}
\caption{The azimuthally-averaged cool gas surface density at 6, 8 and
  10 Gyr as indicated. }
\label{fig:sigmagas}
\end{figure}

\begin{figure}
\centering
\begin{tabular}{c}
\includegraphics[width=\hsize,angle=0]{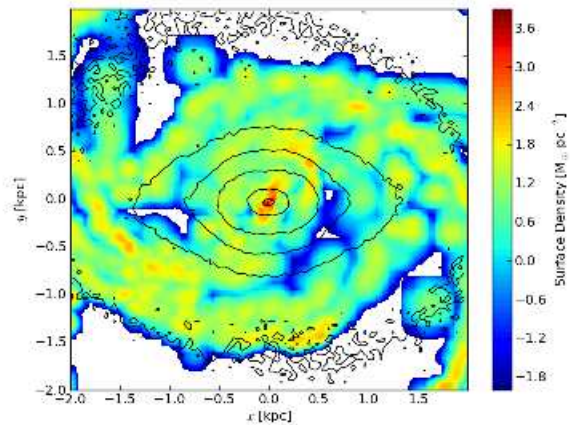} \\
\includegraphics[width=\hsize,angle=0]{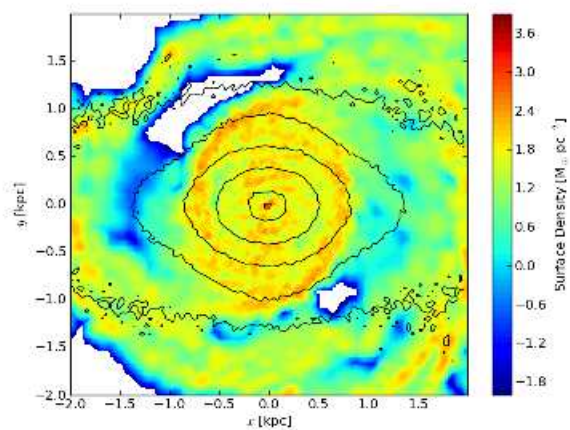} \\
\includegraphics[width=\hsize,angle=0]{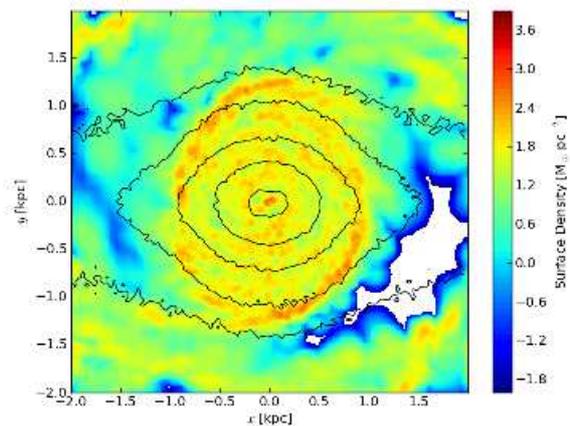} 
\end{tabular}
\caption{The cool gas surface density. The top row is at 6 Gyr,
    the middle is at 8 Gyr and the bottom is at 10 Gyr. The stellar
  surface density is indicated by the contours.}
\label{fig:rhogas}
\end{figure}

Fig. \ref{fig:sigmagas} presents the evolution of the
azimuthally-averaged cool gas (defined as gas that has cooled to below
$50,000$K) density within 3 kpc.  This shows an increase by a
factor of a few inside 1 kpc after 6 Gyr; this gas provides the fuel
from which the stellar ND forms.  Outside the ND, the gas density
remains similar between 6 and 10 Gyr.
Fig. \ref{fig:rhogas} maps the surface density of the cool gas.  The
distribution at 6 Gyr is clumpy with no evidence of a ND, while after
8 Gyr it is more regular, and forms a disc with spirals connecting up
to the gas inflowing along the leading edges of the bar.  Fig.
\ref{fig:barstrength} showed that the bar grows stronger at about 6
Gyr, which is responsible for initiating the gas inflow into the
nucleus.  Except in the inner $\sim 1$ kpc there is no evidence of
enhanced star formation and no external perturbers are present in the
simulation.  The ND in this simulation therefore is seeded by the
strengthening bar.

\begin{figure*}
\centering
\begin{tabular}{c}
\includegraphics[width=0.5\hsize,angle=0]{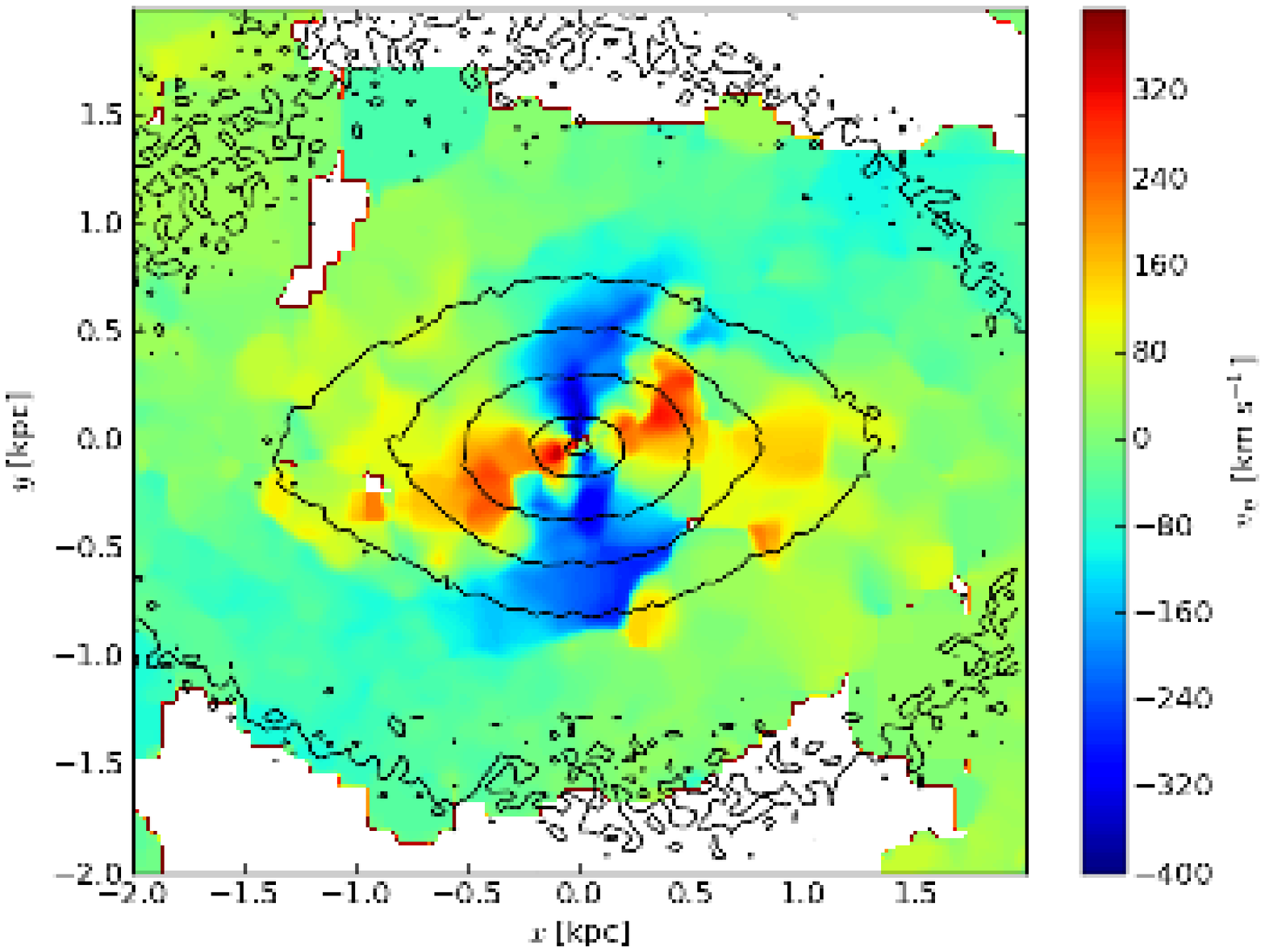} 
\includegraphics[width=0.5\hsize,angle=0]{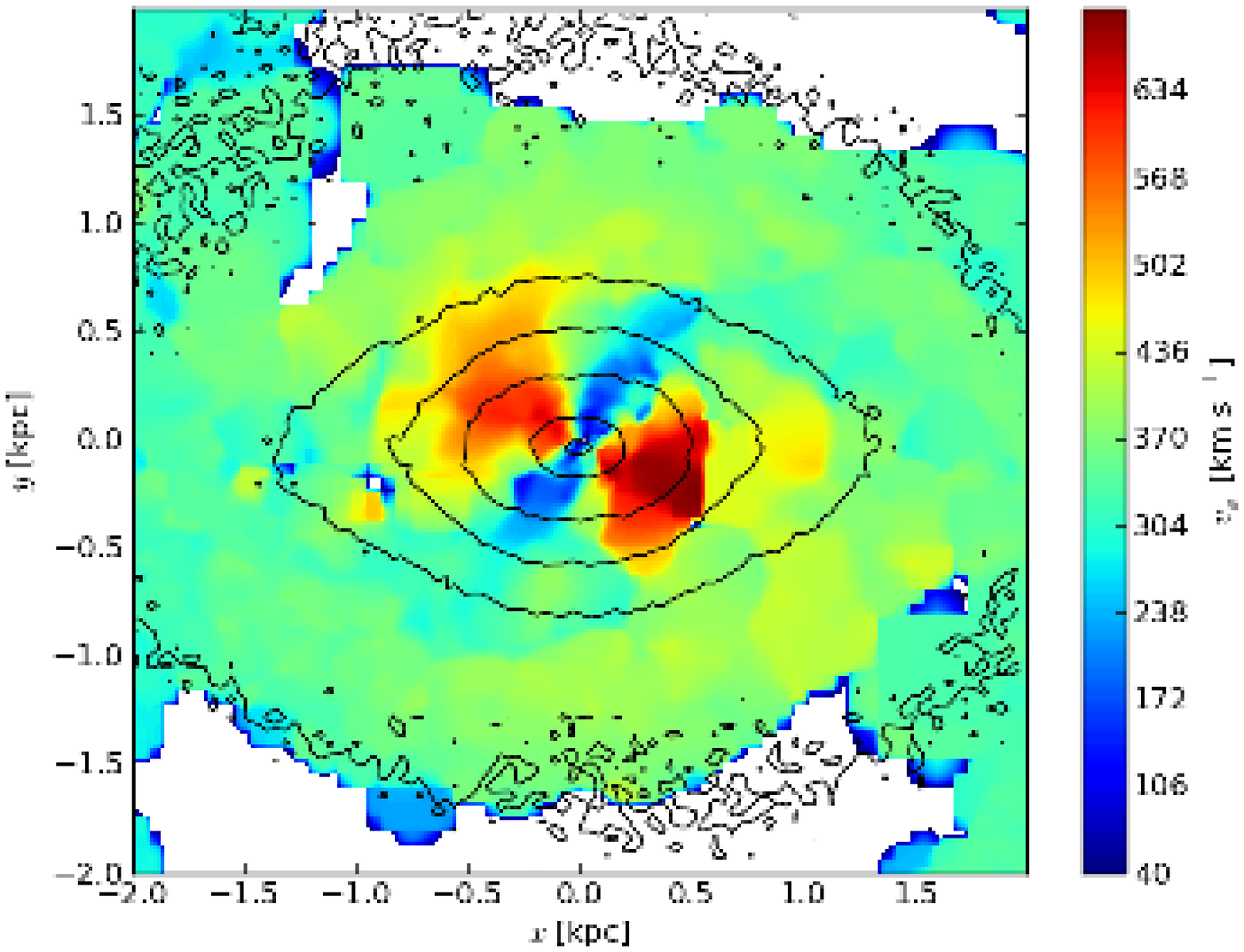} \\
\includegraphics[width=0.5\hsize,angle=0]{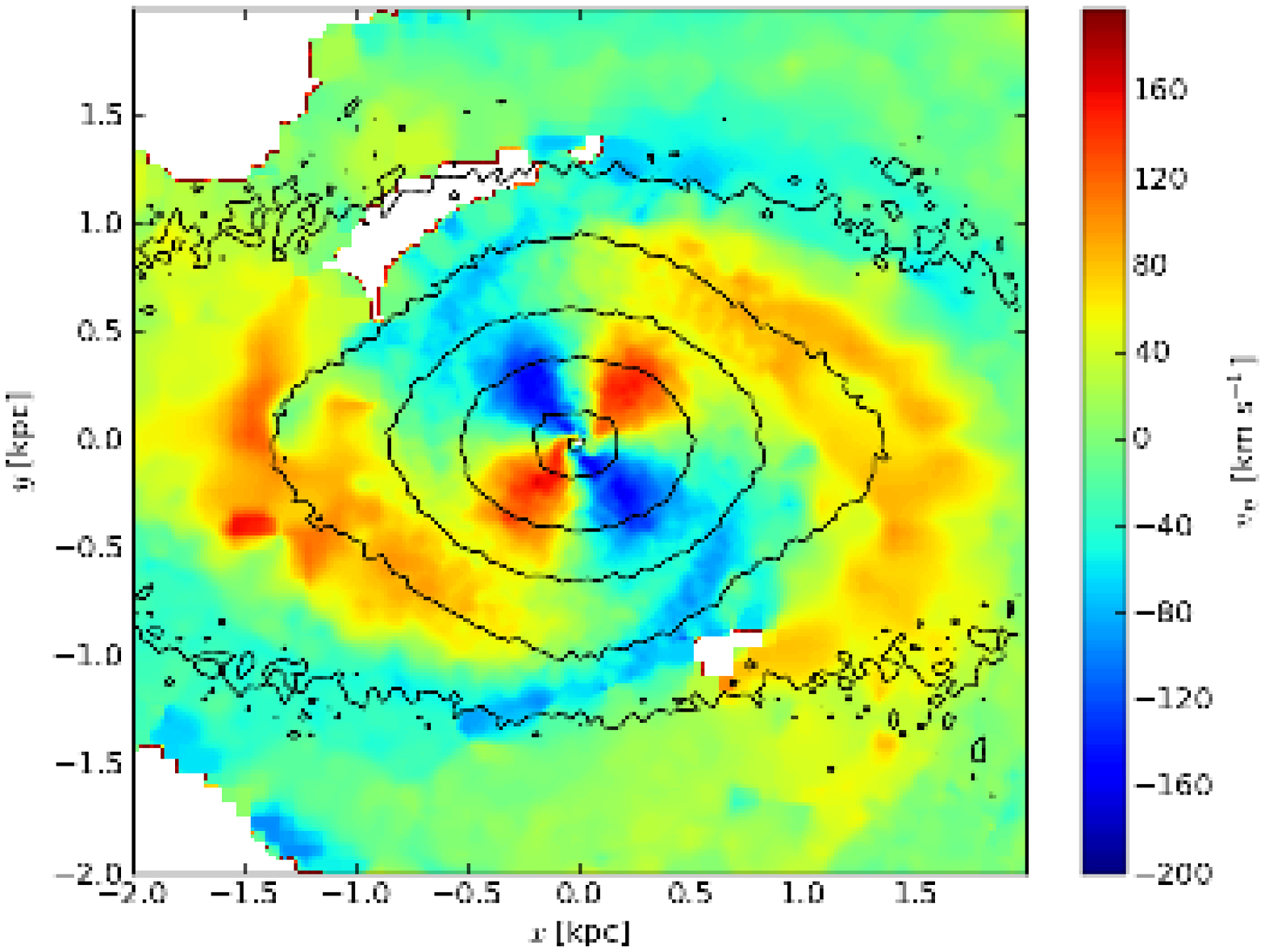} 
\includegraphics[width=0.5\hsize,angle=0]{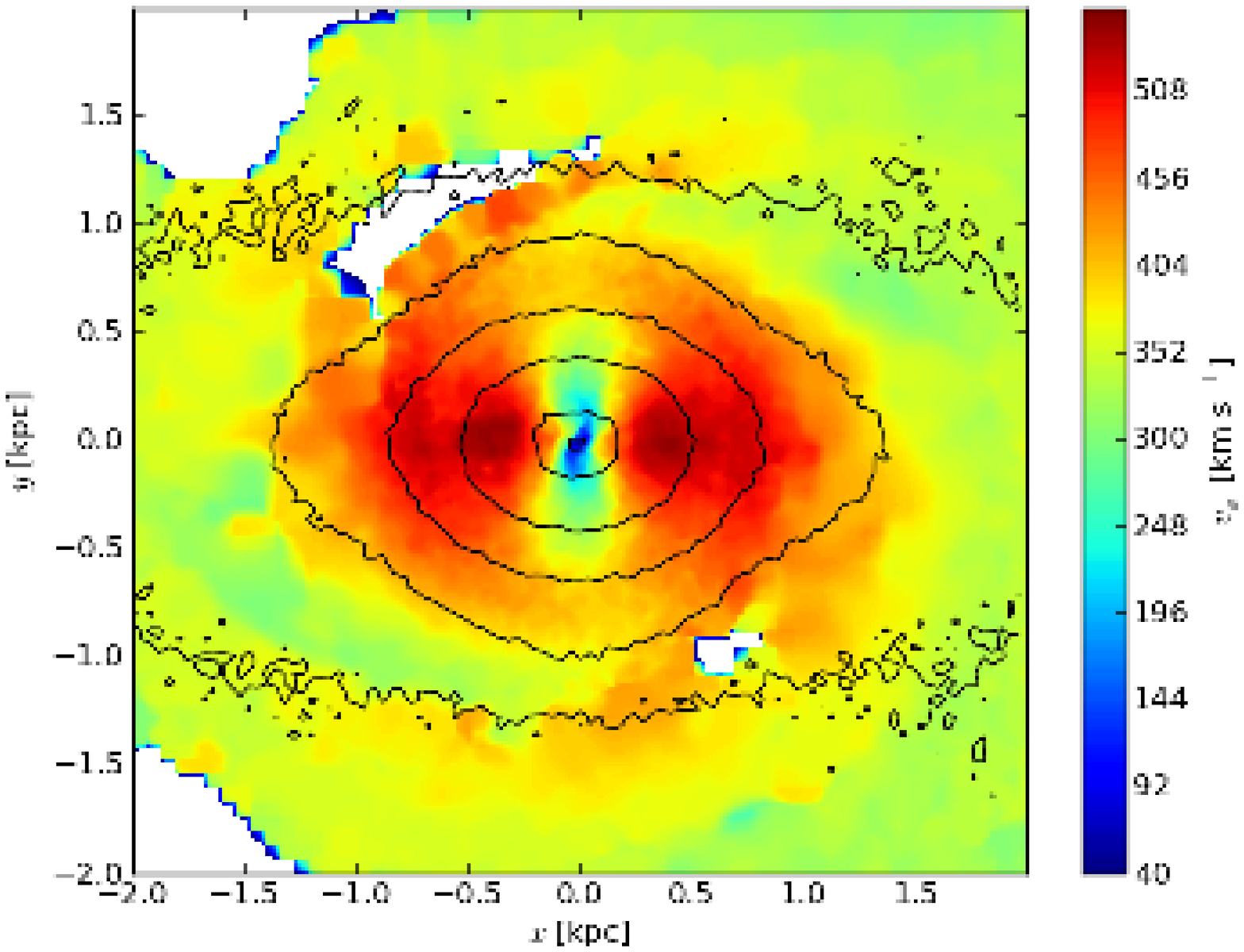} \\
\includegraphics[width=0.5\hsize,angle=0]{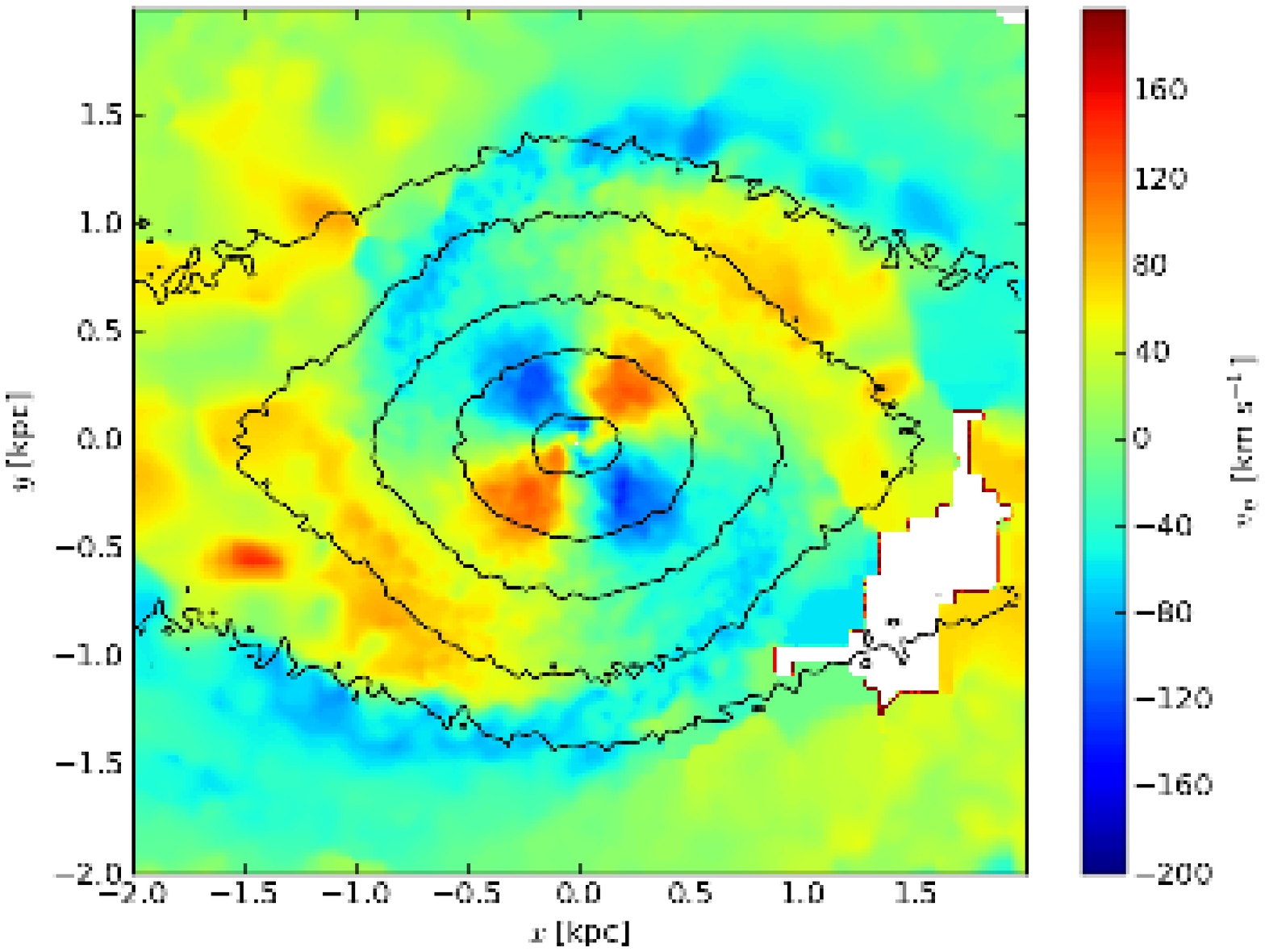} 
\includegraphics[width=0.5\hsize,angle=0]{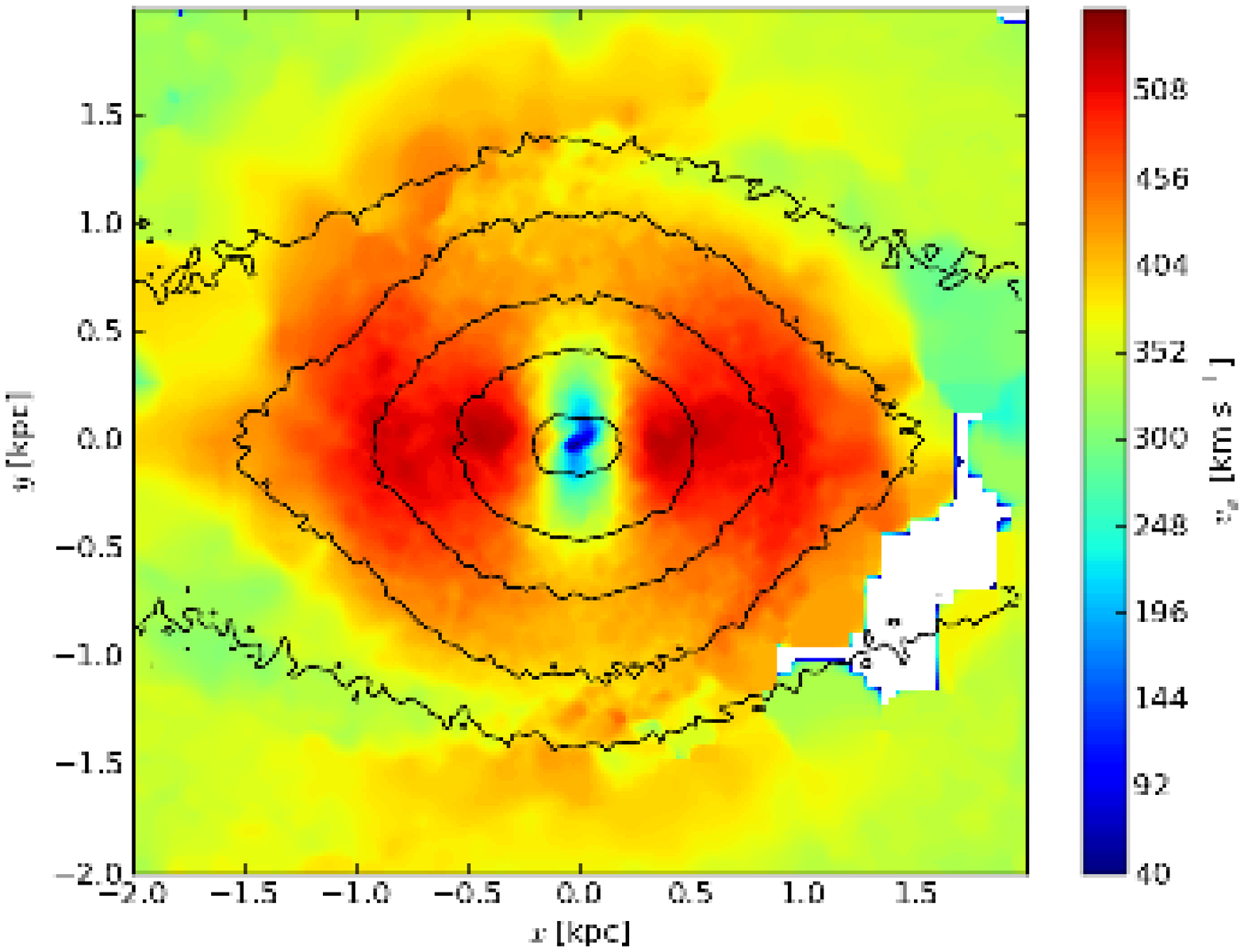} 
\end{tabular}
\caption{The cool gas kinematics. The left column shows $v_R$ while
  the right column shows $v_\phi$. Top is at 6 Gyr, middle is at 8 Gyr
  and bottom is at 10 Gyr.  The stellar surface density is indicated
  by the contours.  Note the change in scale between the 6 Gyr frames
  and the rest.}
\label{fig:gaskin}
\end{figure*}

\begin{figure*}
\centering
\begin{tabular}{c}
\includegraphics[width=0.5\hsize,angle=0]{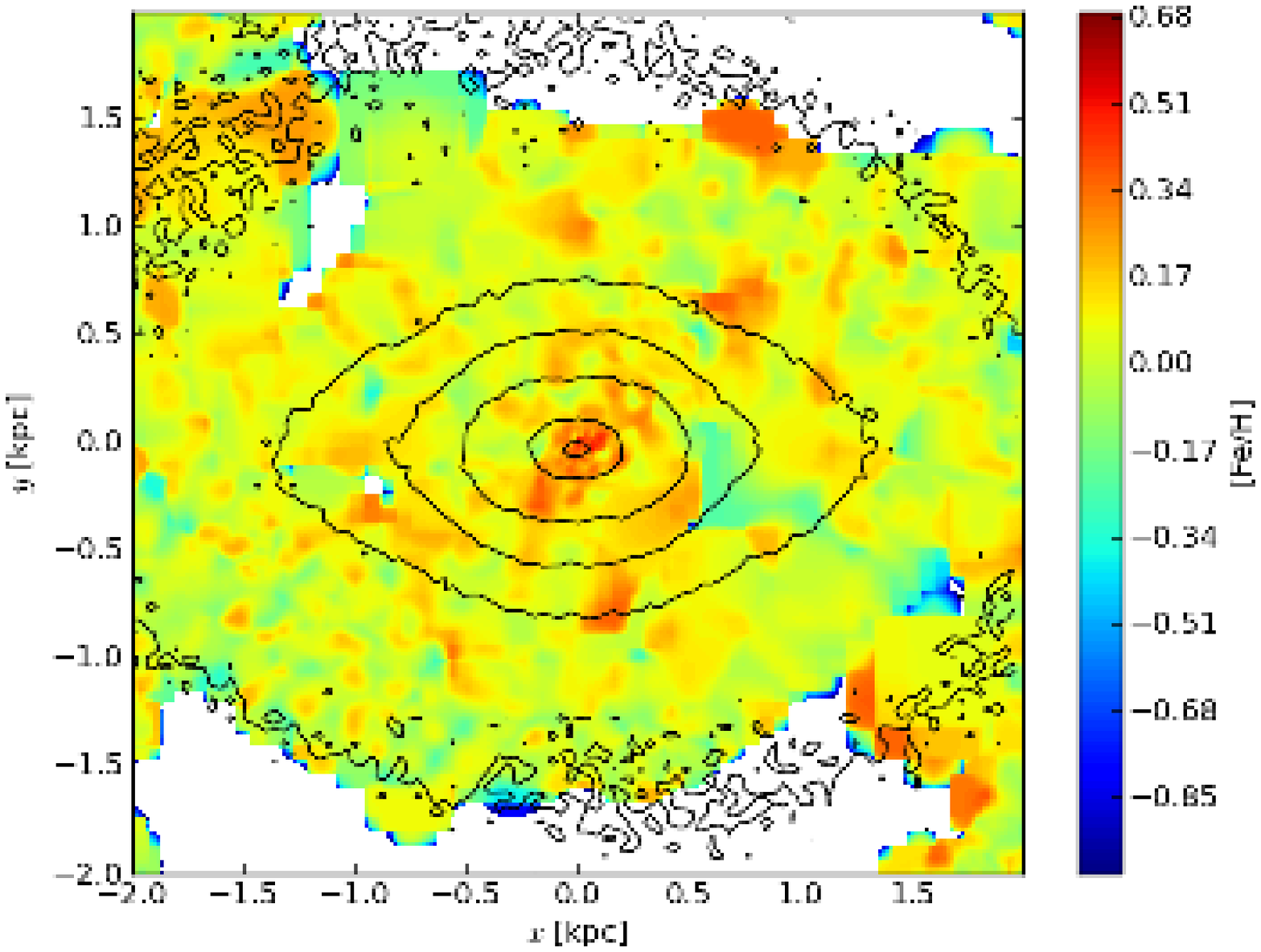} 
\includegraphics[width=0.5\hsize,angle=0]{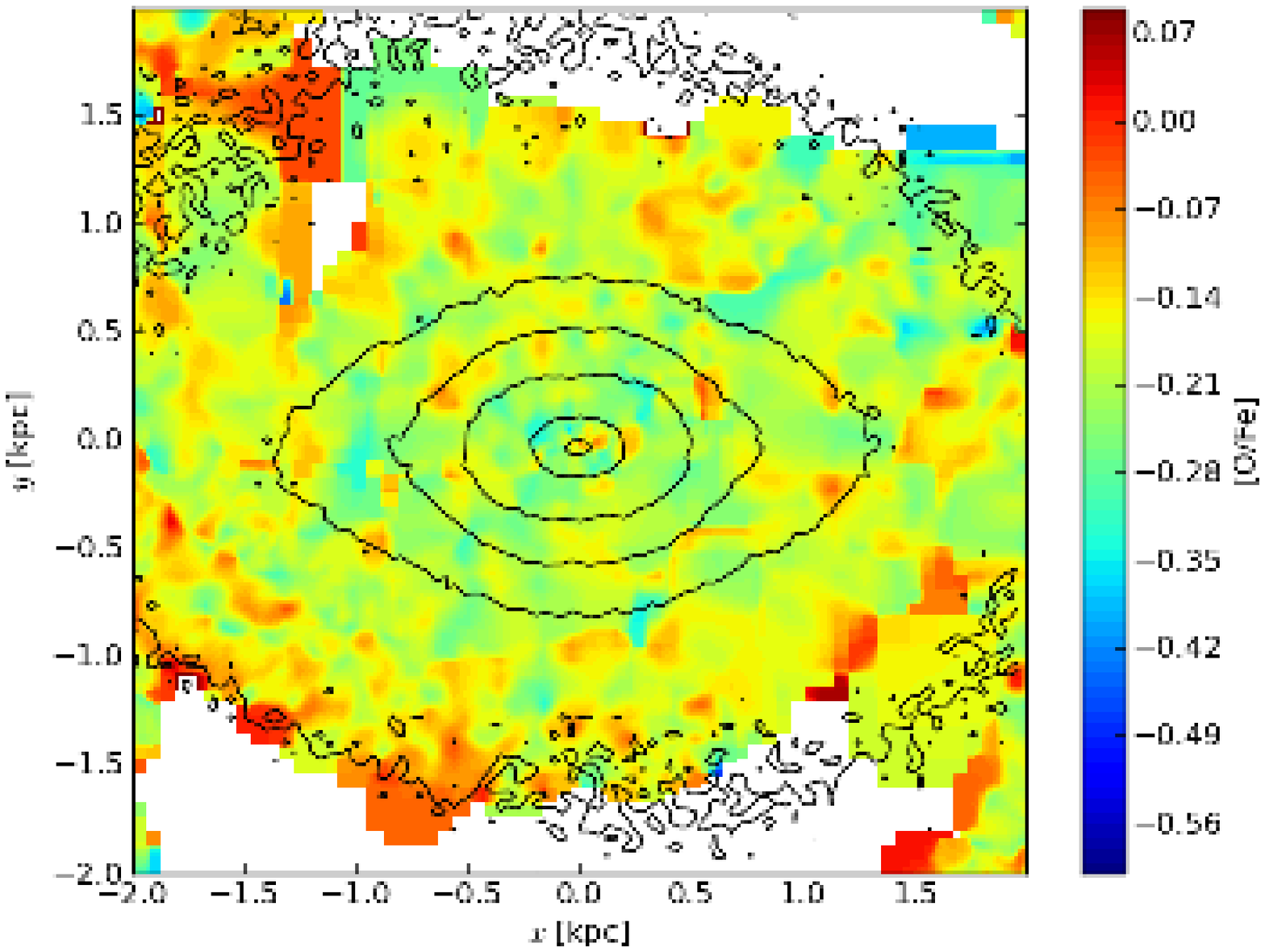} \\
\includegraphics[width=0.5\hsize,angle=0]{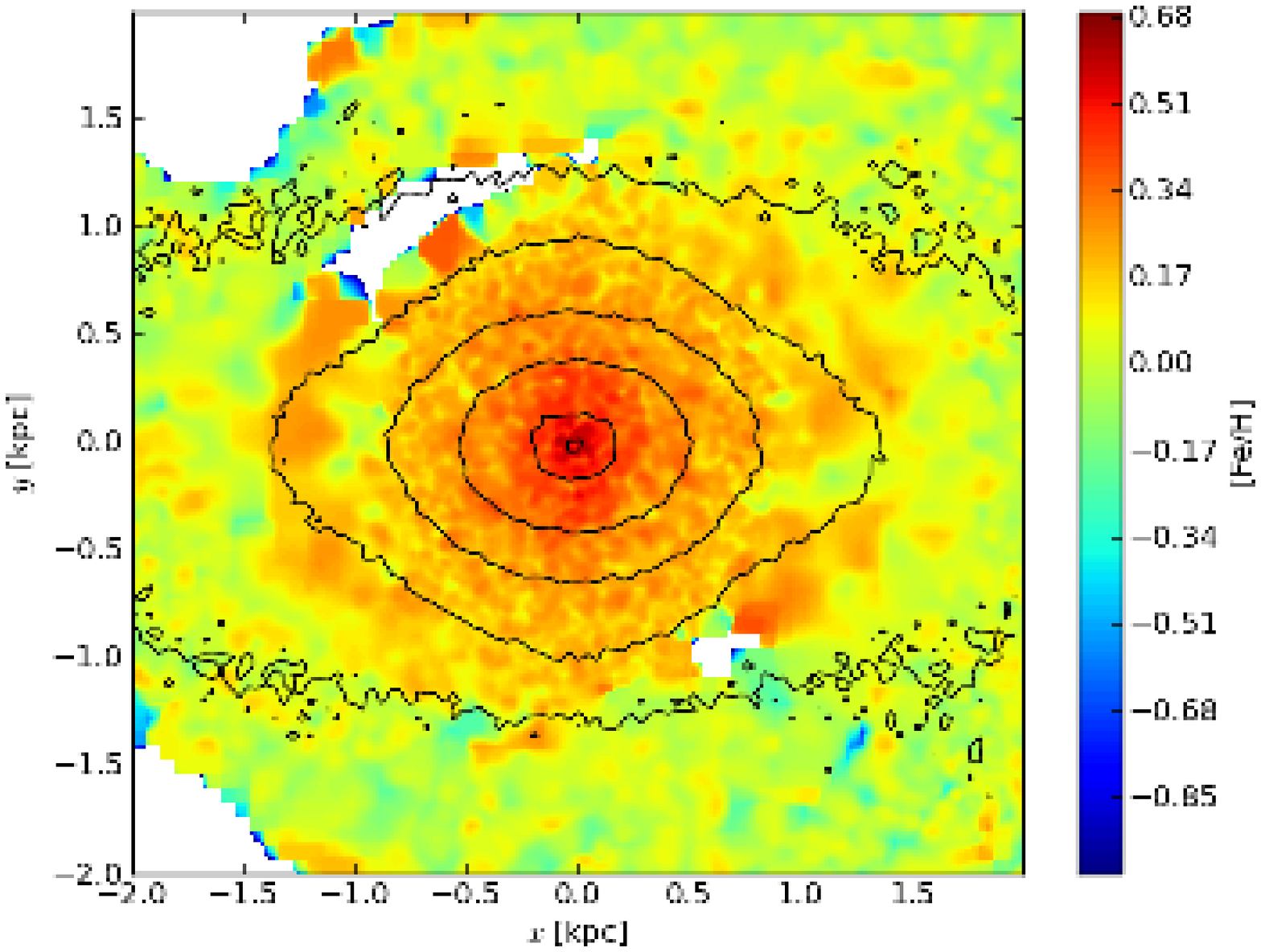} 
\includegraphics[width=0.5\hsize,angle=0]{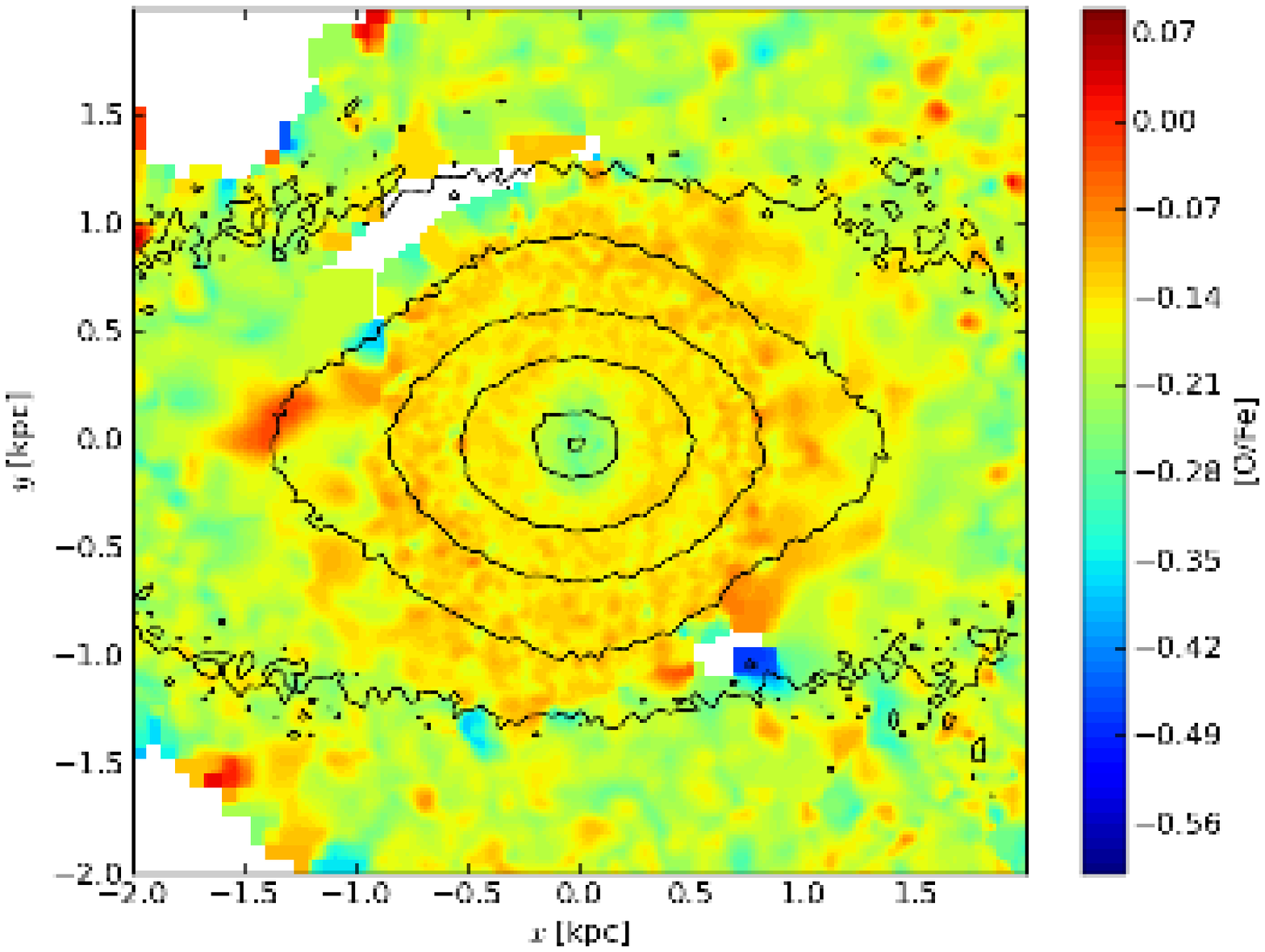} \\
\includegraphics[width=0.5\hsize,angle=0]{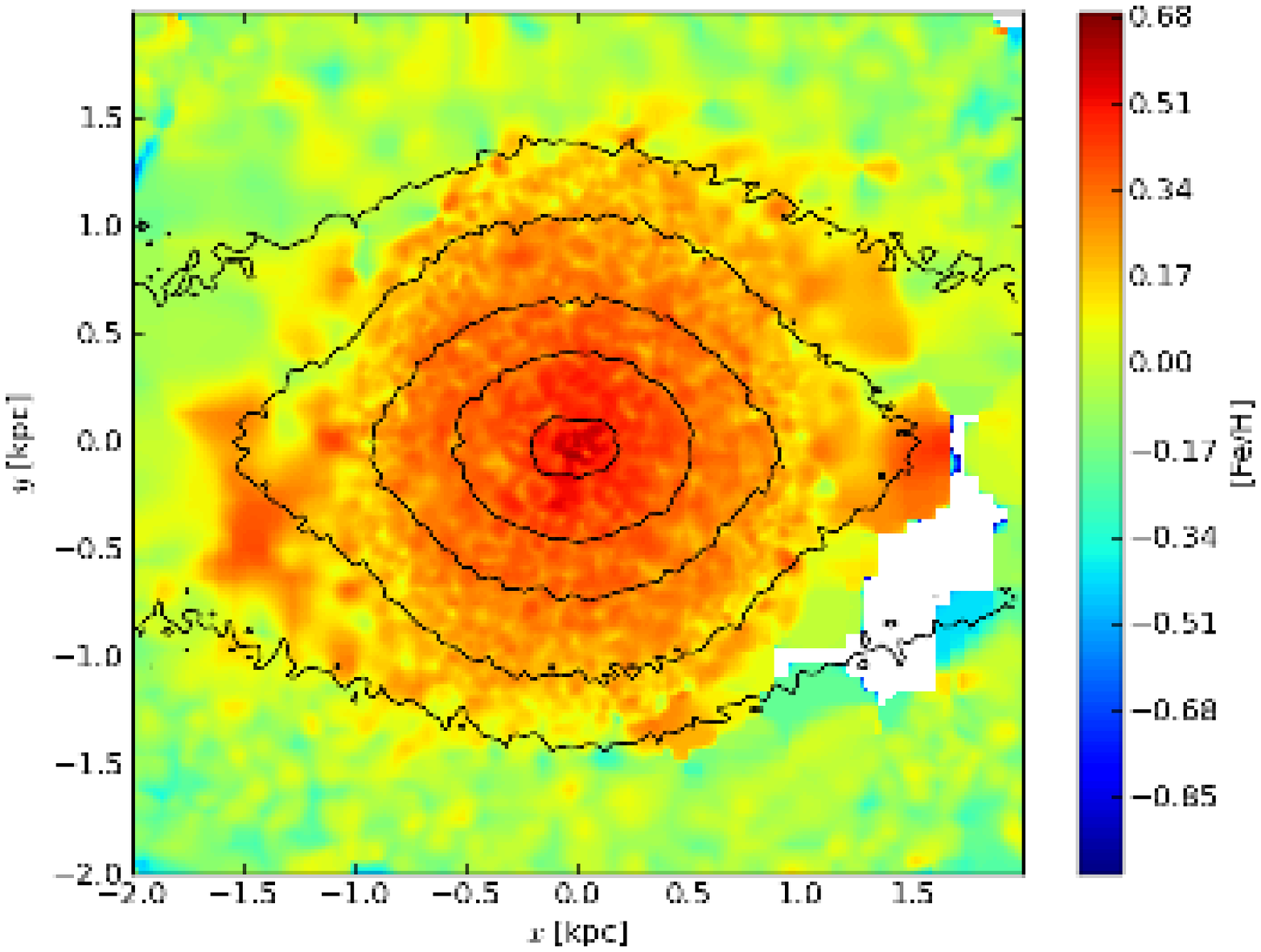} 
\includegraphics[width=0.5\hsize,angle=0]{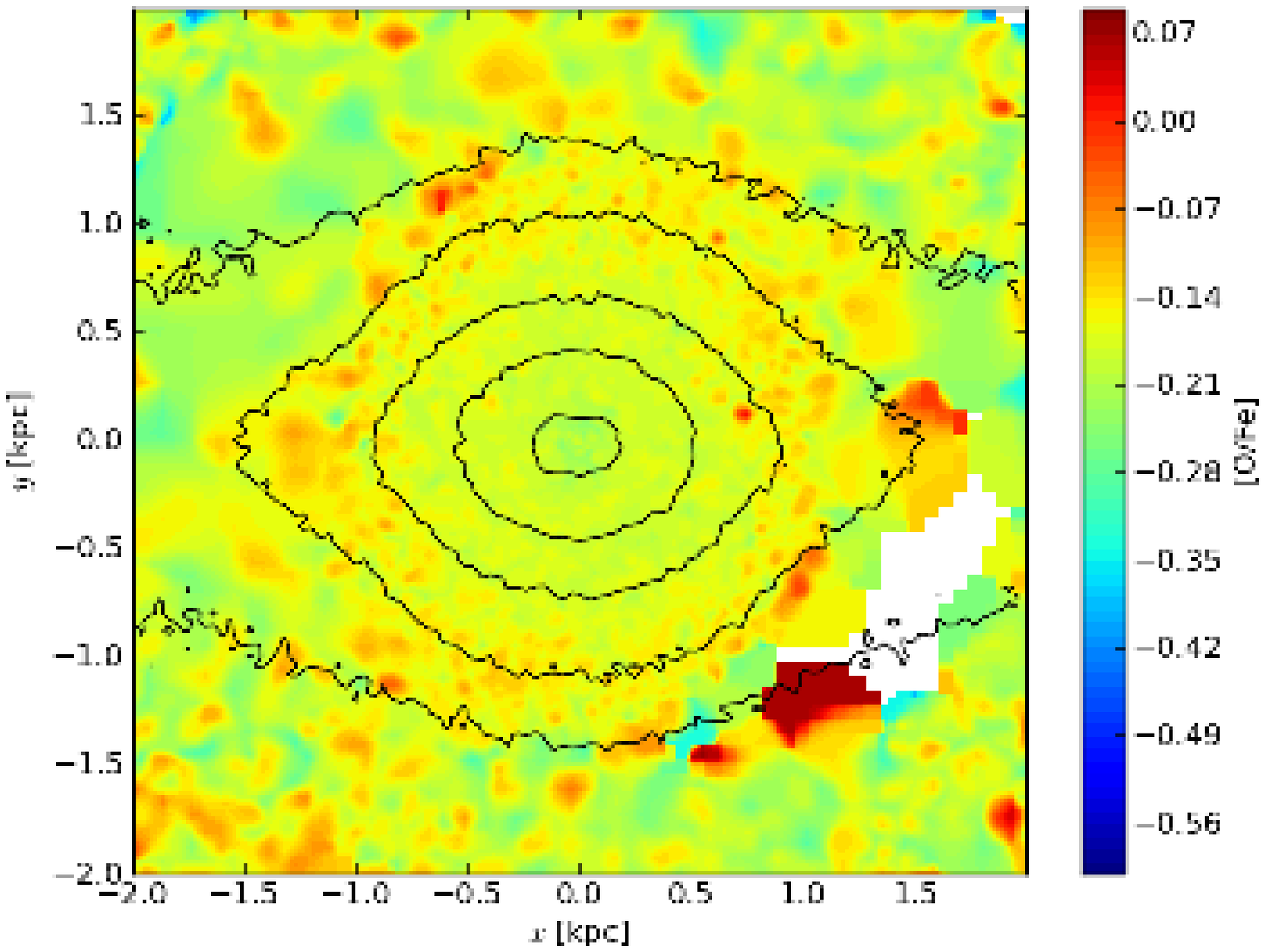} \\
\end{tabular}
\caption{The cool gas chemistry. The left column shows [Fe/H] and the
  right column shows [O/Fe]. The top row is at 6 Gyr, the middle at 8
  Gyr and the bottom at 10 Gyr.  The stellar surface density is
  indicated by the contours.}
\label{fig:gaschemistry}
\end{figure*}

Fig. \ref{fig:gaskin} shows the radial and tangential velocities of
the cool gas. At 6 Gyr there is a gas inflow along the leading sides
of the bar, arriving at the centre with peak inward (outward) velocity
along the bar's minor (major) axis.  The tangential velocity is very
disordered at 6 Gyr.  This is a quite different velocity field from
the stellar one at this time, and is dominated by dissipative inflows
rather than by rotation.  At 8 Gyr $v_R$ has maxima and minima
oriented at approximately $45 \degrees$ to the bar indicating the gas
is now moving under the gravitational field of a massive ND.  The peak
radial velocities are progressively reduced over time but the inflow
and outflow continues.  The $v_\phi$ maps show strong rotation in an
elliptical disc at the later times.  It is worth noting that $v_R$ and
$v_\phi$ for the gas and the stars are rotated by $\sim 90\degrees$
relative to each other (compare figures \ref{fig:starvels} and
  \ref{fig:gaskin}).  The cause of this is that the gas velocities
reflect motions in the ND only, while the stellar velocities are a
superposition of the motions of stars in the bar and in the ND.
Instead the gas kinematics are quite similar to the kinematics of the
young stars, as seen in Fig. \ref{fig:starvels10Gyr}.

Fig. \ref{fig:gaschemistry} shows the mass-weighted nuclear gas
chemistry.  At 6 Gyr the gas has slightly elevated metallicity.  The
irregular distribution of both [Fe/H] and [O/Fe] is caused by gas
inflowing from larger radii.  Once the ND forms the cool gas is
significantly more metal-rich than the surrounding gas.  At 8 Gyr the
gas is $\alpha$-enhanced, a result of the rapid star formation already
ongoing at 6 Gyr (see Fig.  \ref{fig:sfh}), although it is only later
that the star formation settles into an ordered ND.  By 10 Gyr the
[O/Fe] of the gas has dropped to background levels as star formation
continues to pollute the gas with SNIa ejecta.


\section{Comparison to Real Galaxies}
\label{sec:compobs}

\subsection{Photometric Comparisons}

In this Section we compare the simulation to three early-type disc
galaxies where previous study has shown the existence of distinct,
moderately large NDs and/or massive stellar nuclear rings (which could
plausibly be the most visible signature of an extended ND) inside bars
\citep{erwin99, erwin03, erwin04, nowak10, comeron10}.  Further
details concerning the NDs, including their relation to the central
bulges, can be found in \citet{Erwin2014b}.  For each galaxy, we orient
the simulation to match the galaxy as best as possible, first
reproducing the (deprojected) position angle of the bar relative to
the line of nodes and then inclining the simulation by the same angle
as the galaxy.





\subsubsection{Comparison to NGC 4371}
\label{sec:n4371}

\begin{figure*}
\centering
\begin{tabular}{c}
\includegraphics[width=0.8\hsize,angle=0]{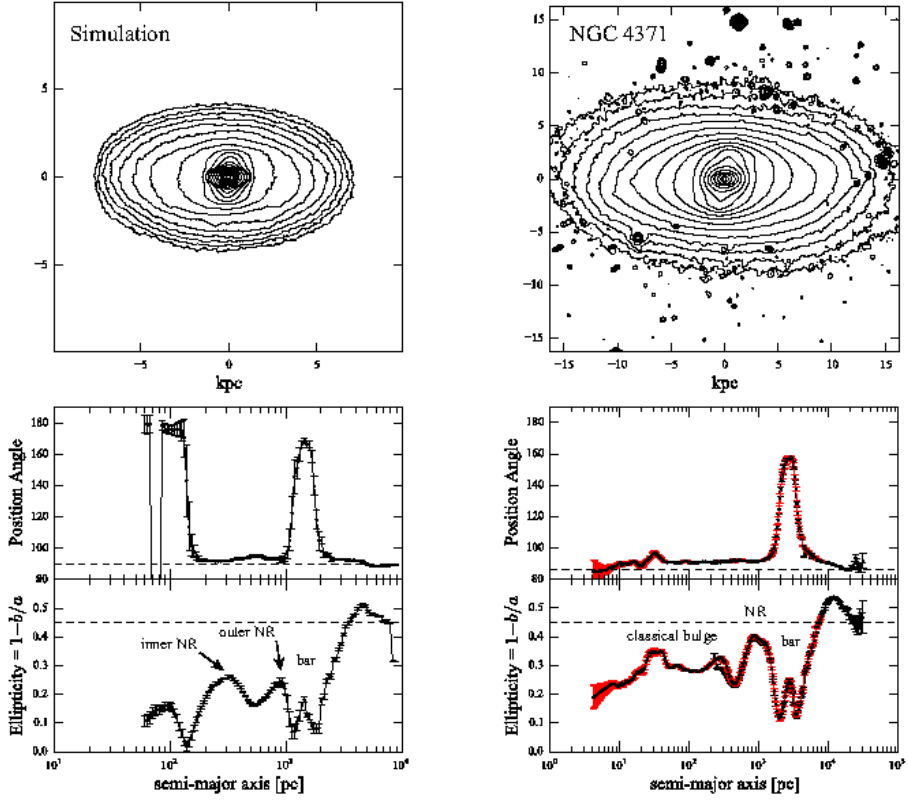} 
\end{tabular}
\caption{ Comparison of simulation with S0 galaxy NGC~4371. Top
  panels: log-scaled isodensity contours (left) and $r$-band isophotes
  (right), median-smoothed, scaled so that the bar is approximately
  the same apparent size. The simulation (left) is oriented to match
  NGC~4371 (bar at deprojected $\Delta$PA = 85\degr{} relative to the
  major axis, galaxy inclination = 58\degr); both have been rotated to
  make the outer-disc major axis horizontal in the panels.  Bottom
  panels: ellipse fits, plotted on a logarithmic semi-major axis
  scale; the NGC~4371 plots (right) combine ellipse fits to the
  INT-WFC $r$-band image from the top right panel (black) and an
  \textit{HST} ACS F850LP image (red). In the simulation plot (left
  panel), the position angle (PA) is plotted so that the major axis is
  at $\approx 90\degr$; the PA for NGC~4371 is the observed value,
  measured CCW from north. Dashed lines indicate estimated outer-disc
  PA and ellipticity.  Labels indicate features corresponding to the
  inner and outer nuclear rings and the bar in the simulation (``inner
  NR'', ``outer NR'', ``bar''), along with the classical bulge,
  nuclear ring, and bar in NGC~4371 (``classical bulge'', ``NR'',
  ``bar''). }
\label{fig:n4371}
\end{figure*}

NGC~4371 is a strongly barred S0 galaxy in the Virgo Cluster with a
surface-brightness-fluctuation distance of 16.9 Mpc
\citep{blakeslee09}.  High-resolution imaging studies have shown that
the interior of the bar is dominated by a very elliptical stellar
nuclear ring with a radius of $\approx 10.5\arcsec$ (860 pc); there is
little evidence for dust except in the very innermost ($r < 1\arcsec$)
regions and no evidence for current star formation, though the stellar
nuclear ring is somewhat blue \citep{erwin99,comeron10}. The bar is
aligned almost along the minor axis of the projected disc; the outer
isophotes have an ellipticity of $\approx 0.45$, indicating an
inclination of $i \approx 58\degr$ (assuming the disc has an intrinsic
thickness of $c/a = 0.2$, typical for S0 galaxies
\citep{Lambas1992,Padilla2008,Rodriguez2013}). For the isophotes
(upper right panel of Fig.~\ref{fig:n4371}), we use an $r$-band image
from the Isaac Newton Telescope's Wide Field Camera \citep{erwin08};
we also use an \textit{HST} ACS/WFC F850LP image from the ACS Virgo
Cluster Survey \citep{cote04} for the ellipse fits (lower panels of
the same figure).

The appearance of both the outer disc and the bar is very similar in
the simulation and the real galaxy. A clear ellipticity peak
\textit{outside} the bar, at a semi-major axis $a$ of 4.6 kpc in the
simulation and 12 kpc in NGC~4371, is most likely due to a slightly
non-circular outer ring; the ellipticity outside this peak approaches
a roughly constant, lower value which is (in the case of the
simulation) consistent with the adopted inclination. The bar itself
manifests as a strong ellipticity minimum with a weak ellipticity peak
in the center of that minimum, accompanied by a strong twist in the
position angle.

Inside the bar, NGC~4371 has a strong inner ellipticity peak at around
800--900 pc, corresponding to the stellar nuclear ring. The
simulation, in contrast, has two inner ellipticity peaks: the outer of
the two corresponds to the outer nuclear ring, at the edge of the ND,
while the inner peak is due to the inner nuclear ring.  \citep[The
apparent inner ellipticity peak at $a \sim 300$ pc in NGC~4371 is
primarily a side effect of the combination of very elliptical
isophotes from the nuclear ring and rounder isophotes from the bulge
inside; see ][]{erwin01}.

The fact that the ND in the simulation is slightly elliptical rather
than circular results in a slight offset between the orientation of
the outer disc (89\degr) and the ellipticity peaks due to the nuclear
disc ($\approx 91.5$--92\degr). A similar offset can be seen in
NGC~4371 (PA $= 86\degr$ for the outer disc and 92\degr{} for the
inner ellipticity peak associated with the nuclear ring), which
suggests that the nuclear disc/ring in NGC~4371 is slightly elliptical
as well.

\citet{Erwin2014b} estimate a ND mass of $7.6 \times 10^9$ \Msun\ for
NGC~4371, or 17 per cent of its total stellar mass, comparable to, but
smaller than, the mass fraction measured in the model at 10 Gyr.  The
size of the ND, as a fraction of the deprojected radius at which the
bar ellipticity peaks, is $0.17$.  If instead we use a bar radius at
which the $m=2$ phase deviates from a constant by more than
$10\degrees$ then the ND size is 0.14.  In comparison, the simulation
has a ND of size 0.34 relative to the radius of peak ellipticity (0.3
relative to the radius at which the bar phase deviates from a constant
by more than $10\degrees$).  Thus the ND in the model is about twice
as large as the one in NGC~4371.



\subsubsection{Comparison to NGC 3945}
\label{sec:n3945}

NGC~3945 is a double-barred S0 galaxy at a distance of approximately
19.8 Mpc (based on the Virgocentric-corrected HyperLeda redshift and a
Hubble constant of 72 \kms~kpc$^{-1}$). Its orientation is very
similar to that of NGC~4371 (bar $\Delta$PA = 88\degr, inclination =
55\degr).  Unlike NGC~4371, NGC~3945 also has a small nuclear bar
\citep{erwin03}, which our model does not.

\begin{figure*}
\centering
\begin{tabular}{c}
\includegraphics[width=0.8\hsize,angle=0]{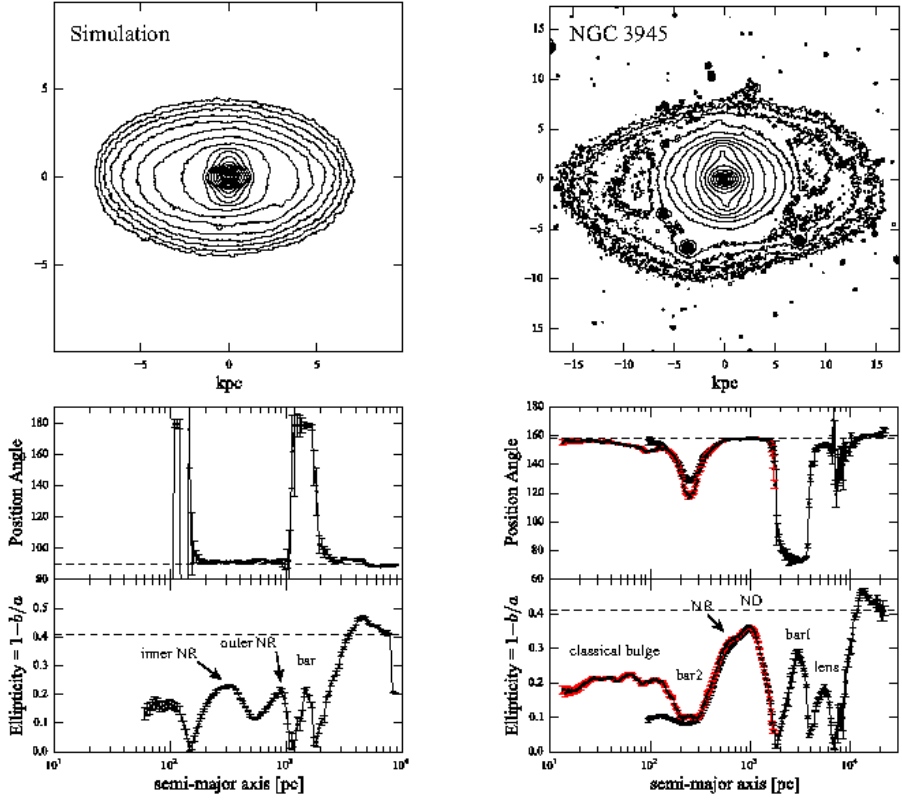} 
\end{tabular}
\caption{ As for Fig.~\ref{fig:n4371}, but now comparing the
  simulation with the double-barred S0 galaxy NGC~3945. Simulation and
  galaxy orientation are: primary bar at $\Delta$PA = 88\degr,
  inclination = 55\degr.  The ellipse-fit plots use data from an
  INT-WFC $r$-band image (black) and an \textit{HST} WFPC2 F814W image
  (red).  In the lower-right panel, labels indicate features
  corresponding to the compact classical bulge, the secondary bar, the
  nuclear ring, the ND, and the primary bar in NGC~3945 (``classical
  bulge'', ``bar2'', ``NR'', ``ND'', `` bar1'').}
\label{fig:n3945}
\end{figure*}

The primary bar in NGC~3945 is aligned almost along the minor axis of
the projected disc; the outer isophotes have an ellipticity of
$\approx 0.41$, indicating an inclination of $i \approx 55\degr$. The
plotted isophotes (upper right panel of Fig.~\ref{fig:n3945}) are
derived from an $r$-band image from the Isaac Newton Telescope's Wide
Field Camera \citep{erwin08}, while the ellipse fits are based on this
image and an \textit{HST} WFPC2 F814W image (PI Carollo, proposal ID
6633).

Because the orientation is almost identical to that of NGC~4371, the
appearance of the projected simulation and the ellipse fits are very
similar to the previous case, and we refer the reader to
Section~\ref{sec:n4371} for more details. As in the case of NGC~4371,
the resemblance between the simulation and the real galaxy is rather
good for the outer isophotes, including a distinct ellipticity peak
corresponding to the outer ring (much more prominent in this galaxy than
in NGC~4371). There is an extra ellipticity peak in NGC~3945 at a
semi-major axis of $a \sim 5.5$ kpc, in between the bar and the outer
ring; this is due to the outer edge of the lens surrounding the bar.

Interior to the bar (which has its maximum ellipticity at $a \sim
3.1$~kpc, corresponding to a deprojected semi-major axis of 5.3~kpc),
NGC~3945 shows a strong ellipticity peak ($a \sim 1$~kpc) due to its
ND\footnote{This is the same structure which was termed an ``inner
  disc'' by \citet{erwin03} and \citet{erwin04}.}. As shown by
\citet{erwin99}, there is a stellar nuclear ring within this disc,
which produces the slight shoulder in the ellipticity profile at $a
\sim 600$ pc. Further inside is a secondary bar oriented almost
parallel to the primary bar and thus close to the galaxy minor axis;
the projection of this produces the ellipticity \textit{minimum} at $a
\sim 250$~pc; a small classical-bulge component dominates the
isophotes at $a \la 100$~pc \citep{erwin03, Erwin2014b}.

\citet{Erwin2014b} estimate a ND mass of $2.7 \times 10^{10}$ \Msun\
for NGC~3945, or 36 per cent of the total stellar mass, larger than
the 29 per cent we estimate for the model at 10 Gyr using the same
method.  \citet{Erwin2014b} measure an exponential scale length of the
ND of 494 pc.  Their estimate of the ND size in terms of the
deprojected radius of peak ellipticity of the bar is $\simeq 0.15$,
comparable to NGC~4371 and smaller than in the simulation.


\subsubsection{Comparison to NGC 3368}
\label{sec:n3368}

\begin{figure*}
\centering
\begin{tabular}{c}
\includegraphics[width=0.8\hsize,angle=0]{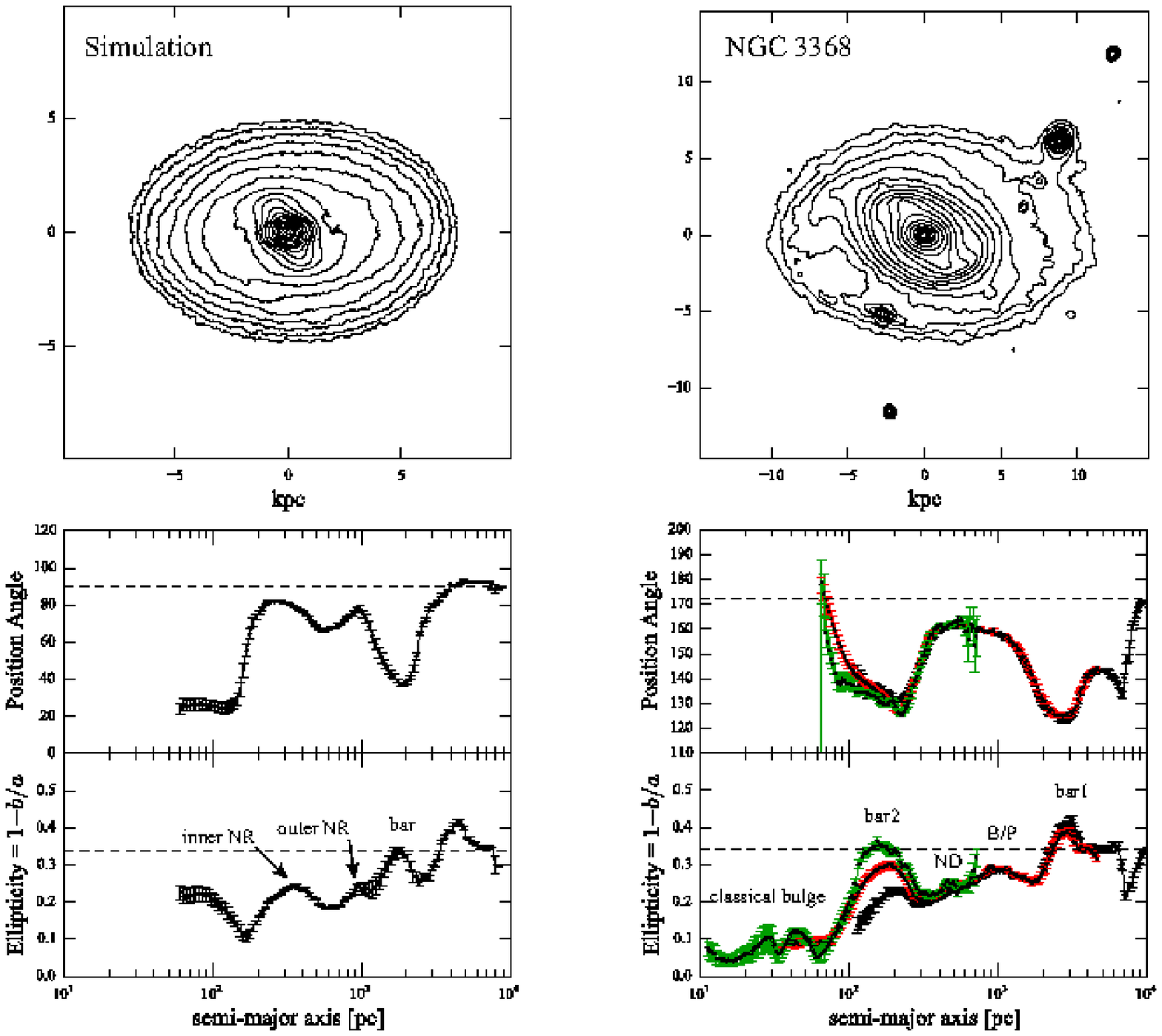}
\end{tabular}
\caption{ As for Fig.~\ref{fig:n4371}, but now comparing the
  simulation with the double-barred Sab galaxy NGC~3368. Simulation
  and galaxy orientation are: primary bar at $\Delta$PA = 67\degr,
  inclination = 50\degr.  The ellipse-fit plots use data from a
  Spitzer IRAC1 image (black), the $K$-band image of \citet{knapen03}
  (red), and an \textit{HST} NICMOS2 F160W image (green). In the
  lower-right panel, labels indicate features corresponding to the
  compact classical bulge, the secondary bar, the ND, the projected
  B/P structure of the primary bar, and the primary bar itself in
  NGC~3368 (``classical bulge'', ``bar2'', ``ND'', ``B/P'', ``
  bar1''). }
\label{fig:n3368}
\end{figure*}

NGC~3368 is a double-barred Sab galaxy in the Leo Group with a Cepheid
distance of 10.05 Mpc \citep{freedman01}. Although it is only slightly
less inclined than NGC~3945 ($i \approx 50\degr$, for an outer-disc
ellipticity of 0.34), its bar is at a more intermediate position angle
(bar $\Delta$PA = 67\degr). The plotted isophotes (upper right panel of
Fig.~\ref{fig:n3368}) are based on a Spitzer IRAC1 image from the
Local Volume Legacy \citep{dale09}; the ellipse fits are based on this
image, the $K$-band image of \citet{knapen03}, and an \textit{HST}
NICMOS F160W image \citep{martini03}.

Superficially, the trend in ellipticity interior to the primary bar in
NGC~3368 is similar to that in the simulation: a weak peak or shoulder
at $a \sim 1$~kpc plus a more distinct peak at a smaller semi-major
axis ($a \sim 150$ pc in NGC~3368 and $a \sim 300$ pc in the
simulation).  However, the underlying structures are different. As
noted above, the two peaks in the simulation's ellipticity profile are
due to the outer and inner nuclear rings of the ND. The outer
peak/shoulder in NGC~3368's ellipticity profile, on the other hand, is
due to the projected B/P structure of the bar
\citep[see][]{erwin-debattista13}, and the inner peak is due to the
secondary bar. The ND in NGC~3368 shows up in the ellipticity profile
as a slight bump in ellipticity at $a \sim 400$--500 pc, and in the
position-angle profile as the local maximum in the same semi-major
axis range\footnote{Note that the ``inner disc'' identified by
  \citet{erwin04} is actually the projected B/P structure.}.  NGC~3368
has a much more prominent B/P structure and a significantly smaller ND
(with an embedded secondary bar) than is the case for the simulation.

\citet{Erwin2014b} estimates a ND mass of $7.1 \times 10^9$ \Msun\ for
NGC~3368, or 11 per cent of the total stellar mass, about a factor of
two lower than in the model at 10 Gyr.  The size of the ND is 0.12
relative to the deprojected radius of maximum ellipticity.  As with
NGC~4371 and NGC~3945, the ND in the model is about twice as large as
in this galaxy.  The scale-length of the ND in NGC3368 is 156 pc.


\subsection{Kinematic comparisons}
\label{sec:n3945kin}

\begin{figure*}
\centering
\begin{tabular}{c}
\includegraphics[width=\hsize,angle=0]{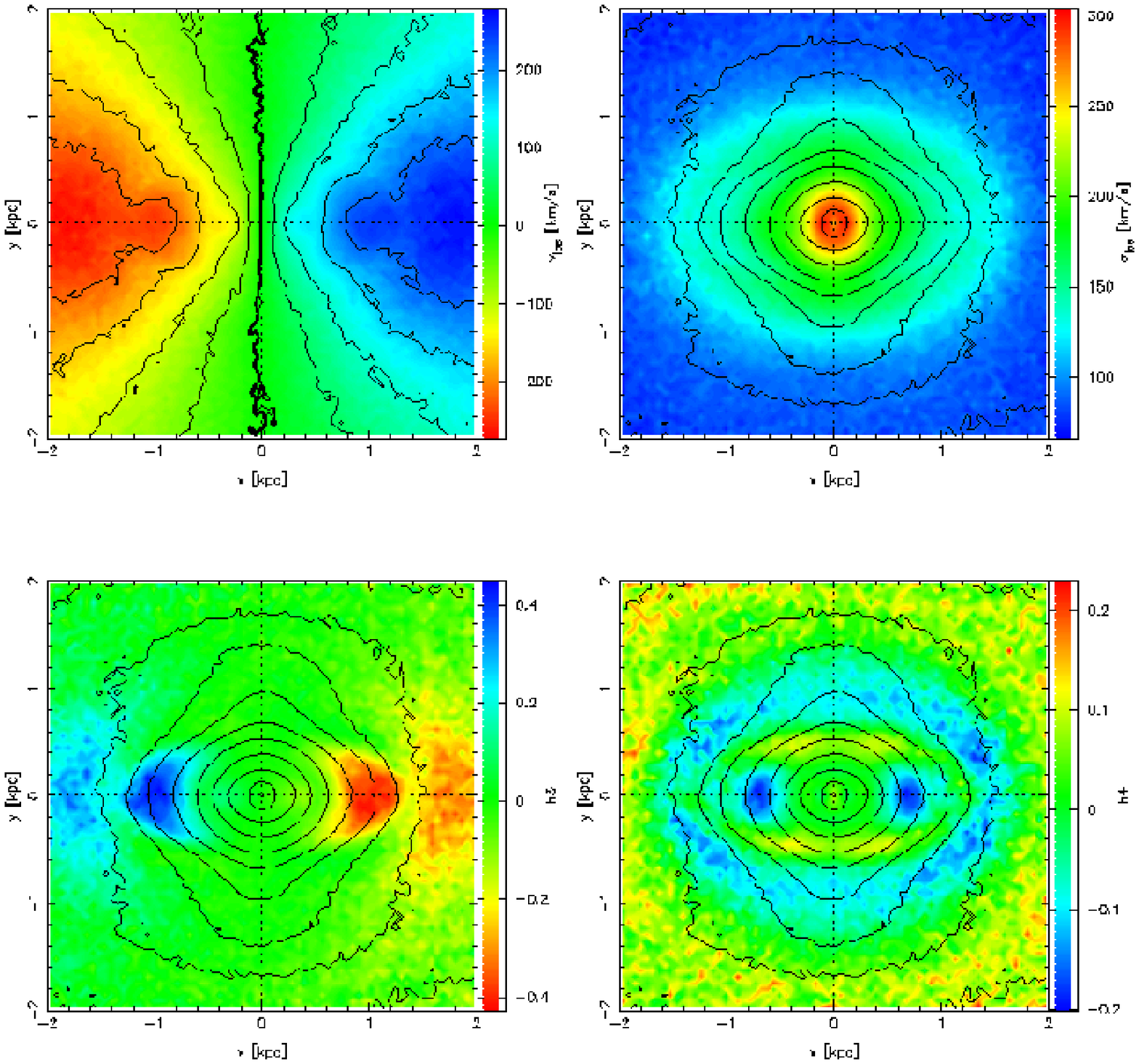} 
\end{tabular}
\caption{Line-of-sight kinematic maps for the model at 10 Gyr
  projected to the orientation of NGC 3945 (bar PA $= 88\degrees$,
  here measured from the $x$-axis, inclination $= 55\degrees$).  Top
  left is $v_{los}$, top right is $\sigma_{los}$, bottom left is h3,
  bottom right is h4, The $v_{los}$ map shows contours of isovelocity,
  with the $v_{los} =0$ shown in bold, while the remaining maps show
  contours of surface density.  The bar is nearly vertical in this
  image while the ND is almost horizontal. }
\label{fig:kin3945map}
\end{figure*}

Integral field kinematic data from the \atlastd\
\citep{Cappellari2011} are available for NGC~4371 and NGC~3945
\citep{Krajnovic2011}.  Fig.  \ref{fig:kin3945map} shows kinematic
maps ($v_{los}, \sigma_{los}$, h3, h4) for the model at 10 Gyr
projected to the same orientation as NGC~3945.  $v_{\rm los}$ has a
peak on the major axis of the ND (i.e. roughly the inclination axis).
$v_{\rm los}$ and h3 are anti-correlated in the simulation, as in the
observational data, providing further evidence for the rapidly
rotating ND.  Within $\sim 100$ pc (which is not resolved in the
model), $\sigma_{\rm los}$ is peaked but is relatively flat beyond on
the major-axis of the ND, in agreement with the observations.  On the
other hand, h4 has a distinct minimum of $\sim -0.2$ while the
observational data have h4 $\sim 0.2$.  The face-on map of Fig.
\ref{fig:h4ages}, however, has a quite strong maximum on the
major-axis of the ND, suggesting that the minimum in h4 at the
orientation of NGC~3945 is too sensitive to orientation to be a useful
diagnostic.  Indeed we find that a negative h4 in the model requires
an inclination $< 25\degrees$.

\begin{figure}
\centering
\begin{tabular}{c}
\includegraphics[width=0.55\hsize,angle=-90]{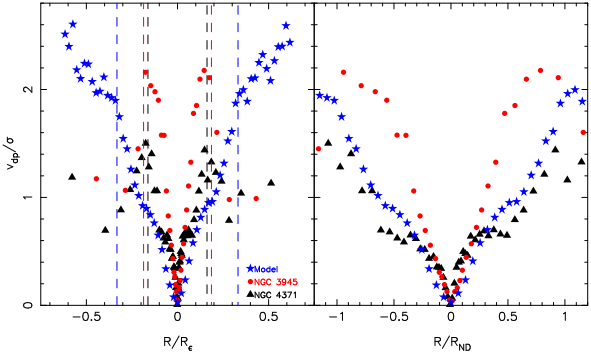} 
\end{tabular}
\caption{$v_{dp}/\sigma$ along the major-axis for NGC~3945 (red
  circles), NGC~4371 (black triangles) and the model at 10 Gyr (blue
  stars).  Left panel: Radii normalised by the the radius of the peak
  ellipticity within the bar, $R_\epsilon$.  The vertical dashed lines
  correspond to the radii of the peak ellipticities of the nuclear
  discs, $R_\mathrm{ND}$.  Right panel: Same as left panel but with
  radii normalised by $R_\mathrm{ND}$.  Real galaxy data from
  \citet{Erwin2014b} (NGC 4371) and \citet{Fabricius2012} (NGC 3945).}
\label{fig:voversigma}
\end{figure}

Following the analysis of \citet{Erwin2014b} Fig. \ref{fig:voversigma}
plots the ratio $v_{dp}/\sigma$, where $v_{dp}$ is the deprojected
velocity along the disc major-axis, and $\sigma$ is the line-of-sight
velocity dispersion.  The observational data are from
\citet{Fabricius2012} (NGC~3945) and \citet{Erwin2014b} (NGC~4371).
Data for NGC~3945 are shown as (red) circles, for NGC~4371 as (black)
triangles and for the model at 10 Gyr as (blue) stars.  The model has
been projected to bar PA $=88\degrees$ and inclination $=55\degrees$,
to match NGC~3945.  In the left panel we normalise the radius by
$R_{\epsilon}$, the deprojected radius at which the bar ellipticity
peaks ($56\arcsec =5.4$ kpc for NGC~3945, $64\arcsec = 5.2$ kpc for
NGC~4371 and $2.65$ kpc for the model).  The right panel instead
normalises the radius by $R_\mathrm{ND}$, the radius of the peak
ellipticity in the ND region.  In the left panel, the $v_{dp}/\sigma$
rises more slowly in the model than in the observations, consistent
with the ND being larger, relative to the bar, than in the
observations, but reaches values intermediate between NGC~3945 and
NGC~4371.  In all cases, $v_{dp}/\sigma > 1$ by the edge of the ND
(right panel).  The dashed vertical lines indicate the radius of peak
ellipticity in the ND region.  In the observations $v_{dp}/\sigma$
then declines again, but no such drop is present in the model although
the slope of $v_{dp}/\sigma$ is considerably shallower.  Given that
the observational data are for lenticular galaxies, while our model is
star forming, this difference between the observations and the model
can be understood as arising from discs in NGC~3945 and NGC~4371 that
become relatively hotter beyond the ND than in the model.


\section{Discussion}
\label{sec:discuss}

We have presented a simulation of an $L_*$ barred spiral galaxy with a
prominent ND.  The ND is elliptical and perpendicular to the bar.  The
disc develops after 6 Gyr, during an episode when the bar strengthens;
the ND therefore contains younger stars.

The stellar kinematics are affected by the presence of the ND,
particularly on the major axis of the ND, i.e. the minor axis of the
bar.  In this area, the mass-weighted mean tangential velocities are
dominated by the ND rotation in an elliptical disc.  The growth of the
ND forces the peak inflow and outflow radial velocities to lie closer
to the bar major axis.  These kinematics can be understood as the
superposition of the motions of (older) stars streaming along the
bar's major axis, and of the (younger) stars in a ND elongated
perpendicular to the bar.  It is also on the minor axis of the bar
that the ND most affects the mass-weighted velocity dispersions, both
radially and vertically.  The ND also alters the mass-weighted h4
kinematic moment of the vertical motion along the major axis of the
ND, producing large peaks, indicative of a peaked line-of-sight
velocity distribution.

The gas kinematics within the central 1 kpc at 6 Gyr show inflows and
outflows.  At the location of the inflows the density is in general
higher, resulting in a net accumulation of gas at the centre.  Once
the ND forms, there is a high gas density throughout the ND region.  Two
spiral arms of gas connect to the major apices of the disc where
radial inward velocities trace dense gas in the spiral arms showing
continuing feeding of the gas to the disc. By 8 Gyr the gas has
settled into the ND but shows continuing inflow along spiral arms
which connect to the gas at the apices of the ND.

The stellar chemistry also shows evidence of the continuing processes
feeding star formation. There is increasing [Fe/H] in the disc from 6
to 8 Gyr with [O/Fe] declining. The ND can be identified clearly in
maps of [Fe/H], but the [O/Fe] maps contain no distinct signature of
the ND.  Gas chemistry also shows a strong increase in [Fe/H] after 8
Gyr, reflecting the enhanced star formation there, even as low
metallicity gas feeds in from outside the central kiloparsec.

We compare the morphological and kinematic properties of the model
with data for the early-type galaxies NGC~4371, NGC~3945 and NGC~3368.
We find broad similarities.  The main difference is in the size of the
ND relative to the bar, which in the model is roughly twice as large
as in the observations.  This may possibly be due to the absence of
gas in the early-type galaxies we have compared. In order to check
this, we re-ran part of our simulation from 7 Gyr with star formation
turned off. By 9 Gyr the resulting ND is smaller relative to the
bar. Using the radius at which the bar phase deviates from a constant
by more than $10\degrees$ to measure the bar length at 9 Gyr, the ND
is 0.32 times the bar length with star formation and 0.24 without, a
25 per cent difference. When gas is absent, the bar can grow more
rapidly while the ND remains largely unchanged.  While our ND
is larger than in those galaxies, its kinematics are generally
similar.  This includes an anti-correlation between h3 and $v_{los}$
along the ND, a relatively flat $\sigma_{los}$ in the ND, and a
$v_{dp}/\sigma$ profile that reaches values intermediate to those
observed by the end of the ND.  This implies that the NDs in these
galaxies grew via gas inflows as in the simulation.  Thus the
simulation clearly demonstrates that gas inflows play an important
role in the continued assembly of structure at the nuclei of galaxies.
Although gas-rich galaxy mergers have been suggested as a mechanism
for driving gas to the centres to fuel the formation of NDs
\citep{Mayer2008, Chapon2013} our model demonstrates that they can
also form due to purely internal evolutionary processes.

\subsection{Implications for nuclear star clusters}

The morphology of our model is able to feed gas to the central 100 pc
of a galaxy and fuel ongoing star formation there. \citet{Seth2006}
find that nuclear star clusters are typically elongated in the plane
of the galaxy disc and are compound structures having a younger thin
disc embedded in an older spheroidal component. Integral Field
Spectroscopy shows rotation in the same sense as the galaxy
\citep{Seth2008}. Our model shows that the creation of such discs is a
natural consequence of secular evolution in a galaxy. Gas falls in to
the nuclear regions due to dynamical evolution of the bar which in
turn causes increased star formation.

Recent observations of the Milky Way's nuclear star cluster show that
it is significantly flattened in the same direction as the Galactic
disc \citep{Schodel2014}, supporting the idea that it has formed via
gas inflows and in-situ star formation.  Our model supports the view
that gas can be fed to such small radii to fuel ongoing episodic star
formation \citep{Pfuhl2011}. Though it is clear that the merger of
globular clusters at the centre is a plausible process for growing
nuclear star clusters \citep{Tremaine1975,
  Capuzzo-Dolcetta1993,Capuzzo-Dolcetta2008,Antonini2012,Antonini2013,Gnedin2014},
helping explain some kinematic kinematic anomalies,
\citep{Hartmann2011, DeLorenzi2013}, it is likely that in-situ star
formation plays a more significant role.


\section{Acknowledgements}

We would like to thank the referee Ronald Buta for his valuable
comments. DRC and VPD are supported by STFC Consolidated grant \#
ST/J001341/1.  RR is partially supported by Marie Curie Career
Integration Grant.  The simulation used in this study was run at the
High Performance Computer Facility of the University of Central
Lancashire.  We made use of pynbody
(https://github.com/pynbody/pynbody) in our analysis for this paper.


\bibliographystyle{mn2e1}
\bibliography{ms}{}


\end{document}